\documentclass[11pt]{article}
\usepackage{amsmath,amssymb,amsfonts,epsfig,amsfonts}
\usepackage{cite}
\usepackage{graphicx}
\usepackage{subfigure}
\usepackage{float}
\usepackage{url}
\usepackage{amsfonts,amssymb,amsmath,bm,paralist,theorem,cite,ifthen,color,amsmath}
\usepackage{theorem}
\usepackage{epsfig}
\usepackage{graphicx}
\usepackage{array}
\usepackage{caption}
\usepackage{calc}
\usepackage{longtable}
\usepackage{mathrsfs}
 \usepackage{txfonts}
\usepackage[titletoc]{appendix}   % to add "Appendix" before "A,B,..." in tabel of contents
\usepackage{sectsty}   % to change the section name
\usepackage[colorlinks,linkcolor=red,citecolor=blue,anchorcolor=green]{hyperref}
\makeatletter

 \DeclareSymbolFont{symbols}{OMS}{cmsy}{m}{n} %% to change the \mathcal font
\makeatletter

\title{ \bf  Interference alignment using finite and dependent channel extensions: the single beam case  \thanks{This work is supported in part by the National Science Foundation, grant number TF-0728676.}
} \vskip 1cm
\author{Ruoyu Sun\thanks{R. Sun is with the Department of Electrical and Computer Engineering, University of Minnesota, Minneapolis, MN 55455. Email: sunxx394@umn.edu.} $\ $ and Zhi-Quan Luo\thanks{Z.-Q. Luo is with the Chinese University of Hong Kong, Shenzhen, China. He is also affiliated with
 the Department of Electrical and Computer Engineering, University of Minnesota, Minneapolis, MN 55455.
 Email: luozq@cuhk.edu.cn. }
}
\date{ } %  July, 2013; Last revised Nov, 2014.}
\def\QED{\hfill {\bf Q.E.D.}}

\newtheorem{lemma}{Lemma}[section]
\newtheorem{proposition}{Proposition}[section]
\newtheorem{theorem}{Theorem}[section]

\newtheorem{coro}{Corollary}[section]

\newtheorem{claim}{Claim}[section]

\newcommand{\st}{{\rm s.t.}}

 \newcommand{\C}{\mathbb{C}}
\newcommand{\Cs}{\mathbb{C}^*}
\newcommand{\black}{ \color{black}}

\begin{document}
%\ninept
%
\maketitle
%
%% \tableofcontents

\begin{abstract}
% The maximum degrees of freedom (DoF) of a K-user SISO interference channel has been shown to be K/2 with an exponential number of generic channel extensions.
Vector space interference alignment (IA) is known to achieve high degrees of freedom (DoF) with infinite independent channel extensions,
but its performance is largely unknown for a finite number of possibly dependent channel extensions.
In this paper, we consider a $K$-user $M_t \times M_r$ MIMO interference channel (IC) with arbitrary number of channel extensions $T$ and arbitrary channel diversity order $L$ (i.e., each channel matrix is a generic linear combination of $L$ fixed basis matrices).
We study the maximum DoF achievable via vector space IA in the single beam case (i.e. each user sends one data stream).
We prove that the total number of users $K$ that can communicate interference-free using linear transceivers is upper bounded by $NL+N^2/4$, where $N = \min\{M_tT, M_rT \}$.
 An immediate consequence of this upper bound is that for a SISO IC the DoF in the single beam case is no more than $\min\left\{\sqrt{ \frac{5}{4}K}, L + \frac{1}{4}T\right\}$.
When the channel extensions are independent, i.e. $ L$ achieves the maximum $M_r M_t T $, we show that this maximum DoF lies in $[M_r+M_t-1, M_r+M_t]$ regardless of $T$.
Unlike the well-studied constant MIMO IC case, the main difficulty is how to deal with a hybrid system of equations (zero-forcing condition) and inequalities (full rank condition).
Our approach combines algebraic tools that deal with equations with an induction analysis that indirectly considers the inequalities.

\end{abstract}

\thispagestyle{empty}
%\newpage
%\tableofcontents
%\newpage
%\newpage
%%%%%%%%%%%%%%%%%%%%%%%%%%%%%%%%%%%%%%%%%%%%%%%%%%%%%%%%%%%%%%%%%%%%%%%%%%%%%
%%%%%%%%%%%----------------   section 1 Introduction --------%%%%%%%%%%%%%%%%
%%%%%%%%%%%%%%%%%%%%%%%%%%%%%%%%%%%%%%%%%%%%%%%%%%%%%%%%%%%%%%%%%%%%%%%%%%%%%
\section{Introduction}
%With the rapid growth of wireless data traffic, multiuser interference has become a major performance limiting factor in today's cellular wireless networks,  making proper interference management absolutely crucial in order to achieve high network throughput. Among various interference management techniques,
{\color{black}
Interference alignment (IA) has recently attracted considerable attention due to its significant potential to achieve high throughput in wireless networks.
 %to mitigate interference in wireless communication networks.
 However, many current IA schemes rely on two rather unrealistic assumptions: an exponentially  (in the number of users) many independent channel extensions and perfect channel state information at the transmitters (CSIT).
 % While the IA scheme under imperfect CSIT has been studied in various scenarios (see \cite{Jafar14TopIA} and reference therein),
  In this paper, we study IA schemes with the first assumption relaxed (while still assuming perfect CSIT). In particular, we consider the following problem: with a finite number of possibly dependent channel extensions, what is the performance limit of any IA scheme?
  }
 % the maximum DoF (Degrees of Freedom) achie

% To achieve the ${K}/{2}$ DoF in a $K$-user interference channel, C-J scheme \cite{JafarIA} relies on two assumptions:

% considered a much needed strategy

\subsection{Prior Work}
{\color{black}
Introduced in \cite{MadIA} for MIMO X channels and in \cite{JafarIA}
for interference channels, interference alignment has been shown to be able to achieve high Degrees of Freedom (DoF).
  % Original in version 6: shown to achieve maximum DoF??
%that grows linearly with the number of users under the assumption that the channel has a large number of independent channel extensions or a high diversity. However, practical wireless channels usually have a finite diversity order which in turn limits the number of available independent channel extensions. For instance, for a wireless network in which each link has a fixed number of propagation paths, the number of independent channel extensions is bounded, regardless of number of users. The goal of this paper is to investigate the effect of finite channel diversity order on the total DoFs that are achievable via vector space interference alignment.
Roughly speaking, the DoF is the first-order approximation of the network capacity in the high SNR regime
 and can be interpreted as the number of data streams that can be transmitted in an interference-free manner.}
 %Because of the difficulty in characterizing the sum capacity
% of even a $3$-user SISO interference channel (IC), people have considered the DoF as a surrogate to evaluate
% the performance of a multiuser network.
% In order to determine the optimal DoF of a network, both the outer bound and the inner bound (achievability schemes) of the DoF are needed.
 % While the DoF outer bounds are usually obtained via information theoretical investigation,
 % While it is difficult to prove nontrivial outer bounds,
 %Interference alignment strategies have provided the best known DoF inner bounds in many networks, and in some cases these inner bounds are precisely the maximum DoF as they match the outer bounds.
 % Here we review some major results on the DoF of the interference channel; see a more comprehensive review for various types of networks in \cite{JafarReview}.
 % that are obtained through much simpler methods.
For a $K$-user time-varying or frequency selective SISO interference channel (IC), Cadambe and Jafar \cite{JafarIA} constructed an asymptotic IA scheme (referred herein as C-J scheme) that achieves ${K}/{2}$ DoF, provided that the number of independent channel extensions grows exponentially in $K^2$.
This surprising result matches the outer bound proven by Host-Madsen and Nosratinia \cite{Host_outerbound}, and implies that each user can get ``half-the-cake'', with the cake representing the maximum achievable DoF in a point-to-point channel, regardless of the number of interfering users present in the system.
This result can be easily extended to a MIMO interference channel where each transmitter/receiver has the same number of antennas \cite{JafarIA}.

The C-J scheme in \cite{JafarIA} belongs to the class of \emph{vector space} IA strategies that apply linear transceivers and align the interference subspaces at each receiver
into a low dimensional space. This is in contrast to \emph{signal level} IA schemes which apply lattice codes at transmitters with the goal of aligning the codes of interference at each receiver into few lattice points. Although signal level IA schemes can theoretically achieve a higher DoF than vector space IA schemes (e.g., for constant SISO interference channel
%), vector space IA can only achieve one DoF, and reference   Host_outerbound,
\cite{RealIA} and constant MIMO interference channel \cite{GMK_MIMO}), they require unrealistic assumptions of infinite precision of the channel state information and exponentially (in terms of the number of users) large codebooks.
In this paper, we focus on vector space IA schemes and study the maximum DoF they can achieve. % if not otherwise specified.

% which may be unrealistic in practice when $K$ is large.
% DELETE 2014/10/29. How will a polynomial (in $K$) number of independent or dependent channel extensions affect the achievable DoF?
% via vector space IA
For a MIMO IC with no channel extension (which is an extreme case of finite extensions),
the references \cite{YetisIA, MeisamIA, TseFeasibility} have analyzed the achievable DoF using algebraic techniques.  In particular, the authors of \cite{YetisIA} formulated the IA condition as a polynomial system of equations with beamforming vectors being the variables, and defined the notion of a ``proper'' system for which the number of equations is no more than the number of variables in every subsystem of equations.
For the single beam case (i.e. every user transmits a single data stream), they applied the Bernstein's theorem to show that the IA condition is feasible only if it is proper, thus obtaining a DoF upper bound by counting the number of equations and number of variables (referred herein as the dimensionality counting argument).
%%% \footnote{The authors of \cite{YetisIA} omitted some important detail when applying the Bernstein theorem in the single data stream per user case; see Section \ref{sec:Bernstein's Theorem}. }.
More recently, Razaviyayn et al.\ \cite{MeisamIA} and Bresler et\ al.\ \cite{TseFeasibility} independently established the same DoF upper bound (based on the counting argument) without the restriction of one data stream per user.
For the special case that each user transmits $d$ data streams in a $M \times M$ MIMO  channel with no channel extension, their result implies that the DoF is upper bounded by $2M$ (more precisely, ${2MK}/{(K+1)}$).
Compared to the $M$ DoF that can be achieved by simply using orthogonalizing strategies,
the gain brought about by interference alignment, called ``alignment gain'' in \cite{TseFeasibility}, is upper bounded by $2$.
This is significantly smaller than the ${K}/{2}$ alignment gain for a $K$-user $M \times M$ MIMO interference channel with $O(e^{K^2})$ independent channel extensions \cite{JafarIA}. The achievability of the DoF upper bound has been established for the case $ d_k = d, \forall\ k$ and $M_k, N_k$ divisible by $d$ in \cite{MeisamIA}, and for the case $M_k = N_k = N$, $d_k =d, \forall\ k$ in \cite{TseFeasibility}, where $d_k$ is the number of data streams user $k$ transmits and $M_k,N_k$ are the number of antennas at transmitter $k$ and receiver $k$ respectively.
The issue of how to design linear transceivers that can achieve interference alignment has been considered in \cite{Raz2012-IAcomplexity}.

%%  If this conjecture can be proved, then the DoF outer bound can be derived by counting the number of equations and the number
%% of variables (we refer to as the   counting argument)

 For the case of an arbitrary number of independent time or frequency extensions, the maximum DoF has been characterized only for the case $K = 3$ \cite{JafarIA,TseDiversity}. In particular, for a $3$-user $M\times M$ MIMO interference channel, reference \cite{JafarIA} showed that the maximum DoF is ${3M}/{2}$, either with an even $M$ and no channel extension or with an odd $M$ and $2$ channel extensions. Reference \cite{TseDiversity} has further considered a $3$-user SISO interference channel with $L$ independent channel extensions for a general $L$,
and characterized the maximum DoF as a function of $L$. However, it is not clear how to extend these results to general $K$.

\subsection{Contributions of This Work}\label{sec: contribution of this work}
\subsubsection{{\color{black}Overview of Contributions}}\label{sec: overview of contribution}
%\textbf{ Remark of modification: this subsubsection I.B.1) combines previous Sec. III.B and the first two paragraphs of previous Sec. I.B, with a few sentences modified. Blue part: copy the original sentences in previous Sec. III.B without modificiation.  Red part: new modifications.}
%{\color{blue}
The existing IA results for the $K$-user IC can be roughly divided into two categories: ``optimistic'' results that achieve $O(K)$ alignment gain \cite{JafarIA,RealIA,GMK_MIMO}, and ``pessimistic'' results that bound the alignment gain by $2$ \cite{YetisIA,MeisamIA,TseFeasibility}.
The optimistic results require impractical assumptions of exponentially many independent channel extensions,
while the pessimistic results only apply to a restricted scenario with no channel extensions.
Then a problem of great interest is how much alignment gain (or DoF) can be achieved for a practical model with polynomially many and possibly dependent channel extensions.
 Little was known before about this problem: it was not even known whether the trivial DoF upper bound $K/2$ can be improved or not.
 The existing works \cite{TseFeasibility} and \cite{JafarACS} slightly improved the lower bound in two scenarios
 from $1$ to a number less than $3/2$ (they proved achievability of certain DoF values for $3$-user SISO IC, which also imply the achievability of these DoF values for $K$-user SISO IC). Still there is a huge gap between $3/2$ and the upper bound $K/2$.
% Little was known before about this problem: it was not even known whether the trivial DoF upper bound $K/2$ can be improved or not.
% }

To model the possible dependency among channel extensions, we introduce the notion of channel diversity order.
Specifically, we say an interference channel has a diversity order $L$ if each channel matrix is a generic linear combination of $L$ fixed basis matrices.
 The benefit of introducing this notion is that many seemingly different practical channel models,
such as the $L$-tap SISO IC, the SISO/MIMO IC with block-independent time or frequency extensions and the SISO/MIMO IC with asymmetric complex signalling,
 can be treated within this unified framework.
% and can be used to describe many practical channels such as
 More examples with diversity order $L$ will be given later.

%As mentioned in the introduction, it is of great interest to study how much alignment gain (or DoF) can be achieved for a practical model with finite dependent channel extensions.
% Moreover, % notice that most existing works only consider independent channel extensions, and
%very little work has studied the important practical scenario that the channel extensions are dependent, and both the trivial lower bound $1$ and
%upper bound $K/2$ are the best known bounds.

Our ultimate goal is to characterize the maximum DoF (achievable via vector space IA) of a $K$-user MIMO IC with any number of channel extensions and any channel diversity order. As an intermediate step, in this work we restrict to the single beam case, i.e. each user sends $d=1$ data stream to its intended receiver. % , and focus on the DoF upper bound.
 Under this restriction, the DoF $Kd/T$ (total DoF per channel use) becomes $K/T$, thus bounding the DoF amounts to bounding the total number of users $K$ for a given $T$.
% Consequently, our results are directly related to the following performance metric of a wireless system:
% Though motivated by the single-beam case,
Note that the number of users that can be accommodated to achieve a given QoS (Quality of Service) threshold is itself an interesting
 performance metric for a wireless system.
Our results characterize this metric in the IA context: given $T$ channel extensions (and any $L$) and the QoS requirement of $1$ aggregate DoF per-user (i.e. $1/T$ DoF per-channel-use per-user), how many users can simultaneously communicate in an interference-free manner?
%While the trivial DoF upper bound $K/2$ does not translate to
We emphasize that no upper bound for the number of supported users with general $T$ was known before (the trivial DoF upper bound $K/2$ does not translate to any bound on $K$), and our results provide the first nontrivial upper bound $K \leq O(T^2)$ for any $T$ and any $L$.

% Our results also shed light on practical questions such as:  How many users can be supported in a given area without interference via IA?

The bound on $K$ can be immediately translated to the bound on the DoF under the single-beam restriction.
 Roughly speaking, our results imply that in the single-beam case, the alignment gain (approximately equal to the DoF divided by the number of antennas per transmitter/receiver) is upper bounded by $O(\sqrt{K})$, regardless of the diversity order or the number of channel extensions. This upper bound
is much smaller than the trivial upper bound $K/2$. For the special case of independent channel extensions (i.e. the diversity order achieves the maximum), we establish
a stronger upper bound of $2$ (Theorem \ref{thm 1: DoF bound for generic extensions}).
Therefore, under the single-beam restriction we can not expect an alignment gain as high as $K/2$ (or even $K^{1/2 + \epsilon}$ where $\epsilon >0$). % , and the most optimal estimate is $O(\sqrt{K})$.
An interesting open question is whether the $O(\sqrt{K})$ DoF can be achieved in the single-beam case with finite diversity order.

% if the answer were yes, this would imply that even with the single-beam restriction IA schemes would provide substantial gains over traditional orthogonalizing schemes.
%Nevertheless, partially due to the lack of powerful IA schemes, it seems difficult to answer this question at this point.

\subsubsection{Detailed Summary of Contributions}

We  consider  a general MIMO IC with an arbitrary number of channel extensions and an arbitrary channel diversity order,
and prove upper bounds on the DoF (achievable via vector space IA) in the single beam case,
as well as the upper bounds on the number of users that can transmit interference-free.
% of a MIMO $K$-user IC with any number of channel extensions and any channel diversity order.
Our main contributions, both theoretical and technical, are summarized as follows.

% In this paper, we study the effect of finite $L$ on the achievable Degrees of Freedom (DoF) via vector space interference alignment.  %The
%We summarize our contributions as follows.

(i) For a general diversity order $L$, we establish a universal DoF upper bound,
{\black which is the first such bound for IA with dependent channel extensions.}
 Specifically, for a $K$-user $M_t \times M_r$ MIMO IC with $T$ channel extensions and channel diversity order $L$, we prove that the total number of users $K$ that can communicate interference-free using linear transceivers is upper bounded by $NL+N^2/4$, where $N = \min\{M_rT, M_tT \}$.
  We emphasize that our result applies to any number of users, any channel diversity order and any number of channel extensions.
 % and $N_r,N_t$ are the ambient transmit dimension and receive dimension respectively.
% To the best of our knowledge, this is the first result that considers
% and to the best of our knowledge, our result is the first of this kind.
 %Remarkably, our result applies to any number of channel extensions, while most previous work in interference alignment only considers either infinite channel extensions or no channel extension.
 This result immediately leads to DoF upper bounds of many practical channels in the single beam case. For instance,
% An immediate consequence of this upper bound is that
 for a $K$-user SISO IC in the single beam case, the maximum DoF is no more than  $\min\left\{\sqrt{ \frac{5}{4}K}, L + \frac{1}{4}T\right\}$, where $T$ is the number of channel extensions (not necessarily independent extensions).
 %  An interesting open question is whether the $O(\sqrt{K})$ DoF can be achieved by vector space IA.

(ii) For the extreme case that the channel extensions are independent (i.e. maximum channel diversity order),
we prove a constant DoF upper bound. % which is close to the achievable lower bound.
Specifically, for a $M_t \times M_r$ MIMO IC with $T$ independent channel extensions in the single beam case, we show that the maximum DoF lies in $[M_r+M_t-1, M_r+M_t]$ regardless of $T$. % where $M_r$ and $M_t$ are the number of receive/transmit antennas (per user) respectively.
The same DoF upper bound has been obtained in \cite{ShiIA} for SISO IC and \cite{BengIA} for MIMO IC;
 however, their results require a strong assumption on the beamformers.
 Our result generalizes the previous DoF bound of MIMO IC with no channel extension in the single beam case \cite{YetisIA}.
 Compared to the ${K}/{2}$ DoF achievable for the multi-beam case \cite{JafarIA}, our result shows that with the single-stream restriction the performance gain provided by vector space IA is very limited.
%  on the system throughput in an interference system.
 % To the best of our knowledge, our result is the first one that d
% under some additional assumptions (e.g., the beamforming solutions are independent and marginally isotropic).

(iii) Our main technical contribution is to develop an induction analysis framework that, combined with algebraic tools, can determine the feasibility of a hybrid system of equations and inequalities. We believe this framework can be of use in other IA contexts.
% a larger class of IA problems than previous techniques.
In particular, it is now well-known that the feasibility of an IA system is equivalent to the solvability of a system of polynomial equations when the direct link channel matrix is generic \cite{YetisIA, MeisamIA,TseFeasibility}.
% First, we emphasize that in general the feasibility of IA systems is not equivalent to the solvability
%of polynomial equations.
We emphasize that this equivalence no longer holds if the direct link matrices are not generic;  in fact, in the latter case, the problem is reduced to the feasibility of a hybrid system of equations and inequalities.
% For the general case that , the feasibility of an IA system is equivalent to .
Since algebraic geometry tools usually cannot deal with hybrid systems directly, other techniques are needed.
% to determine the feasibility of general IA systems
We develop an induction analysis that leverages the recursive structure of the IA system, while indirectly considers the inequalities.
To deal with the equations, we generalize algebraic tools used in the existing IA literature, which may be of independent interest.
Finally, we also provide some clarifications on the use of Bernstein's theorem in the IA context (more specifically, the so called counting argument to determine the feasibility of a system of polynomial equations).

\section{System Model}

\subsection{Channel Diversity Order}\label{subsec: channel diversity}\label{sec: channel model}
To model the possible dependency among channel extensions, we introduce the notion of channel diversity order.
 The benefit of introducing this notion is that many seemingly different practical channel models can be treated within this unified framework.

We say a $K$-user symmetric interference channel has a diversity order $L$ if each channel matrix $H_{ij}, 1\leq i,j \leq K$ is a linear combination of $L$ fixed matrices $A_1,\dots, A_L \in \mathbb{C}^{N_r \times N_t }$:
$H_{ij} = \tau_{ij}^{1} A_1 + \dots + \tau_{ij}^L A_L$, where
%% do not lie in any Zariski closed set (i.e. the set of zeros of a nonzero polynomial,
 $A_1,A_2,\dots, A_L$ are linearly independent.
We call $A_1,\dots, A_L$ the building blocks of this interference channel.
For a symmetric interference channel with $T$ time or frequency extensions where each transmitter (receiver) is equipped with $M_t$ ($M_r$) antennas, the dimensions of the channel matrices $N_t,\ N_r$ are related to $T, \ M_t, \ M_r$ through the relations $N_t = M_t T,\ N_r =  M_r T$.
% each transmit is equipped with the same number of antennas, and each receiver is equipped with the same number of antennas

We are interested in the case that the coefficients $\tau_{ij}^{l}, 1\leq i,j\leq K, 1\leq l \leq L$ are generic (e.g. independently drawn from the same or different continuous random distributions).
Strictly speaking, a property is said to hold for generic $x=(x_1,\dots, x_n) \in \mathbb{C}^n$ if there is a nonzero polynomial $f$ such that
the property holds in set $ \{x \; | \; f(x) \neq 0 \}$. In the so called Zariski topology, such a set $\{x \; | \; f(x) \neq 0\}$ is called a Zariski open set.
% , and has measure $1$.
 In other words, a property holds generically if it holds over a Zariski open set (this implies that it holds with probability one
if $x_i$'s are drawn from continuous random distributions).
%Though there is a slight difference between ``a property holds for generic $x$'' and ``a property holds for independent (continuous) random $x$ with probability one'' (or ``a property holds for almost all $x$''),
% distinguish between these two statements .

% At this point, we do not specify what the property is and just state that the channel coefficients ``are generic''.  }
% the coefficients $\tau_{i,j}^k$ are independently drawn from continuous random distributions.
% For instance, if $\tau_{ij}^{1}, 1\leq i,j\leq K, 1\leq l \leq L$ are ,
% then $\tau_{ij}^{1},\forall i,j,l$ are generic coefficients.

In order to avoid degenerate channel matrices, we impose some mild requirements on the building blocks $A_1, \dots, A_L$. Define
\begin{equation}\label{full rank guarantee}
\begin{split}
 \Psi = \{ (A_1,\dots, A_L)\mid &\ A_{\ell} \in \mathbb{C}^{N_r \times N_t }, \ell=1,\dots, L; \\
                             & \exists \;\; k_1,\dots, k_L \in \mathbb{C}, \; \mathrm{s.t.} \; \text{rank}(k_1 A_1 + k_2 A_2+\dots + k_L A_L) = \min\{ N_r, N_t\} \}.
\end{split}
 \end{equation}
{\color{black}
Then for any $(A_1,\dots, A_L) \in  \Psi$, each channel matrix $H_{ij} = \tau_{ij}^{1} A_1 + \dots + \tau_{ij}^L A_L$  is full rank for generic $(\tau_{ij}^l)_{1\leq i,j\leq K, 1\leq l \leq L}$.} % (i.e. for all $(\tau_{ij}^l)_{1\leq i,j\leq K, 1\leq l \leq L}$ in some Zariski open set).  }

%
%Suppose the number of symbol extensions (time or frequency extensions) is $T$.
%For a SISO interference channel with $T$ time or frequency extensions, the channel matrices are $T \times T$ diagonal matrices, thus $N_r = N_t = T$.  For a $M_t \times M_r $ MIMO interference channel with $T$ time or frequency extensions, the channel matrices are $M_t T \times M_r T$ block diagonal matrices, thus $N_t = M_t T, N_r = M_r T$. Several more detailed examples are given below.

We describe below several channel models with finite diversity order.

\textbf{Example 2.1}: A SISO IC with $L$ generic time or frequency extensions has a diversity order $L$. Indeed, each channel matrix
$H_{ij} = \text{diag}( H_{ij}^{(1)},\dots, H_{ij}^{(L)} )$ is a generic linear combination of $L$ diagonal matrices
$\text{diag}( 1,0,\dots,0  )$, $\text{diag}( 0,1,\dots,0  )$, $\dots,$ $\text{diag}( 0,0,\dots,1  )$, which serve as the building blocks. %.
%The building blocks are $\text{diag}( 1,0,\dots,0  )$, $\text{diag}( 0,1,\dots,0  )$, $\dots,$ $\text{diag}( 0,0,\dots,1  )$.
 Such a channel model is commonly used in the interference alignment studies \cite{JafarIA, TseDiversity}.

\textbf{Example 2.2}: A SISO IC with $L$-tap discrete channels and $N \geq L $ frequency extensions has a diversity order $L$. Indeed,
suppose the channel between all transmitter-receiver pairs $(k,j)$ is an $L$-tap channel with the same time delays
 $\lambda_1 < \lambda_2 < \dots < \lambda_L $, where $\lambda_{\ell} \in \{0,1,\dots, N-1 \}$.
The channel matrix (in the frequency domain) $H_{ij}$ is an $N \times N$ diagonal matrix, and can be expressed as $H_{ij} = \sum_{l=1}^L \tau_{ij}^l A_l $,
where $A_l = diag\left( 1, e^{\frac{2\pi \lambda_l } {N} \sqrt{-1} }, \dots, e^{\frac{2(N-1)\pi \lambda_l }{N} \sqrt{-1} }  \right) $, and $ \tau_{ij}^l $
is the $l$'th tap channel coefficient between transmitter $j$ and receiver $i$.
% generic coefficients.

\textbf{Example 2.3}: A SISO IC with $L$ blocks of generic channel extensions has a diversity order $L$.
More specifically, suppose there are $N$ channel extensions that can be divided into $L$ blocks, where the channel coefficients in the same block are equal, and are independent across blocks. This is the well-known block-fading channel model.
As a simple example, suppose $N_1 + N_2$ channel extensions are divided into two blocks with $N_1,N_2$ channel extensions respectively and the channel matrices can be expressed as
$H_{ij} = diag \left( \tau_{ij}^1 I_{N_1}, \tau_{ij}^2 I_{N_2} \right) $, then the building blocks are
$ A_1 = diag \left( I_{N_1}, 0 \right), A_2 = diag \left( 0, I_{N_2} \right)  $. In general, the building blocks are block diagonal matrices consisting of
identity matrices and zero matrices.
Such a channel model considers a more general ``coherence block structure'' than the model with constant extensions or generic extensions.
{\color{black}Special asymmetric coherence block structures have been considered in blind IA \cite{BlindIA1,BlindIA2} to achieve high DoF. For a general symmetric coherence block structure, the DoF has not been studied in the literature yet. }

%\textbf{Example 4}: MIMO interference channels. We consider three different cases.

\textbf{Example 2.4a}: A constant MIMO IC (``constant'' means flat-fading channel and no channel extension) where
 each transmitter has $M_t$ antennas and each receiver has $M_r$ antennas (we call $M_t \times M_r$ MIMO IC) has a
  diversity order $M_t M_r$. The channel matrix $H_{ij}$ is a $M_r \times M_t$ matrix with generic entries.
  The building blocks are $E_{kl}, 1\leq k \leq M_r, 1\leq l \leq M_t$, where $E_{kl}$ denotes the matrix with only one nonzero entry $1$ in the entry $(k,l)$.
Such a channel model has been studied in references \cite{MeisamIA, TseFeasibility}.
  % which implies that the alignment gain is no more than $2$.

\textbf{Example 2.4b}: An $M_t \times M_r$ MIMO IC with $T$ constant channel extensions has a diversity order $M_t M_r$.
The channel matrix $H_{ij}$ is a $T M_r \times TM_t$ block diagonal matrix
$ \begin{bmatrix}
\bar{H}_{ij} & 0            & \cdots &    0         \\
0            & \bar{H}_{ij} & \cdots &    0         \\
\vdots       & \vdots       & \ddots & \vdots       \\
0 \dots      &  0           & \cdots & \bar{H}_{ij} \\
\end{bmatrix}  , $
 where $\bar{H}_{ij} $ is a $M_r \times M_t$ matrix with generic entries.
The building blocks are block diagonal matrices $ \begin{bmatrix}
E_{k\ell} & 0        & \cdots &    0   \\
0        & E_{k\ell} & \cdots &    0   \\
\vdots   & \vdots   & \ddots & \vdots \\
0 \dots  &  0       & \cdots & E_{k\ell} \\
\end{bmatrix}  . $
     To our knowledge, such a channel model has not been considered in the interference alignment area.

\textbf{Example 2.4c}: An $M_t \times M_r$ MIMO IC with $T$ generic channel extensions has a diversity order $M_t M_r T $.
 The channel matrix $H_{ij}$ is a $M_r T \times M_t T$ block diagonal matrix $ \begin{bmatrix}
H_{ij}^1 & 0        & \cdots &    0   \\
0        & H_{ij}^2 & \cdots &    0   \\
\vdots   & \vdots   & \ddots & \vdots \\
0 \dots  &  0       & \cdots & H_{ij}^T \\
\end{bmatrix}  , $
 where each block $H_{ij}^l$ is a $M_r \times M_t$ matrix with generic entries. Such a channel model is a generalization of the SISO channel in
 Example 2.1, and has been investigated in \cite{GMK_MIMO, GouMIMO} for the case $T \rightarrow \infty$.
 %To answer the question whether spatial extension will provide high alignment gain, we need to explore the DoF for the channel models
 % in Example 2b and Example 2c

 At last, we show an example where asymmetric complex signaling (ACS) \cite{JafarACS} can double the channel diversity order.
 %It is easy to extend the analysis to other aforementioned channel models.

 \textbf{Example 2.5}: Using ACS, a SISO IC with any number of constant channel extensions has a diversity order $2$.
 %%Compared to $1$ diversity order without ACS, using ACS doubles the channel diversity.
 The idea of ACS is to convert each complex channel to two real channels, thus doubling the ambient dimension and the diversity order.
 In a point-to-point channel with input $x \in \mathbb{C}$, channel $h\in \mathbb{C}$, noise $n\in \mathbb{C}$ and output $y\in \mathbb{C}$,
 we have
\begin{align}
 y = hx + n                 \quad
\Longleftrightarrow \quad    \left( \begin{matrix} Re(y) \\ Im(y)  \end{matrix}\right) =
 \left( \begin{matrix} Re(h) & -Im(h) \\ Im(h) & Re(h)  \end{matrix}\right) \left( \begin{matrix} Re(x) \\ Im(x)  \end{matrix}\right) +
  \left( \begin{matrix} Re(n) \\ Im(n)  \end{matrix}\right) ,  \notag
\end{align}
where $Re(\cdot), Im(\cdot)$ denote the real part and the imaginary part of a complex number respectively.
In a $K$-user SISO IC with $T$ constant channel extensions, suppose the channel matrix between transmitter $j$ and receiver $k$ is a $T\times T$ diagonal matrix $\text{diag}( h_{kj},\dots, h_{kj} )$.
The channel diversity order is $1$, with the identity matrix $I$ being the building block. %%  where $\{h_{kj}\}$ are generic.
Using ACS, the channel matrix becomes a $2T \times 2T $ block diagonal matrix
$$ H_{kj} = \begin{bmatrix}
Re(h_{kj}) & -Im(h_{kj})          & \cdots &      0 & 0        \\
Im(h_{kj}) & Re(h_{kj})             & \cdots &      0 & 0        \\
\vdots & \vdots  & \ddots & \vdots & \vdots       \\
0 & 0 \dots      & \cdots &      Re(h_{kj}) & -Im(h_{kj}) \\
0 & 0 \dots      & \cdots &      Im(h_{kj}) & Re(h_{kj})
\end{bmatrix}    .$$     %% where $\bar{H}_{ij} = \left( \begin{matrix} Re(h_{kj}) & -Im(h_{kj}) \\ Im(h_{kj}) & Re(h_{kj})  \end{matrix}\right)  $.
The two basis matrices (or building blocks)   are
$$I =  \begin{bmatrix}
1 & 0           & \cdots &      0 & 0        \\
0 & 1            & \cdots &      0 & 0        \\
\vdots & \vdots  & \ddots & \vdots & \vdots       \\
0 & 0 \dots      & \cdots &      1 & 0 \\
0 & 0 \dots      & \cdots &      0 & 1
\end{bmatrix}  \quad  \mbox{  and } \quad
\begin{bmatrix}
0 & -1           & \cdots &      0 & 0        \\
1 & 0            & \cdots &      0 & 0        \\
\vdots & \vdots  & \ddots & \vdots & \vdots       \\
0 & 0 \dots      & \cdots &      0 & -1 \\
0 & 0 \dots      & \cdots &      1 & 0
\end{bmatrix}. $$
This gives a diversity order of $L=2$.
%and the coefficients corresponding to $H_{kj}$ are $Re(h_{kj})$ and $Im(h_{kj})$.
%Assume $Re(h_{kj}), Im(h_{kj}), \; k,j=1,\dots, K$ are independently drawn from continuous random distributions (such as standard Gaussian distribution for the Rayleigh fading model), then $Re(h_{kj}), Im(h_{kj}), \forall k,j$ are generic.

We remark that the notion of channel diversity order we define here is different from the number of channel extensions. Although in Example 2.1 with generic channel extensions, channel diversity order coincides with the number of channel extensions, in general this is not
the case.  For instance, the networks in Example 2.4a and Example 2.4b have the same diversity order, but different numbers of channel extensions. In Example 2.3,
the number of channel extensions is larger than the channel diversity order. In general,
channel diversity is a more limited resource than the number of channel extensions.

%{\color{black}We intend to capture the similarities between various channel models by defining the channel diversity order as the number of building blocks
%(i.e. the dimension of the space spanned by all channel matrices).
%We have derived a unified DoF upper bound for the class of channel models with a given diversity order (see Section \ref{sec: upper bound}).
%Nevertheless, in general the maximum DoF may be related to the specific building blocks. }
%% regardless of the structure of the building blocks.

\subsection{System Model and IA Condition}
Consider a $K$-user interference channel with diversity order $L$, i.e. each channel matrix $H_{ij}$
is the linear combination of $L$ fixed matrices $A_1, \dots, A_L \in \mathbb{C}^{ N_r \times N_t}$, where $A_1,\dots, A_L \in \Psi$ as defined in (\ref{full rank guarantee}).

%One motivating example is the $2$-tap-channel network with $N$ frequency extensions.
% More specifically, suppose there are $N$ frequency bands, and the time delay of the second path is $\frac{\rho}{N}$, where $\rho \in \{1,2,\dots, N-1 \}$.
% The channel matrix between the $j$th transmitter and
%the $i$th receiver can be expressed (in the frequency domain) as $H_{ij} = \tau_{ij}^1 I + \tau_{ij}^2 Z $,
% where $I \in \mathbb{C}^{N\times N}$ is the identity matrix,
%$Z = diag\left( 1, e^{\frac{2 \rho \pi}{N} \sqrt{-1} }, \dots, e^{\frac{2 \rho (N-1)\pi}{N} \sqrt{-1} }  \right)$, and $\tau_{ij}^1, \tau_{ij}^2$ are generic.
% For simplicity, we assume $N_r = N_t = N$, then $H_{ij}$ is a square matrix with dimension $N$.
In this work, we focus on vector space interference alignment strategies which deploy linear transmit and receive beamformers
 and align the subspaces of interference into a low dimensional space at each receiver.
Specifically, suppose transmitter $k$ intends to transmit $d_k$ independent data streams $s_k \in \mathbb{C}^{d_k \times 1}$ to receiver $k$.
 In the transmitter side, a linear beamforming matrix $V_k \in \mathbb{C}^{N_t \times d_k}$ is used to encode $s_k$, i.e. transmitter $k$ sends a signal $x_k = V_k s_k$.
 Receiver $k$ receives a signal $$
  y_k = H_{kk}x_k + \sum_{j\neq k} H_{kj}x_j + n_k= H_{kk} V_k s_k + \sum_{j\neq k} H_{kj} V_j s_j +  n_k ,
 $$
 where $n_k \in C^{N_r \times 1}$ is a zero mean additive Gaussian white noise.
The receiver $k$ applies a receive beamforming matrix $U_k \in \mathbb{C}^{N_r \times d_k}$ to process the received signal to obtain an estimate of the signal
\begin{equation}\label{eqn: receive signal}
 \hat{s}_k = U_k^H y_k =  U_k^H H_{kk} V_k s_k + \sum_{j\neq k} U_k^H H_{kj} V_j s_j +  U_k^H n_k .
 \end{equation}

In vector space interference alignment, we want to design beamforming matrices $\{V_k\}$ such that  %% The goal of interference alignment is to align
 all interference is aligned into a small space that is linearly independent of the signal space, thus the interference can be eliminated by multiplying
 zero-forcing matrices $\{U_k\}$.
 More formally, $\{V_k, U_k\}$ must satisfy the following IA condition \cite{JafarAlgo, YetisIA, MeisamIA, TseFeasibility}:
\begin{equation}\label{eqn: IA condition with U}
\begin{split}
& U_k^H H_{kj} V_j =0, \quad \quad \forall\ 1\leq k\neq j \leq K, \\
& \text{rank}( U_k^H H_{kk} V_k ) = d_k, \quad \forall\ 1\leq k \leq K .  \\
\end{split}
\end{equation}
% where $U_k $ is the receive beamforming matrix of receiver $k$.
The first equation implies that the interference from any user can be eliminated at the receiver side by using the zero-forcing beamforming matrix $U_k$.
The second equation ensures that no information in $s_k$ is lost when multiplied by $U_k^H H_{kk} V_k$.

We say that a DoF $\bar{d}$ is achievable by a vector space IA scheme if there exist
$V_k \in \mathbb{C}^{N_t \times d_k},\ U_k \in \mathbb{C}^{N_r \times d_k}, \ k=1,\dots, K,$ {such that}
(\ref{eqn: IA condition with U}) \text{ holds}  and $\bar{d} = {(d_1 + \dots + d_K)}/{T}$,
where $T$ is the number of channel uses.
 %The focus of this paper is on the
% DoF of the network (or sometimes called the total DoF), not the DoF per user, thus we use the term ``the DoF'' as a shorthand of ``the DoF of the network''.
%
Throughout this paper, we only consider the total DoF (or DoF for short) that is achievable by a vector space IA scheme.
%In this work, we will study the maximum DoF of a $K$-user interference channel with arbitrary diversity order and arbitrary number of channel extensions.

%%%%%%%%%%%%%%%%%%%%%%%%%%%%%%%%%%%%%%%%%%%%%%%%%%%%%%%%%%%%%%%%%%%%%%%%%%%%%%%%%%%%%%%%%%%%%%%%%%%%%%%%
\section{Main Results}\label{sec: upper bound}
%%It is well known that the maximum $\text{DoF}$ is $\frac{K}{2}$ when $L \rightarrow \infty$ \cite{JafarIA}. This result indicates that
%%each user can get ``half the cake'' regardless of the number of total users (the cake represents the DoF achievable in a point-to-point
%%channel), which is a great improvement over standard orthogonalization techniques where each user only gets $\frac{1}{K}$ of the cake.
%%However, the techniques used in \cite{JafarIA} only show that $\text{DoF} = O(K)$ is achievable when the channel diversity and the number of
%%channel extensions are $L = N = O(e^{K^2})$. It is thus interesting to study how many DoF can be achieved (by vector space IA)
%%when the channel diversity and/or the number of channel extensions is not too large.

Consider a symmetric interference channel with a given diversity order. In this section, we present several DoF bounds for interference channels in which each user transmits a single data stream (i.e. single-beam case, $d_k=1,\forall k$),
as well as the bounds on the number of users that can transmit interference-free using linear transceivers.

% \subsection{Main Results and Implications}
The first result shows that in a $M_t \times M_r$ MIMO IC with generic channel extensions (in this case the channel diversity order achieves the maximum), the DoF is upper bounded by
$ M_r + M_t  $ in the single beam case. As a special case, the DoF of a SISO IC in the single beam case is upper bounded by $2$, regardless of the number of generic channel extensions. The proof of this result will be provided in Section \ref{sec: proof of Thm 1, generic extension}.

\begin{theorem}\label{thm 1: DoF bound for generic extensions}
Consider a $K$-user $M_t \times M_r$ MIMO interference channel with $T$ independent channel extensions (all the channel coefficients are independently drawn from continuous random distributions).
Suppose each user transmits single data stream. Then the following results hold with probability one.
% The following results hold for for almost all channel coefficients.
%% A necessary condition for the DoF tuple $(d_1,\dots, d_K) = (1,1,\dots, 1) $ to be achievable via vector space IA is  and $d_k=1$ for all $k$.
\begin{enumerate}
\item [\text{(a)}] If a vector space IA scheme exists, then
 \begin{equation}\label{eqn in thm: thm 1, MIMO}
 K \leq (M_r + M_t)T -1.
\end{equation}
% In addition,
\item [\text{(b)}] If $K = (M_r + M_t - 1)T,$ then a vector space IA scheme exists. % for almost all channel coefficients.
% Therefore,
\item [\text{(c)}]  The maximum DoF achievable (via vector space IA) in this single beam case is no more than $M_r + M_t$, and no less than $ M_r + M_t - 1$.
\end{enumerate}
\end{theorem}

% {\color{red} Revise note: Upon one reviewer's request. Previously: for almost all channels; now: for independent channels, w.p. 1.  }

 {\color{black}
Remark 1: The bound \eqref{eqn in thm: thm 1, MIMO} can be slightly improved to $K \leq (M_r + M_t)T -2$ if $T \geq 2 $, by using the techniques of \cite{ShiIA,BengIA}.
Since we are mainly interested in the asymptotic DoF bound in this work (as discussed in Section \ref{sec: contribution of this work}), we choose to just present \eqref{eqn in thm: thm 1, MIMO}.
}
% and the improved bound on $K$ does not change the DoF bound in part (c)
% Nevertheless, if one intends to find the maximal $K$, the improved bound

 % and we only focus on the DoF

%Remark: Theorem \ref{thm 1: DoF bound for generic extensions} states that some property (this ``property" includes the three statements (a),(b),(c)) holds for almost all channel coefficients, i.e. for all channel coefficients that lie in a set with measure one.
%%-----------------------Delete----------------
% Theorem \ref{thm 1: DoF bound for generic extensions} is equivalent to the following result: if the channel coefficients are independently drawn from continuous random distributions, then the properties (a), (b) and (c) hold with probability one.
 %------------------------------------------------
 {\color{black}
Remark 2: By a variant of our proof, it is possible to show a slightly stronger result that the same properties in fact hold for generic channel coefficients (i.e. for all channel coefficients in a Zariski open set).
Similarly, Theorem \ref{thm: main result, bound on K} can be shown to hold for ``generic'' coefficients.
}
 % can be easily modified

%Remark: Theorem \ref{thm 1: DoF bound for generic extensions} states that some property (this ``property" includes all statements after ``$T$ generic extensions'') holds for generic channel coefficients, i.e. this property holds for all channel coefficients in a Zariski open set.
%In Section \ref{sec: proof of Thm 1, generic extension}, we will prove a slightly weaker result that this property holds almost surely, i.e. this property holds for all channel coefficients that lie in a set with measure one.
%Thus for the case

 Notice that
part (c) of Theorem \ref{thm 1: DoF bound for generic extensions} is a direct consequence of part (a) and part (b) since for the single-beam case the DoF equals $K/T$.
Part (b) of Theorem \ref{thm 1: DoF bound for generic extensions} can be proved by the result in \cite{MeisamIA}.
In fact, \cite[Theorem 2]{MeisamIA} implies that when $T=1$ and $K=M_r + M_t -1$, the DoF tuple $(1,\dots, 1)$ is achievable (since the number of transmit/receive antennas are divisible by $d=1$).
For general $T$, we can use an orthogonalizing scheme to achieve $(1,\dots, 1)$ for $K = (M_r + M_t -1)T$ users:
for each of the $T$ channel uses, ($M_r + M_t -1$) users transmit a single data stream interference-free while other users do not transmit; in total, $K = (M_r + M_t -1)T$ users can transmit a single data stream interference-free. It remains to prove part (a) of Theorem \ref{thm 1: DoF bound for generic extensions} and Section \ref{sec: proof of Thm 1, generic extension} is devoted to such a proof.

{\color{black}
Part (a) of Theorem \ref{thm 1: DoF bound for generic extensions}, although seems simple, is not easy to prove.
It has been studied in \cite{ShiIA} for SISO IC and \cite{BengIA} for MIMO IC under strong assumptions (e.g., the beamforming solutions are independent and marginally isotropic). Essentially their results require the assumption that the beamforming solutions do not have any zero entries,
which is an artificial assumption being added so that the proof in \cite{YetisIA} can be directly applied.
% Under this assumption, the proof of part (a), as adopted in \cite{ShiIA,BengIA}, is a simple extension of the proof in \cite{YetisIA}.
However, as shown later, the difficulty of proving part (a) is exactly due to the possibility
  that the beamforming solutions may contain zero entries. The new technique developed herein to tackle this difficulty is one of the main contributions of this paper. It is also crucial to prove other results of this paper. % Theorem \ref{thm: main result, bound on K}.
   }
    %However, both proofs are problematic because of the inappropriate use of Bernstein's theorem. The source of this confusion can be attributed to the dimensionality counting argument first used in \cite{YetisIA}. We will clarify these confusions in Section \ref{subsec: correct form of Bernstein thm}.  }
%%While the proof in \cite{YetisIA} can be easily corrected, a rigourous proof to Theorem \ref{thm 1: DoF bound for generic extensions} is much more
%%complicated than the argument used in \cite{ShiIA,BengIA}.

Theorem \ref{thm 1: DoF bound for generic extensions} states that for the single-beam case, the maximum DoF lies in the region $[M_r + M_t -1, M_r + M_t]$
for any $T$. For the special case $T=1$ (i.e. constant MIMO IC), this result has been proven in \cite{MeisamIA, TseFeasibility}, and \cite{MeisamIA} shows that the maximal DoF is indeed $M_r + M_t -1 $.
 Therefore, for the single-beam case, increasing the number of generic channel extensions does not lead to a significant improvement of the DoF.
%%%% the number of generic channel extensions has little effect on the DoF of a MIMO IC.

%In a SISO IC ($M_t = M_r = 1$) with arbitrary number of generic extensions,
When $M_t=M_r = M$, Theorem \ref{thm 1: DoF bound for generic extensions} implies that the DoF is no more than $2M$ for the single-beam case. This result is in sharp contrast to $\frac{K}{2}M$ DoF in a $M \times M$ MIMO IC when each user is allowed to transmit multiple data streams \cite{JafarIA}. Therefore, the single data stream restriction appears to be a throughput limiting factor even in the presence of an arbitrary number of generic channel extensions.
It would be interesting to find how the DoF scales with the number of data streams that each user can transmit.

The second result considers a general channel model with any
channel diversity order and any number of users, antennas and channel extensions. We provide a universal upper bound on the number of users that
can be accommodated to achieve IA.
This upper bound is a function of the diversity order $L$ and the minimum of the transmit dimension and the receive dimension $\min\{ N_t, N_r \}$.
The proof of this result will be presented in Section \ref{sec: proof of main result, diversity L bound}.
\begin{theorem}\label{thm: main result, bound on K}
Consider a $K$-user $M_t \times M_r$ MIMO interference channel with $T$ channel extensions {\color{black}(possibly dependent extensions)}. Let $N_t = M_t T, N_r = M_r T$ and $$N = \min \{N_t, N_r \}.$$
Suppose the channel diversity order is $L$, i.e.\ the channel matrices $ H_{ij} = \tau_{ij}^1 A_1 + \dots + \tau_{ij}^L A_L  \in \mathbb{C}^{N_r \times N_t}$,
 with $(A_1,\dots, A_L) \in \Psi $ defined in (\ref{full rank guarantee}). Then
 for almost all coefficients $\left( \tau_{ij}^l \right)_{1\leq i,j \leq K, 1\leq l \leq L}$, the DoF tuple $(d_1,\dots, d_K) = (1,1,\dots, 1) $ is achievable via a vector space IA scheme only if
 \begin{equation}\label{thm 2's bound}
 K \leq N L + \frac{N^2 }{4 }.
\end{equation}

\end{theorem}

% {\color{black}
Theorem \ref{thm: main result, bound on K} gives an upper bound on $K_{\rm max}$, where $K_{\rm max}$ is the maximal number of users that can be accommodated to achieve IA for almost all coefficients $\left( \tau_{ij}^l \right)$.
Note that it was not even known before whether an upper bound on $K_{\rm max}$ exists; in other words, we have improved the bound of $K_{\rm max}$
from $K_{\rm max} \leq \infty$ to $K \leq N L + \frac{N^2 }{4 }.$

 How tight is our bound \eqref{thm 2's bound}?
A trivial lower bound that can be achieved by any channel model is $K_{\rm max} \geq N$, simply using orthogonizing schemes.
Orthogonizing schemes can achieve better lower bound
 \begin{equation}\label{lower bound of K}
  K_{\mathrm{max}} \geq NL
   \end{equation}
  for some channel models, such as $1 \times M_r $ SIMO IC with $T$ constant channel extensions ($N_t = T ,N_r = T M_r$ and $N=T, L=M_r$ for this setting).
 For this channel model, when $T = M_r$, i.e. the number of channel extensions equals the number of receive antennas per-user,
 the lower bound $NL = M_r^2$ is of the same order as the upper bound $NL + \frac{N^2}{4} = \frac{5}{4}M_r^2$.
 In other words, there exists a very special case in which the bound \eqref{thm 2's bound} is almost optimal.
 The bound \eqref{thm 2's bound} can be very loose in some cases, such as the SISO IC with indepedent channel extensions.
 In this case Theorem 3.1 establishes a bound $K \leq 2T-1$, while the bound \eqref{thm 2's bound} $K \leq 2T + \frac{T^2}{4}$ is much looser.
 In general, there is a gap of $N^2/4$ between the bound \eqref{thm 2's bound} and the lower bound for a special case \eqref{lower bound of K}
(for general channel models whereby only a lower bound $K_{\rm max} \geq N$ is known, the gap increases to $N^2/4 + (L-1)N$).
Nevertheless, since the lower bound is only achieved by simple orthogonizing schemes, it is possible that more sophisticated IA schemes can achieve
better lower bounds. Further evaluation of the tightness of our bound \eqref{thm 2's bound} is left as future work.

Theorem \ref{thm: main result, bound on K} can be used to derive DoF upper bounds of various channel models for the single-beam case.
We first consider the SISO IC, and provide two upper bounds: one in terms of $K$ and another in terms of $T$ and $L$.

\begin{coro}\label{coro: dof bound for SISO}
Consider a $K$-user SISO interference channel with a diversity order $L$ and $T$ channel extensions.
If $d_k=1,\forall k$, then the maximum DoF achievable via a vector space IA scheme is at most $\min\left\{\sqrt{ \frac{5}{4}K}, L + \frac{1}{4}T \right\}$.
\end{coro}

\emph{Proof of Corollary \ref{coro: dof bound for SISO}}: In the SISO IC, the transmit dimension $N_t$ and receive dimension $N_r$ are both equal to $T$,
thus $N = \min\{N_t, N_r \} =  T$.
 According to Theorem \ref{thm: main result, bound on K}, we
have
\begin{equation}\label{bound K in Coro 1}
 K \leq T L + \frac{T^2}{4} ,
 \end{equation}
thus the DoF for the single-beam case $\mathrm{DoF_s}$ can be bounded as
$$ \mathrm{DoF_s} = \frac{ K d }{T} = \frac{K}{T} \leq L + \frac{1}{4}T . $$

Since the channel matrices are $T \times T$ diagonal matrices, the diversity order $L $ should be no more than $T$. Then (\ref{bound K in Coro 1}) implies
% Furthermore, using the fact $L \leq T$, we have
$$ K \leq T^2 + \frac{T^2}{4} = \frac{5}{4} T^2 .$$ Thus the DoF for the single-beam case
$$ \mathrm{DoF_s} = \frac{K}{T} \leq \sqrt{ \frac{5}{4} K}.  $$
\QED

Corollary \ref{coro: dof bound for SISO} applies to Example 2.1-2.3 in Section \ref{sec: channel model}. Specifically, the achievable DoF is upper bounded by
 $\min\left\{\sqrt{ \frac{5}{4}K}, L + \frac{1}{4}T \right\}$ for the single-beam case in the following networks: the SISO IC with $T=L$ generic time or frequency extensions (Example 2.1), $L$-tap SISO IC with $T$ frequency extensions (Example 2.2), the SISO IC with $L$ generic blocks of $T$ channel extensions (Example 2.3).
 {\color{black}
 For Example 2.1, this bound is weaker than the bound of $2$ provided by Theorem \ref{thm 1: DoF bound for generic extensions}.
 Nevertheless, for Example 2.2 and Example 2.3, this is the first general DoF upper bound to our knowledge.
  % Therefore, Corollary \ref{coro: dof bound for SISO} just provides a loose upper bound.
}

Compared with the ${K}/{2}$ achievable DoF for a SISO IC in the multi-beam case, Corollary \ref{coro: dof bound for SISO} proves a $O(\sqrt{K})$ upper bound in a SISO IC  for the single-beam case, regardless of the channel diversity order or the number of channel extensions.
Corollary \ref{coro: dof bound for SISO} also shows that the DoF for the single-beam case is bounded by $L + \frac{1}{4}T $, which increases linearly both with $L$ and $T$.
% When the $T$ channel extensions are generic, Theorem \ref{thm 1: DoF bound for generic extensions} proves a stronger result that the DoF is bounded by $2$.
When the channel extensions are not generic, it is not known whether the $O(\sqrt{K})$ bound or the $L + \frac{1}{4}T$ bound can be improved to a constant bound.

Applying Theorem \ref{thm: main result, bound on K} to the MIMO IC also obtains a $O(\sqrt{K})$ DoF bound in the single beam case.
Specifically, for an $M_t \times M_r$ MIMO IC with $T$ possibly dependent channel extensions, the channel diversity order $L \leq M_r M_t T$.
Assume $M_t \leq M_r$, then $N = M_t T$, and (\ref{thm 2's bound}) becomes
$$ K \leq NL + \frac{ N^2 }{4} = M_t^2 M_r T^2 + \frac{ 1}{4} M_t^2 T^2 \leq \frac{5}{4}M_t^2 M_r T^2,$$
which further implies that the DoF in the single-beam case $\mathrm{DoF_s}$ can be bounded as
$$
 \mathrm{DoF_s} = \frac{K}{T} \leq M_t \sqrt{\frac{5}{4}M_r K}.
$$
 This bound is possibly loose; in the special case with no channel extension (i.e. constant MIMO IC in Example 2.4a), this bound is weaker than the constant DoF bound (see \cite{YetisIA,MeisamIA,TseFeasibility} or Theorem \ref{thm 1: DoF bound for generic extensions}).
  Nevertheless, it provides the first nontrivial DoF upper bound both for Example 2.4b (MIMO IC with constant channel extensions), and for Example 2.4c (MIMO IC with an arbitrary number generic extensions) with a general $K$.

\section{Algebraic Tools } % Solvability of Polynomial Equations and }
\label{sec:Bernstein's Theorem}
{\color{black}
The zero-forcing conditions for interference alignment can be interpreted as a system of polynomial equations with the beamforming matrices being the variables.
In this section, we present some technical tools from algebraic geometry related to the solvability of polynomial systems.
% that provide conditions for
%will be used in the proofs of Theorem \ref{thm 1: DoF bound for generic extensions} and \ref{thm 2's bound}.
% Thus the solvability of a system of polynomial equations is important for showing the feasibility of IA systems.

}
% the existence of an interference alignment scheme.
 % In this section, we clarify some misconceptions related to the dimensionality counting argument.

\subsection{Review of Field Theory and a Useful Lemma }\label{subsec: prelim}
 {\color{black}
 The goal of this subsection is to prove Lemma \ref{lemma: rational count} that determines the infeasibility of a class of polynomial systems.
 Lemma \ref{lemma: rational count} will be used in the proof of Theorem \ref{thm: main result, bound on K} in Section \ref{sec: proof of main result, diversity L bound}.
To derive this Lemma, we first review the theory of transcendence degree (see Chapter 1 and Chapter 8 of \cite{Field} and \cite{MeisamIA}).
}

% can be used to establish
Let $\Omega/F$ be a field extension, i.e. ${F}$ and $\Omega$ are two fields such that $F\subseteq \Omega$. Denote $F[x_1,\dots, x_n]$ as the
polynomial ring which consists of all polynomials in variables $x_1, \dots, x_n$ with coefficients in $F$.
 We say $\alpha_1, \dots, \alpha_n \in \Omega$ are
\emph{algebraically independent} over $F$ if the only polynomial $P$ in $F[x_1,\dots, x_n]$ that satisfies $P(\alpha_1, \dots, \alpha_n) = 0$
is $P = 0$. Otherwise, we say $\alpha_1, \dots, \alpha_n $ are \emph{algebraically dependent} (i.e. there exists a nonzero polynomial $P$ such
that $P(\alpha_1, \dots, \alpha_n) = 0$). The notion of algebraically independence/dependence is analogous to linear independence/dependence
in linear algebra.
% For the special case $n=1$, an element $\alpha \in \Omega$ is said to be algebraic over $F$ if $\alpha$ is algebraically dependent over $F$;
% if $\alpha$ is algebraically independent over $F$, we say $\alpha$ is transcendent over $F$.

\textbf{Example 4.1}: Consider the field extension $\mathbb{R}/\mathbb{Q}$, where $\mathbb{R}$ is the field of real numbers, and
$\mathbb{Q}$ is the field of rational numbers. For any rational number $q$,
$\pi$ and $q$ are algebraically independent over $\mathbb{Q}$ since $\pi$ is not the root
of any polynomial with rational coefficients. On the other hand,
$\pi$ and $2\sqrt{\pi} + 1$ are algebraically dependent over $\mathbb{Q}$ because $P(\pi, 2\sqrt{\pi} + 1 ) = 0$, where
 $P(z_1, z_2) = 4 z_1 - (z_2 - 1)^2$.

\textbf{Example 4.2}: Consider the field extension $\mathbb{C}(x_1, x_2)/\mathbb{C}$, where $\mathbb{C}(x_1, x_2)$ is
 the field of rational functions in variables $(x_1, x_2)$ with complex coefficients.
 A rational function has the form ${f_1}/{f_2}$, where $f_1, f_2$ are polynomials and $f_2 \neq 0$.
    Note that the field of rational functions $\mathbb{C}(x_1,x_2)$ is different from the
 ring of polynomials $\mathbb{C}[x_1, x_2]$. The three rational functions
 $$ g_1(x_1,x_2) = \frac{x_2^2}{x_1+1}, \quad g_2(x_1,x_2) = x_1, \quad g_3(x_1,x_2) = x_1 x_2  $$
 are algebraically dependent over $\mathbb{C}$. This is because $P( g_1, g_2, g_3 ) = 0 $
 where $P(z_1, z_2, z_3) = z_1 z_2^2( z_2 + 1) - z_3^2$.

% In linear algebra, we say a vector $x \in \mathbb{C}^m$ is linearly dependent on $A \subset \mathbb{C}^m$ if $x$ can be expressed as a linear combination
% of a finite number of vectors in $A$, or equivalently, there exists
% $y_1, \dots, y_n \in A,  \lambda_1 \dots, \lambda_n \in \mathbb{C}, \lambda \in \mathbb{C}\\{0 \}$ such
% that $\lambda_0 x + \sum_{i=1}^n \lambda_i y_i = 0$.
%For any subset $A \subseteq \Omega$, let $\mathcal{F}(A) $ denote the field of rational functions $P(x_1, \dots, x_n),$ where the variables $x_1,x_2,\dots, x_n \in A$
% and the coefficients are drawn from $\mathcal{F}$.
We say $\alpha \in \Omega$ is \emph{algebraically dependent} on $A \subseteq \Omega$ if there exists $\beta_1, \dots, \beta_m \in A$ and a polynomial
$P(x_1, \dots, x_m, y) $ such that $P( \beta_1, \dots, \beta_m ,\alpha ) = 0$ (i.e. $ \beta_1, \dots, \beta_m, \alpha$ are algebraically dependent)
and $P( \beta_1, \dots, \beta_m, y ) $ is not a zero polynomial in the variable $y$. We say a set $B$ is algebraically dependent on $A$ if every element of
$B$ is algebraically dependent on $A$. Note that the statement ``$B$ is algebraically dependent on $A$'' is analogous to ``$B$ can be spanned by $A$'' in
linear algebra.

In linear algebra, a basis of a linear space is defined as a minimum set of vectors that can span all vectors in the linear space.
In field theory, the \emph{transcendence basis} is defined as follows: we say $A$ is a transcendental basis for the filed extension $\Omega/F$
 if $A \in \Omega$ is an algebraically independent set such that $\Omega$ is algebraically dependent on $A$.
 The two definitions are both consistent with the notion of ``minimum spanning set'': a ``basis'' should be
  a minimum set of elements that can ``span'' all elements.

 It can be shown that the transcendence basis exists and any two transcendence bases have the same cardinality (possibly infinite).
Based on these results, we can define the \emph{transcendence degree} of $\Omega/F$ as the cardinality of a transcendence basis for $\Omega/F$.

\textbf{Example 4.3}: Denote $\mathbb{C}(x_1, \dots, x_n)  $ as the field of rational functions in variables $x_1, \dots, x_n$ with coefficients from $\mathbb{C}$.
It can be shown that $g_1(x_1,\dots, x_n) = x_1,  \dots, g_n(x_1, \dots, x_n) = x_n$ is a transcendence basis of  $\mathbb{C}(x_1, \dots, x_n)/\mathbb{C}  $.
Hence the transcendence degree of $\mathbb{C}(x_1, \dots, x_n)/\mathbb{C}  $ is $n$.

The correspondences between linear algebra and the theory of transcendence degree are summarized in the following table \cite{Field}:

\begin{center}
\begin{tabular}{|l|l|}
  \hline
  Linear algebra & Transcendence \\
  \hline
  linearly independent & algebraically independent \\
  \hline
  A can be spanned by B &  A is algebraically dependent on B \\
  \hline
  basis & transcendence basis \\
  \hline
  dimension & transcendence degree \\
  \hline
\end{tabular}
\end{center}

% any maximal linearly independent set of vectors is a basis. A corollary is that
%any maximal algebraically independent set $A$ is a transcendence basis
%(sometimes used as the definition of the transcendence basis).
%A corollary is that
In linear algebra, it can be shown that any $n+1$ vectors in a $n$-dimensional linear space
are linearly dependent.
In field theory, a similar result holds:
for a field extension $\Omega/ {F}$ with transcendence degree $n$, any $n +1$ elements in $\Omega$ are algebraically dependent
over $ {F}$ (a simple proof is to use Theorem 8.5 in \cite{Field}). The result for the case that $ {F} = \mathbb{C}, \Omega = \mathbb{C}(x_1, \dots, x_n)$ is useful for our problem, and
we restate it in the following proposition.
% if $A$ is maximal, i.e. $A$ is not properly contained in any other algebraically independent set
\begin{proposition}\label{prop: n+1 rational is dep}
Consider the field extension $ \mathbb{C}(x_1,\dots, x_n)/\mathbb{C} $. Any $n+1$ rational functions
$$ g_1(x_1,\dots,x_n), \dots, g_{n+1}(x_1,\dots, x_n)$$
are algebraically dependent, i.e. there exists a nonzero polynomial $P \in \mathbb{C}[z_1, \dots, z_{n+1}]  $ such that
$ P(g_1,\dots, g_{n+1}) = 0$, or equivalently, $P\left( g_1(x_1,\dots, x_n), \dots, g_{n+1}(x_1, \dots, x_n) \right) = 0, \forall\ x_1, \dots, x_n$.
\end{proposition}

In Example $6.2$, we have shown that the three rational functions $g_1, g_2, g_3 \in \mathbb{C}(x_1, x_2)$ are algebraically dependent.
Proposition \ref{prop: n+1 rational is dep} implies that any three rational functions in $x_1, x_2$ are algebraically dependent.

Proposition \ref{prop: n+1 rational is dep} is more general than Example 4 in Section III of \cite{MeisamIA}, which states that
any $n+1$ polynomial functions in $\mathbb{C}[x_1,\dots, x_n]$ are algebraically dependent.
Reference \cite{MeisamIA} used this fact to show that for a certain type of polynomial
systems of equations, a necessary condition for the solvability is that the number of variables is no less than the number of equations.
Using the more general result Proposition \ref{prop: n+1 rational is dep}, we show below that
the dimensionality counting argument works for a broader class of systems.   % of polynomial equations.
%In fact, we can prove the following result, which states that the number of variables is no less than the number of some special equations.

\begin{lemma}\label{lemma: rational count}
Consider a system of polynomial equations and inequalities
\begin{subequations}\label{eqn: poly system, rational func}
\begin{align}
\alpha_i f_i( x) + g_i( x) = 0, & \quad i=1,\dots, m,  \label{part a of eqn: poly system, rat}  \\
f_i( x) \neq 0 , & \quad i=1,\dots, m,  \label{part b of eqn: poly system, rat}
\end{align}
\end{subequations}
where $ \alpha = (\alpha_1, \dots, \alpha_m) $ is a parameter vector,
 $x = (x_1,\dots, x_n)$ is the variable and $f_i, g_i,i=1,\dots,m$ are polynomial functions of $x$.
If for a positive measure of $\alpha$, the above polynomial system (\ref{eqn: poly system, rational func}) has a solution $x$,
then the number of equations cannot exceed the number of variables, i.e. $m \leq n$.
\end{lemma}
% If a positive measure of $(\alpha, \beta)$ satisfy the following property for some $\bar{m} \leq m$ :
% there exists a solution $x$ to the system of equations (\ref{eqn: poly system, rational func})
% such that the cardinality of set $ \{ i | f_i(\beta, x) \neq 0 \}$ is greater than or equal to $ \bar{m} $, then $\bar{m} \leq n$.

Remark: Lemma \ref{lemma: rational count} can be stated in a slightly different way: for generic $\alpha$, a necessary condition for \eqref{eqn: poly system, rational func} to be solvable is $m \leq n$.

Besides the polynomial equations in (\ref{part a of eqn: poly system, rat}), Lemma \ref{lemma: rational count} adds an requirement that the coefficients of $\alpha_i$ are nonzero. This requirement guarantees that $\alpha_i$ can be expressed as a rational function
$- {g_i(x)}/{f_i(x)}$ if $x$ is a solution to (\ref{eqn: poly system, rational func})
((\ref{part b of eqn: poly system, rat}) ensures that these rational functions are well defined).
In the special case of $f_i=1$ for all $i$, Lemma \ref{lemma: rational count} reduces to the key result proved in \cite{MeisamIA}.
%\begin{coro}\label{lemma: Meisam count}
%If for a positive measure of $\alpha = (\alpha_1, \dots, \alpha_m)\in \mathbb{C}^{m}$, the system of polynomial equations
%\begin{equation}\label{eqn: poly system, poly func}
%\alpha_i + g_i(x) = 0, \; i=1,2,\dots, m,
%\end{equation}
%has a solution $x \in \mathbb{C}^n$, then $ m\leq n $.
%\end{coro}
The difference of Lemma \ref{lemma: rational count} and the corresponding result in
 \cite{MeisamIA} only lies in how $\alpha_i$ is expressed:
 $\alpha_i$ equals  a rational function in the former case and a polynomial function in the latter case.
The proof of Lemma \ref{lemma: rational count} is given below.

\noindent
\emph{Proof of Lemma \ref{lemma: rational count}:}
We prove by contradiction. Assume the contrary that $m > n$.
Denote $\mathcal{I}$ as the set of $\alpha$ such that (\ref{eqn: poly system, rational func}) has a solution $x$.
The assumption of Lemma \ref{lemma: rational count} is that $\mathcal{I}$ has a positive measure.
By Proposition \ref{prop: n+1 rational is dep}, the $m$ rational functions $- { g_i( x )}/{ f_i( x ) }, i=1,\dots, m$,
in $n$ variable $x_1, \dots, x_n$ are algebraically dependent.
Thus, there exists a nonzero polynomial $P \in \mathbb{C}[z_1, \dots, z_m]$ such that
\begin{equation}\label{proof of ration count}
P\left( -\frac{ g_1( x )}{ f_1( x ) }, \dots, - \frac{ g_m( x )}{ f_m( x ) } \right) = 0, \quad\forall\ x .
\end{equation}
We claim that
\begin{equation}\label{rationl; P=0}
 P(\alpha_1, \dots, \alpha_m) = 0, \quad \forall\ \alpha \in \mathcal{I}.
\end{equation}
In fact, for any $\alpha \in \mathcal{I}$, there exists a solution $x$ to (\ref{eqn: poly system, rational func}), i.e. $ \alpha_i f_i( x) = -g_i( x)$ and $f_i(x) \neq 0$, $i=1,\dots, m $. Thus for any $\alpha \in \mathcal{I}$, there exists an $x$ such that $ \alpha_i = -  { g_i( x) }/{ f_i( x)}, i=1,\dots, m$.
Applying (\ref{proof of ration count}), we have
$$P(\alpha_1, \dots, \alpha_m) = P\left( -\frac{ g_1( x )}{ f_1( x ) }, \dots, - \frac{ g_m( x )}{ f_m( x ) } \right) = 0.$$
Thus, $P(\alpha_1, \dots, \alpha_m) = 0$ for a positive measure of $\alpha$, implying that $P$ is a zero polynomial, which is a contradiction.  \QED

\subsection{ Solvability of a System with Generic Coefficients }   % Bernstein's Theorem and
In this subsection, we consider a more restricted class of polynomial systems in which all coefficients are generic (in Lemma \ref{lemma: rational count} only one parameter in each equation is generic). The results presented in this subsection can be derived from either Lemma \ref{lemma: rational count} or Bernstein's theorem.

Consider a system of polynomial equations with $m$ equations and $n$ variables:
\begin{equation}\label{eqn: general poly system}
\mathcal{P}: \quad f_k(x_1, \dots, x_n) = 0,\quad k=1,\dots, m,
\end{equation}
where $f_i, i=1,\dots, m$ are nonzero polynomials in variables $x_1, \dots, x_n$.
Define
$$ \mathbb{C}^* \triangleq \mathbb{C}\backslash \{0 \}. $$
The following lemma states that a generic overdetermined system has no strictly nonzero solution (i.e. solution in $(\mathbb{C}^*)^n$).  %  (i.e. solution vector that each element is nonzero).
\begin{lemma}\label{claim: generic system doesn't have a no-zero-component-solution}
Consider a system of polynomial equations $\mathcal{P}$ as in (\ref{eqn: general poly system}). If $m \geq n + 1$ and the coefficients of all polynomials
$f_k$ are generic, then $\mathcal{P}$ has no solution in $(\mathbb{C}^*)^{n}$.
% If $|\mathcal{P}_J| \geq |J|+1$ and $\mathcal{P}_J$ has generic coefficients, then $\mathcal{P}_J$ has no solution in $(\mathbb{C}^*)^{|J|}$.
\end{lemma}
\emph{Proof of Lemma \ref{claim: generic system doesn't have a no-zero-component-solution} (by Bernstein's theorem)}: Since $m \geq n+1$, we can pick a subsystem of $\mathcal{P}$ with $n$ polynomial equations. According to Corollary \ref{coro: finiteness in C*n} in Appendix \ref{appen: Bernstein's thm} (a corollary of Bernstein's theorem), this subsystem has only a finite number of solutions in $(\mathbb{C}^*)^{n}$. These finite number of solutions in $(\mathbb{C}^*)^{n}$ can not satisfy the remaining $(m-n)$ generic equations in $\mathcal{P}$, thus $\mathcal{P}$ has no solution in $(\mathbb{C}^*)^{n}$, which proves Lemma \ref{claim: generic system doesn't have a no-zero-component-solution}.

\noindent \emph{Another Proof of Lemma \ref{claim: generic system doesn't have a no-zero-component-solution} (by Lemma \ref{lemma: rational count})}:
Each nonzero polynomial $f_{k}$ contains at least one nonzero monomial, thus % $\alpha_k h_k(x)$, where $\alpha_k$ is a generic coefficient, and $h_k(x)$ is a monomial.
 $f_{k}=0$ can be written as $ \alpha_k h_k(x) + g_k(x) = 0 $, where $\alpha_k$ is a generic coefficient, $h_k(x)$ is a nonzero monomial, and $g_k$ is a polynomial.
Since $ m \geq n+1$, by Lemma \ref{lemma: rational count}, $\alpha_k h_k(x) + g_k(x) = 0, h_k(x) \neq 0,k=1,\dots, m $ has no solution for generic $\alpha_k$'s. Note that $x \in (\mathbb{C}^*)^{n} $ implies that all monomials $h_k(x) \neq 0$,
thus $\mathcal{P}: \alpha_k h_k(x) + g_k(x) = 0, x \in (\mathbb{C}^*)^{n}, k=1,\dots, m $ has no solution either. \QED

Remark: The proof of Lemma \ref{claim: generic system doesn't have a no-zero-component-solution} by Bernstein's theorem has appeared in \cite[Section VI]{YetisIA}, though it is not explicitly stated that the term ``solution'' to a polynomial system should be interpreted as a strictly non-zero solution in  $(\mathbb{C}^*)^{n}$. % Corollary \ref{a sufficient condition for the counting argument}

Lemma \ref{claim: generic system doesn't have a no-zero-component-solution} only considers solutions in $(\mathbb{C}^*)^n$. In the following, we use Lemma \ref{claim: generic system doesn't have a no-zero-component-solution} to derive a sufficient condition under which an overdetermined system has no solution in
$\mathbb{C}^n$.
Define the support of a vector $x = (x_1,\dots, x_n) \in \mathbb{C}^n$ as the set $\{j \;|\; x_j \neq 0  \}$.
% The number of solutions in $\mathbb{C}^n$ is zero is equivalent to: for any $J \subseteq \{1,\dots,n \}$, the number of solutions with support $J$ is zero.
For any support $J$, define a truncated variable $x_J = (x_j)_{j \in J}$ and its complement variable $\bar{x}_J = (x_i)_{i\notin J}$.
We define a ``partial-system'' $\mathcal{P}_J$ by restricting the system $\mathcal{P}$ to the support $J$ as follows.
  In the system $\mathcal{P}$
set the variables $x_i, i \notin J $ to be $0$, then the monomials with any variable drawn from $\{x_i \mid i\notin J \}$ vanish and we obtain a new system
\begin{equation}\label{eqn: truncated poly system}
\begin{split}
\mathcal{P}_J: & f_{kJ}(x_J) = 0,\quad k=1,\dots, m,   \\
\text{where }  f_{kJ}(x_J) =  &   f_k(x_J, \bar{x}_J ) \ \ \ \text{ with } \bar{x}_J = 0,\quad \; k=1,\dots, m.
\end{split}
\end{equation}

Note that the partial-system $\mathcal{P}_J$ is different from a subsystem $\mathcal{Q}$ which consists of a subset of equations: in (\ref{eqn: truncated poly system}), $\mathcal{P}_J$ is defined by restricting $\mathcal{P}$ to a subset of variables $J$.
%The partial-system $\mathcal{P}_J$ may contain some trivial equations $0=0$. For example, if we set $x_1, x_2$ to be $0$, the equation $x_1x_3^2 + x_2x_4 =0$ becomes $0=0$. We should not count the trivial equation $0=0$ when we compute the number of equations in $\mathcal{P}_J$, since adding or removing an equation $0=0$ does not affect the solvability of the system. % (an implicit assumption of Bernstein's theorem is that all polynomials have at least one monomial).
We define
$$ |\mathcal{P}_J| \triangleq \text{number of nonzero polynomials in the set } \{ f_{kJ}, \; k=1,\dots,m \} . $$

%With these definitions, we present below an immediate corollary of Lemma \ref{claim: generic system doesn't have a no-zero-component-solution}.
Now we are ready to show that a generic system $\mathcal{P}$ has no solution in $\mathbb{C}^n$ if every partial-system $\mathcal{P}_J$ (including $\mathcal{P}$ itself) is overdetermined.
\begin{coro}\label{a sufficient condition for the counting argument}
Consider a system of polynomial equations $\mathcal{P}$ as in (\ref{eqn: general poly system}). Suppose the coefficients of all polynomials
$f_k$ are generic and
\begin{equation}\label{sufficient condition, in Lemma}
  |\mathcal{P}_J| \geq |J| +1, \quad \forall J \subseteq \{1,\dots, n \},
  \end{equation}
where $\mathcal{P}_J$ is defined in \eqref{eqn: truncated poly system}.
Then the system $\mathcal{P}$ has no solution in $\mathbb{C}^{n}$.
\end{coro}

\noindent\emph{Proof of Corollary \ref{a sufficient condition for the counting argument}:}
The system $\mathcal{P}$ has no solution in $\mathbb{C}^{n}$ is equivalent to: for any subset $J \in \{1,\dots, n \}$, $\mathcal{P}$ has no solution with support $J$, or equivalently, the partial-system $\mathcal{P}_J$ has no solution in $(\mathbb{C}^*)^{|J|}$.
Since $|\mathcal{P}_J| \geq |J| +1$, by applying
 Lemma \ref{claim: generic system doesn't have a no-zero-component-solution} to the system $\mathcal{P}_J$, we obtain that $\mathcal{P}_J$ has no solution in $(\mathbb{C}^*)^{|J|}$.
% Claim \ref{coro: overdetermined generic has no solution},
% which is an immediate corollary of Lemma \ref{claim: generic system doesn't have a no-zero-component-solution}:
%\begin{claim}\label{coro: overdetermined generic has no solution}
%If $|\mathcal{P}_J| \geq |J|+1$ and $\mathcal{P}_J$ has generic coefficients, then $\mathcal{P}_J$ has no solution in $(\mathbb{C}^*)^{|J|}$.
% \end{claim}
 \QED

{\color{black}
Remark 1: Corollary \ref{a sufficient condition for the counting argument} will be used to prove Proposition \ref{prop: constant MIMO, single-beam}, a known DoF bound for constant MIMO IC, in Section
\ref{subsec: proper and feasible}.
It will not be used to prove Theorem \ref{thm 1: DoF bound for generic extensions}, but the idea of considering the solutions with specific supports is crucial for the proof of Theorem \ref{thm 1: DoF bound for generic extensions}.
% (or Theorem \ref{thm: main result, bound on K})

Remark 2: The idea of considering solutions with all possible supports has also been used in other related studies such as \cite[Lemma 5]{sturmfels};
%Although Corollary \ref{a sufficient condition for the counting argument} is just a simple corollary of Bernstein's theorem, we are not aware of explicit statement of it in the literature. % (his is partially because the area of algebraic geometry has different focus).
%Some results, such as \cite[Lemma 5]{sturmfels}, has also adopted the idea of considering solutions with all possible supports,
%but they are different from Corollary \ref{a sufficient condition for the counting argument};
see more discussions in Appendix \ref{appen: Bernstein's thm}.
%\cite{SturmCn,Lemma 5}.

}

Corollary \ref{a sufficient condition for the counting argument} implies that a necessary condition for a generic system $\mathcal{P}$ to have a solution in $\mathbb{C}^n$ is:
\begin{equation}\label{condition of ours}
 \text{There exists a subset } J\in \{1,\dots, n \} \text{ such that } |\mathcal{P}_J| \leq |J|.
\end{equation}

In contrast, Lemma \ref{claim: generic system doesn't have a no-zero-component-solution} implies that a necessary condition for a generic system $\mathcal{P}$ to have a solution in $(\mathbb{C}^*)^n$ is
%(let $J = \{1,\dots, n \}$ in  % a strictly nonzero solution (i.e.
 %(see an argument in \cite{YetisIA}):
\begin{equation}\label{condition of misconcep 2}
|\mathcal{P}| \leq n. % \quad \text{ where } J = \{1,\dots, n \}.
\end{equation}

%%%----------------------------------Deleted --------------------------------------------------------------------
%This condition means that a feasible generic system, possibly overdetermined, must have a partial-system that is not overdetermined. In fact, the partial-system corresponding to the support of the solution of a generic system is not overdetermined.
%We conjecture that condition \eqref{condition of ours} is not sufficient for a generic system $\mathcal{P}$ to have a solution in $\mathbb{C}^n$. It seems not easy to construct a counterexample to this conjecture. If $|\mathcal{P}_J| \leq |J| = 0$ for $J = \emptyset$, then $(0,\dots, 0)$ is a solution to $\mathcal{P}$ since no polynomial in $\mathcal{P}$ has a nonzero constant term. If $|\mathcal{P}_J| \leq |J| = 1$ for $J = \{ x_j\}$, it is easy to show that $\mathcal{P}$ is also solvable.
% For example, suppose $|\mathcal{P}_J| = |J| = 0$ for $J = \emptyset$, then no polynomial in $\mathcal{P}$ has a constant term, thus $(0,\dots, 0)$ is a solution to $\mathcal{P}$).
%%%----------------------------------Deleted End --------------------------------------------------------------------

We give an example to illustrate the subtle and yet important difference of ``solutions in $(\mathbb{C}^*)^n$'' and ``solutions in $\mathbb{C}^n$''.
This example also shows that $ |\mathcal{P}| \leq n$ is not a necessary condition for the feasibility of $\mathcal{P}$.
% Note that the restriction of solutions in $(\mathbb{C}^*)^{|J|}$ is necessary.

\noindent\textbf{Example 4.4} Consider the following system with $n = 2$ variables and $\ m =3$ equations:
\begin{equation}\label{counterexample to misconcept 2}
\begin{split}
f_1(x_1,x_2) =  a_1 x_1^2 + a_2 x_1  = 0, \\
f_2(x_1,x_2) = b_1 x_1 x_2 + b_2 x_1 = 0, \\
f_3(x_1,x_2)=c_1x_1x_2^3+c_2x_1=0,\\
\end{split}
\end{equation}
where $a_i,b_i, c_i,\ i=1,2$ are generic coefficients.
Obviously, $(x_1,x_2) = (0,x_2)$ is a solution to (\ref{counterexample to misconcept 2}) for any $x_2$; thus (\ref{counterexample to misconcept 2}) has infinitely many solutions in $\mathbb{C}^2$.
{\color{black}
In other words, $m > n$ does not imply the infeasibility of a generic system.
 If we require $x_1 , x_2 \in \mathbb{C}^*$, then $\mathcal{P}$ has no solution, i.e. $m>n$ implies that no strictly non-zero solution exists.
 }
 % only one solution $(x_1, x_2) = (-{a_2}/{a_1}, -{b_2}/{b_1} )$; thus $\mathcal{P}$ has only one solution in $(\mathbb{C}^*)^n$.

% (more precisely, the number of solutions equals the mixed volume of the Newton polytope of the polynomials in this subsystem)

%--------------------Deleted---------------------------------------------
%The above argument of Lemma \ref{claim: generic system doesn't have a no-zero-component-solution} also works for the case $J = \emptyset$. For example, consider the system $\mathcal{P}: f_1(x_1) = a x_1 + b = 0$, where $a,b$ are generic.
%Then $|\mathcal{P}_{\emptyset}| \geq |\emptyset|+1 = 1$, since there is one nontrivial equation $ f_{1 \emptyset}(x_{1 \emptyset}) =0$ (or equivalently, $b=0$). However, $b = 0$ does not hold for generic $b$, thus $\mathcal{P}$ does not have a solution with support $\emptyset$ (i.e. $x_1=0$ is not a solution).
%-----------------------------------------------------------------
 % In contrast, if $\mathcal{P}$ is $ax_1 = 0$, then $|\mathcal{P}_{\emptyset}| = 0 = |\emptyset|$ and $x_1 = 0$ is a solution to $\mathcal{P}$.
% Note that $f_{k \emptyset}(x_{\emptyset})$ is the constant term in $f_k(x)$, and should be nonzero since
% $f_{k \emptyset}(x_{\emptyset}) = 0$ is nontrivial. $f_{k \emptyset}(x_{\emptyset}) = 0$ does not hold implies that $\mathcal{P}$ has no solution in % $\mathbb{C}^{0}$, i.e., $(0,\dots,0)$ is not a solution to $\mathcal{P}$.

{\color{black}
\subsection{ Applications to the constant MIMO IC}\label{subsec: proper and feasible} % :   }   % in Constant MIMO IC  Dimensionality Counting Argument
% In this subsection, we provide the application of the algebraic tools
% new interpretations on the feasibility of IA systems.

A polynomial system of equations is called proper if in every subsystem the number of equations does not exceed the number of variables (see \cite{YetisIA}).
% (strictly speaking, \cite{YetisIA} only defines the properness for the IA system of a constant MIMO IC).
Researchers have tried to identify IA systems for which improperness implies infeasibility, since for such systems the DoF upper bound can be obtained
by a dimensionality counting argument. The constant MIMO IC represents one such IA system.
We emphasize that improperness does not imply infeasibility in general; even for the simple case that the coefficients are generic, counterexamples exist (see Example 4.4).
Therefore, the counting argument works only when the problem exhibits some special structure.
Below we will explain how the special structure of the constant MIMO IC makes the counting argument work.
}
% Our explanation also helps understand why the techniques for constant MIMO IC cannot be easily extended to more general channel models.

%%%%----------------------------------Deleted from Version 6 --------------------------------------------------------------------
%The condition \eqref{condition of ours} and \eqref{condition of misconcep 2} have different choices of the subset $J$, which is due to the different restrictions on the support of the solution.
%%%----------------------------------Deleted End --------------------------------------------------------------------

% The following proposition shows that $m \leq n$ is a necessary condition for (\ref{eqn: part a, Yetis, IA condition with U}) to have a nonzero solution.
\begin{proposition}\label{prop: constant MIMO, single-beam}
Consider the IA condition for the constant MIMO IC in the single beam case:
\begin{subequations}\label{eqn: Yetis, IA condition with U}
\begin{align}
& u_k^H H_{kj} v_j =0,\quad \forall\ 1\leq k\neq j \leq K, \label{eqn: part a, Yetis, IA condition with U}\\
& u_k^H H_{kk} v_k  \neq 0,\quad \forall\ k, \label{eqn: part b, Yetis, IA condition with U}
\end{align}
\end{subequations}
where $H_{kj}$ is a $M_r \times M_t $ matrix with generic entries, and $v_k \in \mathbb{C}^{ M_t \times 1}, u_k \in \mathbb{C}^{M_r \times 1}, k=1,\dots, K$ are beamformers.
A necessary condition for (\ref{eqn: part a, Yetis, IA condition with U}) to have a solution $\{ u_k, v_k \}_{k=1}^K$ is $K(K-1)\leq K(M_t+M_r-2)$,
i.e. $K \leq M_t + M_r - 1$. Hence, the DoF in the single-beam case is no more than $M_t + M_r -1$.
\end{proposition}

Proposition \ref{prop: constant MIMO, single-beam} is a special case of Theorem \ref{thm 1: DoF bound for generic extensions}, and also a special case of the results in \cite{YetisIA, MeisamIA, TseFeasibility}.
%They use Bernstein's Theorem to prove (\ref{misconcept 2}).

{\color{black}
We describe in high-level terms how the special structure makes the counting argument possible.
Lemma \ref{claim: generic system doesn't have a no-zero-component-solution} and Corollary \ref{a sufficient condition for the counting argument} show that
for a generic system $\mathcal{P}$,
\begin{subequations}\label{compare generic solvability}
\begin{align}
     \mathcal{P} \text{ has a solution in } \mathbb{C}^n \;  & \Longrightarrow  \exists \; J \in \{1,\dots, n \} \text{ such that } |\mathcal{P}_J| \leq |J| ;           \label{compare a} \\
      \mathcal{P} \text{ has a solution in } (\mathbb{C}^*)^n  & \Longrightarrow   |\mathcal{P}| \leq n.        \label{compare b}
\end{align}
\end{subequations}
To apply the counting argument, we need to prove
\begin{equation}\label{counting argument}
 \mathcal{P} \text{ has a solution in } \mathbb{C}^n  \Longrightarrow   |\mathcal{P}| \leq n.
\end{equation}
There are two simple methods to prove \eqref{counting argument} by \eqref{compare generic solvability}.
The first one is to prove the equivalence of LHS (Left-Hand-Side) of \eqref{compare a} and \eqref{compare b} using the special structure of the problem, i.e.
\begin{equation}\label{LHS equivalence}
\mathcal{P} \text{ has a solution in } \mathbb{C}^n \Longleftrightarrow  \mathcal{P} \text{ has a solution in } (\mathbb{C}^*)^n.
\end{equation}
Then combining \eqref{LHS equivalence} and \eqref{compare b} proves \eqref{counting argument}.
The second method is to prove the equivalence of RHS (Right-Hand-Side) of \eqref{compare a} and \eqref{compare b} using the special structure of the problem, i.e. \begin{equation}\label{RHS equivalence}
\exists \; J \in \{1,\dots, n \} \text{ such that } |\mathcal{P}_J| \leq |J|  \Longleftrightarrow |\mathcal{P}| \leq n.
\end{equation}
Then combining \eqref{RHS equivalence} and \eqref{compare a} proves \eqref{counting argument}.

The first method was adopted in \cite{YetisIA}; see the last paragraph of Section VI in \cite{YetisIA} for a proof sketch. It should be noted that certain details were not explicitly stated in the proof.
The proof relies on the following structureless property of the channel: the product of a generic matrix and a unitary basis matrix is still a generic matrix.
This structureless property implies that a solution in $\mathbb{C}^n$ can be mapped to
a solution in $(\mathbb{C}^*)^n$ via a unitary transformation without altering the genericity of the channel matrix.
Thus there is no loss of generality in restricting
the solutions to $(\mathbb{C}^*)^n$ as opposed to $\mathbb{C}^n$, which establishes \eqref{LHS equivalence}.
This argument can be formalized by a probability computation presented by
the authors of \cite{YetisIA} in private communication.
% In version 6: The structureless property of the channel, i.e., that the channel is not partial to any particular basis representation,

 Below we provide a different proof that follows the second method.
We will prove \eqref{RHS equivalence} by showing $|\mathcal{P}_J| = |\mathcal{P}_{\{1,\dots, n \}} |, \forall J $, which is also a consequence of the structureless property of MIMO channel matrices.
}

%%------------------------------------------------------------------------------------------------------

\emph{A new proof of Proposition \ref{prop: constant MIMO, single-beam}:}
First, notice that the nonzero solution to (\ref{eqn: part a, Yetis, IA condition with U}) $u_k\neq 0, v_k \neq 0, \forall\ k$ will satisfy (\ref{eqn: part b, Yetis, IA condition with U}) automatically since $H_{kk}$ is a matrix with generic entries.
Therefore, (\ref{eqn: Yetis, IA condition with U}) is solvable iff (\ref{eqn: part a, Yetis, IA condition with U}) has a nonzero solution.
We need to prove: if (\ref{eqn: part a, Yetis, IA condition with U}) has a nonzero solution $u_k  \in \mathbb{C}^n\backslash \{ 0\}, v_j \in \mathbb{C}^n\backslash \{ 0\}, \forall\ k,j$, then $ K(K-1)\leq K(M_r + M_t-2)$.

Suppose $u_{k p_k} \neq 0, v_{j q_j } \neq 0, 1 \leq p_k \leq M_r, 1\leq q_j \leq M_t, \forall\ k,j $.
We can scale $u_k, v_j$ such that $u_{k p_k} = 1, v_{j q_j} = 1$ \footnote{In \cite{MeisamIA}, the authors assume that $u_{k1} = v_{j1} = 1, \forall k,j$. Such an assumption is valid when $H_{kj}$ is a full generic matrix, but one should use caution when applying this assumption to other problems. In particular, for
diagonal channel matrices considered in Theorem \ref{thm 1: DoF bound for generic extensions}, the assumption $u_{k1} = v_{j1} = 1, \forall k,j$ is no longer valid. Here, we choose to present a more complicated assumption $u_{k p_k} = 1, v_{j q_j} = 1$ since it does not require any property of the channel matrices.
} and $(u_k,v_k)_{1\leq k \leq K}$ is still a solution to (\ref{eqn: part a, Yetis, IA condition with U}). After scaling, (\ref{eqn: part a, Yetis, IA condition with U}) becomes
\begin{equation}\label{transformed system of MIMO constant}
\mathcal{P}: \; 0 = u_k^H H_{kj} v_j = \sum_{ (p,q)  } u_{k p}^* H_{kj}(p,q) v_{j q}
= H_{kj}(p_k, q_j) + \sum_{ (p,q) \neq (p_k, q_j) } u_{k p}^* H_{kj}(p ,q) v_{q s}, \;\;  \forall\ k\neq j,
\end{equation}
where  $H_{kj}(p,q)$ denotes the $(p,q)$'th entry of $H_{kj}$.
$\mathcal{P}$ is a system of polynomial equations with generic coefficients, since the coefficients of different equations come from different channel matrices.
The number of variables in $\mathcal{P}$ is $n = K(M_r + M_t -2)$ and the number of equations in $\mathcal{P}$ is $m = K(K-1)$.

Note that $|\mathcal{P}_J| = m$ for any support $J \subseteq \{1, \dots, n \}$, since each equation has a nonzero constant term $  H_{kj}(p_k, q_j) $ and will not become a trivial equation $0=0$ by restricting to any support $J$. According to Corollary \ref{a sufficient condition for the counting argument}, a necessary condition for $\mathcal{P}$ to have a solution in $\mathbb{C}^n$
is $|\mathcal{P}_J| \leq |J| $ for some $J$. Thus we have $m = |\mathcal{P}_J| \leq |J| $. Combining with the fact $|J| \leq n$, we obtain
$m \leq n$, i.e. $K(K-1) \leq K(M_r + M_t -2)$. \QED

{\black
In the above proof, a generic channel coefficient is separated from other terms in the expression \eqref{transformed system of MIMO constant},
which is possible due to the structureless of the channel (in fact, if $H_{kj}$ is diagonal, then the separated term $H_{kj}(p_k, q_j)$ will be zero if $p_k \neq q_j$, thus we may not be able to separate a generic channel coefficient).
 The expression \eqref{transformed system of MIMO constant} is the key step in the above proof.
Interestingly, seperating a generic channel coefficient is also a critical step in the proofs of \cite{MeisamIA,TseFeasibility}, though they apply different algebraic tools. Specifically, the proof of \cite{MeisamIA} applies Lemma \ref{lemma: rational count} to the expression \eqref{transformed system of MIMO constant} with $\alpha$ being $H_{kj}(p_k, q_j)$'s and $f_i$ being $1$. A benefit of using expression \eqref{transformed system of MIMO constant}, compared with the other way of utilizing the structureless property (i.e. multiplying a nonsingular matrix does not change genericity), is that it can be generalized to the multi-beam case (see \cite{MeisamIA,TseFeasibility}).

The above two methods for proving \eqref{counting argument} are based on the structureless property of the channel, thus it is not surprising that they can not be simply applied to structured channels, even with generic extensions.
We will explain the difficulty of extending them to the MIMO IC with generic channel extensions (see Section \ref{subsec: thm 1 analysis}).
}

% \section{Proof of Theorem \ref{thm 1: DoF bound for generic extensions} }\label{sec: proof of Thm 1, generic extension}
\section{Proof of Theorem \ref{thm 1: DoF bound for generic extensions} }\label{sec: proof of Thm 1, generic extension}

This section is devoted to the proof of Theorem \ref{thm 1: DoF bound for generic extensions} %. The organization of this section is as follows.
%In section \ref{subsec: thm 1 analysis}, we discuss the difficulties in proving Theorem \ref{thm 1: DoF bound for generic extensions} and how we resolve these difficulties.
%Then we prove Theorem \ref{thm 1: DoF bound for generic extensions}
for the SISO IC in Section \ref{subsec: thm 1 SISO proof}. The proof of Theorem \ref{thm 1: DoF bound for generic extensions} for the MIMO IC is a similar to the proof for the SISO IC, and is given in Appendix \ref{subsec: thm 1 MIMO proof}.
The major difference is that we only consider the supports of $u_k, v_k$ in SISO IC, while in MIMO IC we consider both the supports and the ``block-supports'' of $u_k ,v_k$.

% This section is devoted to the proof of Theorem \ref{thm 1: DoF bound for generic extensions} in Section \ref{subsec: thm 1 MIMO proof}.
  %
  % We first analyze the difficulty in Section \ref{subsec: thm 1 analysis}
 %. The organization of this section is as follows.
%In section \ref{subsec: thm 1 analysis}, we discuss the difficulties in proving Theorem \ref{thm 1: DoF bound for generic extensions} and how we resolve these difficulties.
%Then we prove Theorem \ref{thm 1: DoF bound for generic extensions}

%
%The proof for the MIMO IC is given in Appendix \ref{subsec: thm 1 MIMO proof}.

\subsection{Preliminary Analysis}\label{subsec: thm 1 analysis}
% For simplicity of analysis,
Consider a SISO IC with $T$ generic channel extensions. The IA condition is
\begin{subequations}\label{SISO generic IA}
\begin{align}
& u_k^H H_{kj} v_j =0, \quad \forall\ 1\leq k\neq j \leq K, \label{SISO generic IA, part a}\\
& u_k^H H_{kk} v_k  \neq 0, \quad \forall\ 1 \leq k \leq K, \label{SISO generic IA, part b}
\end{align}
\end{subequations}
where $H_{kj} = \text{diag}\left( h_{kj}(1),\dots, h_{kj}(T) \right) $ is a $T \times T$ diagonal matrix with generic diagonal entries, and $v_j \in \mathbb{C}^{ T \times 1}, u_k \in \mathbb{C}^{ T \times 1}$ are the beamforming vectors.
%Suppose (\ref{SISO generic IA}) has a solution $u_k \neq 0, v_j \neq 0, \forall\ k,j$.
% Suppose $u_{k p_k} \neq 0, v_{j q_j } \neq 0, 1 \leq p_k, q_j \leq N, \forall\ k,j $.
%We can scale $u_k, v_j$ such that $u_{k p_k} = 1, v_{j q_j} = 1$, and $(u_k,v_k)_{1\leq k \leq k}$ is still a solution to
%(\ref{SISO generic IA}).
%Then (\ref{SISO generic IA, part a}) becomes
%\begin{equation}\label{transformed system of SISO generic IC}
%\mathcal{P}: \; 0 = u_k^H H_{kj} v_j = \sum_{ (p,q)  } u_{k p}^* H_{kj}(p,q) v_{j q}
%= H_{kj}(p_k, q_j) + \sum_{ (p,q) \neq (p_k, q_j) } u_{k p}^* H_{kj}(p ,q) v_{q s}, \;\;  \forall\ k\neq j,
%\end{equation}
%where  $H_{kj}(p,q)$ denotes the $(p,q)$'th entry of $H_{kj}$.
After a simple scaling of $\{u_k, v_k\}$ (i.e., setting one of the entries of these vectors to 1), (\ref{SISO generic IA, part a}) becomes a system of polynomial equations $\mathcal{P}$ with
$n = 2K(T-1)$ variables and $m = K(K-1)$ equations.
% provides a sufficient condition for the infeasibility of a system of generic polynomial equations, it is still not

{\black
Unlike the generic (full) channel matrices in the constant MIMO IC, $ H_{kj}$'s here are generic diagonal matrices.
% Due to this diagonal structure,
Due to the lack of the structureless property, the two methods described in Section \ref{subsec: proper and feasible} cannot be directly applied to
this problem.
%due to the following two reasons.
% More specifically, there are two reasons.
% First, both \eqref{LHS equivalence} and \eqref{RHS equivalence} do not hold for the diagonal structured channel matrices.
% Specifically, the first method does not work for the following reason.   a crucial property to \eqref{LHS equivalence} is
Specifically, the first method proves \eqref{LHS equivalence} by using the invariance of genericity under unitary transformation: the product of a generic matrix and a unitary matrix $\Theta$ is still generic.
Note that the purpose of introducing $\Theta$ is to map a beamformer in $(\Cs)^n$ to $(\C)^n$, thus $\Theta$ should be a full matrix.
 For the IA system \eqref{SISO generic IA}, the invariance property does not hold for generic diagonal matrices: the product of a generic diagonal matrix and a unitary matrix $\Theta$ is no longer diagonal (unless $\Theta$ itself is diagonal, which is not the case here). Thus \eqref{LHS equivalence} does not hold for
our problem, i.e. we can not assume that the solutions are in $(\Cs)^n$.

The second method (``a new proof of Proposition \ref{prop: constant MIMO, single-beam}'') cannot be directly applied for our problem due to the following reason.
% We then show that \eqref{RHS equivalence} do not hold for the diagonal structured channel matrices.
To prove \eqref{RHS equivalence} for constant MIMO IC,
we use the fact that $u_k^H H_{kj} v_j$ does not become zero when $u_k, v_j$ are nonzero vectors and $H_{kj}$ is a full generic matrix.
For the IA system \eqref{SISO generic IA}, $u_k^H H_{kj} v_j$ may become zero even if $u_k, v_j$ are nonzero:
simply setting $ u_{k1}=0, v_{j2}=\dots = v_{jT} = 0 $ makes $u_k^H H_{kj} v_j$ vanish for any diagonal matrix $H_{kj}$.
Thus unlike the constant MIMO IC case, here we can make $|\mathcal{P}_{J}|$ smaller than $m = |\mathcal{P}_{\{1,\dots, n\}}|$.
Moreover, $|\mathcal{P}_{J}|$ can be made as small as $0$: setting $u_{k1}=0, v_{j2}=\dots = v_{jT} = 0, \forall\ k,j$ makes all $u_k^H H_{kj} v_j$ vanish, which means $|\mathcal{P}_J| = 0 < |J|$ for the corresponding support $J$ when $T \geq 3$ (when $T\geq 3$, the variable vector has $|J| = K(T-2)>0$ variables after scaling).
% Thus the property in the constant MIMO IC $|\mathcal{P}_{\{1,\dots, n\}}| = |\mathcal{P}_{\{J\}}|, \forall J$ does not hold
% for the IA system \eqref{SISO generic IA}.
This example disproves \eqref{RHS equivalence} for the IA system \eqref{SISO generic IA}.
}

{\black
Another more serious challenge arises when we further explore the example
\begin{equation}\label{solution of nonoveralpping support}
 u_k = (0,1,\dots, 1),\quad v_j = (1,0,\dots, 0), \quad \forall\ k,j.
 \end{equation}
We notice that \eqref{solution of nonoveralpping support} satisfies the zero-forcing condition (\ref{SISO generic IA, part a}) for any $K$,
thus it seems that no meaningful DoF upper bound can be derived from (\ref{SISO generic IA, part a}).
One may argue that \eqref{solution of nonoveralpping support} is not a valid example % to \eqref{RHS equivalence}
since it does not satisfy the full rank condition (\ref{SISO generic IA, part b}).
This observation reveals an important fact that may have not been explicitly recognized before:
\emph{for structured channels, the (nonzero) solution to the zero-forcing condition does not satisfy
 the full rank condition automatically.}
 This fact poses a new challenge to our problem: we need to take the inequalities (\ref{SISO generic IA, part b}) into account (in contrast, in the constant MIMO IC the full rank condition (\ref{eqn: part b, Yetis, IA condition with U}) can be discarded). % as long as each beamformer is nonzero)
   %in establishing the DoF bound
 In other words, we need to consider a hybrid system of equations and inequalities, rather than just a system of equations.
 Unfortunately, the solvability of hybrid systems is not the focus of algebraic geometry, thus conventional algebraic geometry tools cannot be directly applied.
 }

Our proof technique can be briefly described as follows. % (see more details in Section \ref{subsec: thm 1 MIMO proof}).
We develop an induction analysis that leverages the recursive structure of the IA system,
 % (i.e. \eqref{MIMO generic IA} implies \eqref{MIMO, reduced dimension IA condition})
which implicitly utilizes the inequalities (i.e. the full rank condition). The algebraic tool Lemma \ref{claim: generic system doesn't have a no-zero-component-solution} is
used to prove a crucial intermediate bound (\ref{bound on Omega^c, MIMO}).

%%%%%%%%%%%%%%%%%%%%%%%% subsection 5.2 %%%%%%%%%%%%%%%%%%%%%%%%%%%%%%%%%%%%%%%%%%%%%%%%%%%%%%%5
\subsection{Proof of Theorem \ref{thm 1: DoF bound for generic extensions} for a SISO IC}\label{subsec: thm 1 SISO proof}

We define the maximal $K$ as $f(T)$, i.e.
\begin{equation}\label{definition of f(T), max K}
\begin{split}
f(T) \triangleq & \max \left\{K \;|\; (\ref{SISO generic IA}) \text{ is solvable for a positive measure of } \left(h_{kj}(t)\right)_{1\leq k,j \leq K, \; 1\leq t \leq T} \right\}.
       %%  & \max \left\{K \;|\; \Gamma(T, K) \text{ has a positive measure} \right\}  \\
\end{split}
\end{equation}
When $T=0$, we define $f(0)=0$.

To prove part (a) of Theorem \ref{thm 1: DoF bound for generic extensions}, we only need to prove the following bound:
\begin{equation}\label{SISO IC bound}
f(T) \leq 2T -1, \quad \forall \; T \geq 1.
\end{equation}
We will do so using an induction argument.

For the basis of the induction ($T=1$), it is easy to show that $f(1) \leq 2T-1 =1$. In fact, any two nonzero $1$-dimensional vectors (i.e. scalars) are linearly dependent, thus when $K\geq 2$ the IA condition \eqref{SISO generic IA} has no solution.

Now suppose (\ref{SISO IC bound}) holds for each positive integer that is smaller than $T$. We prove that (\ref{SISO IC bound}) holds for $T$.
Suppose $K$ satisfies that the IA condition (\ref{SISO generic IA}) has a solution $\{ \tilde{u}_k, \tilde{v}_k \}_{k=1,\dots, K}$ for a positive measure of $\left( h_{kj}(t) \right)$. For a vector $x = \left( x_1, \ldots, x_T  \right)^T\in \mathbb{C}^T$, denote the support of $x$ as
$$\text{supp}(x) = \{j \;|\; x_j \neq 0 \}. $$
Each $\left(h_{kj}(t) \right)$ corresponds to (at least) one collection of supports $ \{ \text{supp}(\tilde{u}_k), \text{supp}(\tilde{v}_k) \}_{k=1,\dots,K}$. Since there are finitely many possible collections of supports, it follows that there exists a positive measure of $\left(h_{kj}(t) \right)$ which corresponds to the same collection of supports  $\{ R_k, S_k \}_{k=1,\dots, K}$, where $R_k, S_k \subseteq \{1,\dots, T \}$.
Denote this set of $\left(h_{kj}(t) \right)$ as $ \mathcal{H} $ which has a positive measure.

%%  Suppose $ \tilde{u}_{k r_k} = 1, \tilde{v}_{k s_k}=1 $, where $r_k \in R_k, s_k \in S_k$.
%% Since scaling does not affect the solutions of (\ref{SISO generic IA, part a}),

Define a set of transmitter-receiver pairs as
\begin{align}
\notag \Omega \triangleq & \left\{ (k,j) \;|\; 1\leq k\neq j \leq K, \; \text{supp}(\tilde{u}_k) \cap \text{supp}(\tilde{v}_j) = \emptyset  \right\}  \\
              = & \left\{ (k,j) \;|\; 1\leq k\neq j \leq K, \; R_k \cap S_j = \emptyset  \right\}.
\end{align}
 The complement of $\Omega$ in the set
$ \{(k,j)  \;|\; 1\leq k\neq j \leq K \}$ is
$$\Omega^c = \left\{ (k,j) \;|\; 1\leq k\neq j \leq K,  \; R_k \cap S_j  \neq \emptyset  \right\} . $$
Furthermore, we denote $a_k, \ b_k$ as the number of nonzero entries in $\tilde{u}_k, \tilde{v}_k$ respectively, i.e.
$$a_k \triangleq |\text{supp}(\tilde{u}_k)| = |R_k|, \;\; b_k \triangleq |\text{supp}(\tilde{v}_k)| = |S_k|, \quad k=1,\dots, K.$$
Since $\tilde{u}_k\neq 0, \tilde{v}_k\neq 0,\ \forall\; k$, it follows that $a_k\ge1$ and $b_k\ge1$.

We will bound $|\Omega^c |$ by Claim \ref{claim: generic system doesn't have a no-zero-component-solution} and bound $|\Omega|$ by the induction hypothesis.
We first provide an upper bound on $|\Omega^c |$.
Fix $r_k \in R_k, s_k \in S_k, \; k=1,\dots, K$. %% Since scaling does not affect the solutions of (\ref{SISO generic IA}),
In the system of equations (\ref{SISO generic IA, part a}), scale each $u_k$ by $1/u_{k r_k} $ and each $v_k$ by $1/v_{k s_k} $. Then we obtain a new system of polynomial equations,  denoted as $\mathcal{P}$, with $K(K-1)$ equations and $ 2K(T-1)$ variables.

For any $\left( h_{kj}(t) \right) \in \mathcal{H}  $, the IA condition (\ref{SISO generic IA}) has a solution $\{ \tilde{u}_k, \tilde{v}_k \}_{k=1,\dots, K}$ with supports
 \begin{equation}
 \text{supp}(\tilde{u}_k) = R_k, \; \text{supp}(\tilde{v}_k) = S_k, \quad k=1, \dots, K.
  \end{equation}
 Since $r_k \in R_k, s_k \in S_k$, we have $\tilde{u}_{k r_k} \neq 0, \tilde{v}_{k s_k} \neq 0$. Scale each $\tilde{u}_k$ by $1/\tilde{u}_{k r_k}$ and each $\tilde{v}_k$ by $1/\tilde{v}_{k s_k}$, we obtain a new set of beamformers $ \{  \tilde{u}_k /\tilde{u}_{k r_k},  \tilde{v}_k /\tilde{v}_{k s_k} \}_{k=1,\dots, K} $, which is a solution to the system $\mathcal{P}$.
  For simplicity, we still denote the scaled version  $ \{  \tilde{u}_k /\tilde{u}_{k r_k},  \tilde{v}_k /\tilde{v}_{k s_k} \}_{k=1,\dots, K} $ as
  $\{ \tilde{u}_k, \tilde{v}_k \}_{k=1,\dots, K}$ (i.e. assume $ \tilde{u}_{k r_k} = 1, \tilde{v}_{k s_k}=1 , \; \forall k$).

The beamforming vectors $ \tilde{u}_k, \tilde{v}_k, k=1,\dots, K$ can be concatenated to form a vector in $\mathbb{C}^{2K(T-1)}$
(after discarding the $2K$ entries $\tilde{u}_{kr_k}, \tilde{v}_{ks_k}, k=1,\dots, K$ since they have been normalized to 1).
Let $J$ be the support of this concatenated vector. Then $J$ is a fixed subset of $\{ 1,\dots, 2K(T-1)\}$ determined by $R_k \backslash\{ r_k\}, S_k\backslash\{s_k \}, \; k=1,\dots, K$. The size of $J$ is
% the supports of $u_k, v_k, k=1,\dots, K$ and
\begin{equation}\label{count J}
|J| = \sum_k (a_k -1) + \sum_k (b_k -1) = \sum_k (a_k + b_k) -2K.
\end{equation}

Consider the system of equations $\mathcal{P}_J$, obtained by restricting $\mathcal{P}$ to the subset of variables indexed by $J$ (i.e. set $u_{k r} = 0$ if $ r \notin R_k  $ and $v_{k s} = 0$ if $s \notin S_k$ in $\mathcal{P}$. See (\ref{eqn: truncated poly system}) for a formal definition of $\mathcal{P}_J$).
% becomes a new system of polynomial equations $\mathcal{P}_J$.
We claim that the number of nonzero polynomial equations in $\mathcal{P}_J$ is exactly $|\Omega^c|$, i.e.
\begin{equation}\label{count PJ}
|\mathcal{P}_J| = |\Omega^c|.
\end{equation}
Note that $u_k^H H_{kj} v_j = \sum_t u_{kt}^* h_{kj}(t) v_{jt}$. Then we have
%For $(k,j) \in \Omega$, $\text{supp}(\tilde{u}_k) \cap \text{supp}(\tilde{v}_j) = \emptyset $ implies $\text{supp}(\tilde{u}_k)^c \cup \text{supp}(\tilde{v}_k)  = \{1,\dots, N \}$.
%Then each $t$ is either in $(\text{supp}(\tilde{u}_k))^c $ or $(\text{supp}(\tilde{v}_j))^c $, thus each term $u_{kt}^* h_{kj}(t) v_{jt}$ will become zero in
%$\mathcal{P}_J$. Therefore,  $u_k^H H_{kj} v_j =0 $ becomes a zero equation $0=0$ in $ \mathcal{P}_{J}$ if $(k,j) \in \Omega$.
%For $(k,j) \in \Omega^c$, there exists one $t_{kj} \in \text{supp}(\tilde{u}_k) \cap \text{supp}(\tilde{v}_j)$, which implies $ u_{kt}^* h_{kj}(t) v_{jt} \neq 0 $ in $ \mathcal{P}_{J}$. Thus, $u_k^H H_{kj} v_j=0$ does not become a zero equation in $\mathcal{P}_{J}$ if $(k,j) \in \Omega^c$.
%Thus, (\ref{count PJ}) is proved.
\begin{align}
\notag       &    u_k^H H_{kj} v_j=0  \text{ does not become a zero equation in } \mathcal{P}_{J}  \\
\notag  \Longleftrightarrow \quad      &    \tilde{u}_k^H H_{kj} \tilde{v}_j \neq 0 \text{ (since the support of } \{\tilde{u}_k, \tilde{v}_k\} \text{ is } J ) \\
\notag  \Longleftrightarrow \quad  &  \exists \; t \text{ such that } \tilde{u}_{kt}^* h_{kj}(t) \tilde{v}_{jt} \neq 0  \\
\notag  \Longleftrightarrow  \quad & \exists  \; t \in \text{supp}(\tilde{u}_k) \cap \text{supp}(\tilde{v}_j)   \\
\notag  \Longleftrightarrow  \quad & \text{supp}(\tilde{u}_k) \cap \text{supp}(\tilde{v}_j) \neq \emptyset  \\
\notag  \Longleftrightarrow  \quad & (k,j) \in \Omega^c.
\end{align}
Therefore, $|\Omega^c|$ equals the number of nonzero equations in $\mathcal{P}_J$, which proves (\ref{count PJ}).

We then prove the following bound on $|\mathcal{P}_J|$ by Claim \ref{claim: generic system doesn't have a no-zero-component-solution}:
\begin{equation}\label{PJ bounded by J}
 |\mathcal{P}_J|  \leq |J|.
 \end{equation}
%We prove (\ref{PJ bounded by J}) by contradiction.
Assume the contrary, that $ |\mathcal{P}_J|  \geq |J| +1 $. According to Claim \ref{claim: generic system doesn't have a no-zero-component-solution}, the system
$\mathcal{P}_J$ has no solution in $(\mathbb{C}^*)^{|J|}$ for generic $\left( h_{kj}(t) \right)$.  %%% (or: for almost all $\left( h_{kj}(t) \right)$)
This contradicts the fact that for a positive measure of $\left( h_{kj}(t) \right)$ (in $\mathcal{H}  $), the system $\mathcal{P}_J$ has a solution $\{ \tilde{u}_k, \tilde{v}_k \}_{k=1,\dots, K}$ in $(\mathbb{C}^*)^{|J|}$. Thus (\ref{PJ bounded by J}) is proved.
 Plugging (\ref{count J}) and (\ref{count PJ}) into (\ref{PJ bounded by J}), we obtain an upper bound on $|\Omega^c|$:
\begin{equation}\label{bound on Omega^c, SISO}
|\Omega^c| \leq \sum_k (a_k + b_k) -2K.
\end{equation}

Next, we provide an upper bound on $|\Omega|$. Define
$$ \Omega_k = \left\{j \;| \; (k,j) \in \Omega \right\} =
\left\{ j \mid j\in \{1,\dots, K \}\backslash \{ k \}, \;  \text{supp}(\tilde{u}_k) \cap \text{supp}(\tilde{v}_j) = \emptyset   \right\}.$$
Then $\Omega = \bigcup_{k=1,\dots, K} \Omega_k $.
We will prove the following bound on $|\Omega_k|$:
\begin{equation}\label{SISO, bound Omega_k}
|\Omega_k | \leq f( T- a_k), \quad k=1,\dots, K.
\end{equation}
Without loss of generality, assume $\text{supp}(\tilde{u}_k) = \{1,2,\dots, a_k \}$.
Since $ \text{supp}(\tilde{v}_j)\cap \text{supp}(\tilde{u}_k) = \emptyset, \forall j\in \Omega_k$, we have
\begin{equation}\label{v outside support of u equals zero}
\tilde{v}_{j1} =\dots =\tilde{v}_{j a_k}=0, \quad \forall j\in \Omega_k.
 \end{equation}
 Consider a restriction of the beamforming vectors and channel matrices to the $(T-a_k)$ dimensional space. Specifically, we define
\begin{align}
\notag   x_j = ( \tilde{u}_{j, a_k+1}, \dots, \tilde{u}_{jT}  ), \;\;  y_j = ( \tilde{v}_{j, a_k+1}, \dots, \tilde{v}_{jT}  ), \quad \forall\ j\in \Omega_k , \\
\notag   \hat{H}_{ij} = diag\left( h_{ij}(a_k + 1), \dots, h_{ij}( T )  \right) , \quad \forall\ i,j \in \Omega_k .
\end{align}
It follows from (\ref{v outside support of u equals zero}) that $ \tilde{u}_i^H H_{ij} \tilde{v}_j = x_i^H \hat{H}_{ij} y_j , \forall\ i,j \in \Omega_k$. Since $\{ \tilde{u}_k, \tilde{v}_k \}_{k=1,\dots, K}$ satisfies the IA condition (\ref{SISO generic IA}), we have that $\{ x_k,y_k \}_{k=1,\dots, K}$ satisfies
\begin{equation}\label{reduced dimension IA condition}
\begin{split}
x_i^H \hat{H}_{ij} y_j = 0,    &   \quad\forall\ i\neq j \in \Omega_k,  \\
x_j^H \hat{H}_{jj} y_j \neq 0, & \quad \forall\ j \in \Omega_k.
\end{split}
\end{equation}
Note that (\ref{reduced dimension IA condition}) is the IA condition for a SISO IC with $(T-a_k)$ channel extensions and $|\Omega_k|$ users.
For a positive measure of $\left( h_{kj}(t) \right)$ (in $\mathcal{H}  $), (\ref{reduced dimension IA condition}) has a solution $\{ x_k,y_k \}_{k=1,\dots, K}$.
By the definition of $f(\cdot)$ in (\ref{definition of f(T), max K}), we have $|\Omega_k| \leq f(T-a_k)$, which proves (\ref{SISO, bound Omega_k}).

Since $a_k\ge1$, it follows from the induction hypothesis that $f(T-a_k) \leq 2(T-a_k)-1$ when $a_k < T$. Since $f(0)=0$, we have $f(T-a_k) = 2(T-a_k)$ when $a_k = T$. In summary, we have
\begin{equation}\label{bound of f(T-a_k)}
f(T-a_k) \leq 2(T-a_k), \quad k=1,\dots, K.
\end{equation}
Combining (\ref{SISO, bound Omega_k}) and (\ref{bound of f(T-a_k)}), we obtain
\begin{equation}\label{SISO, further bound Omega_k}
|\Omega_k | \leq 2( T- a_k), \quad k=1,\dots, K.
\end{equation}
Summing up (\ref{SISO, further bound Omega_k}) for $k=1,\dots, K$, we obtain $ |\Omega| \leq \sum_k 2(T-a_k) $.
Similarly, we can prove $|\Omega| \leq \sum_k 2(T-b_k)$. Combining these two bounds on $|\Omega|$, we get
\begin{equation}\label{bound on Omega, SISO}
| \Omega | \leq \frac{1}{2} \left( \sum_k 2(T-a_k) + \sum_k 2(T- b_k) \right) = 2KT - \sum_k (a_k + b_k).
\end{equation}
Finally, combining the bounds on $|\Omega|$ and $|\Omega^c|$ (cf.\ (\ref{bound on Omega^c, SISO}) and (\ref{bound on Omega, SISO})),
we can obtain the following bound on $K$
\[
              K(K-1) = |\Omega^c| + |\Omega| \leq   \sum_k (a_k + b_k) -2K + 2KT - \sum_k (a_k + b_k)
\]
which simplifies to $K-1\le -2+2T$ or equivalently $K\leq 2T -1$. Thus $f(T) \leq 2T -1 $ holds for $T$.
This completes the induction step, so that (\ref{SISO IC bound}) holds for any $T$, as desired.  % \QED

\section{Proof of Theorem \ref{thm: main result, bound on K} }\label{sec: proof of main result, diversity L bound}
%This section is devoted to the proof of Theorem \ref{thm: main result, bound on K}. %. The organization of this section is as follows.
%In section \ref{subsec: thm 1 analysis}, we discuss the difficulties in proving Theorem \ref{thm 1: DoF bound for generic extensions} and how we resolve these difficulties.
%Then we prove Theorem \ref{thm 1: DoF bound for generic extensions}
In this section, we will first analyze the difficulties of proving Theorem \ref{thm: main result, bound on K}, and then present the formal proof for $L=2$ in Section \ref{sec: proof of L=2, K bound by (LN)square}. The proof of Theorem \ref{thm: main result, bound on K} for general $L$ is similar to the proof for $L=2$, and is given in Appendix \ref{appen: proof of Thm 2 for general L}.
%The major difference is that we only consider the supports of $u_k, v_k$ in SISO IC, while in MIMO IC we consider both the supports and the ``block-supports'' of $u_k ,v_k$.
%This section is devoted to the proof of Theorem \ref{thm: main result, bound on K}.

\subsection{Preliminary Analysis}\label{subsec: thm 2 analysis}
Consider a $K$-user interference channel of diversity order $L$,  with each channel matrix given by $H_{ij} = \tau_{ij}^{1} A_1 + \dots + \tau_{ij}^L A_L$, where $A_1,A_2,\dots, A_L \in \Psi \subseteq C^{N_r \times N_t}$ (see the definition of $\Psi$ in (\ref{full rank guarantee})).  %% the coefficients $\tau_{i,j}^k$ are generic,
In the single-beam case (i.e., $d_k=1$ for each $k$), the IA condition (\ref{eqn: IA condition with U}) becomes
\begin{subequations}\label{diversity L IA}
\begin{align}
& u_k^H H_{kj} v_j =0, \quad \forall\ 1\leq k\neq j \leq K, \label{diversity L IA, part a}\\
& u_k^H H_{kk} v_k  \neq 0, \quad \forall\ 1 \leq k \leq K, \label{diversity L IA, part b}
\end{align}
\end{subequations}
where $v_j \in \mathbb{C}^{ N_t \times 1}, u_k \in \mathbb{C}^{ N_r \times 1}$ are transmit and receive beamforming vectors respectively.

%By the duality of interference alignment, we can assume that the transmit dimension $N_t$ is no more than the receive dimension $N_r$. More specifically, if $N_t > N_r$, we define a dual interference channel as follows.
%Define channel matrices
%　$\tilde{H}_{kj} = H_{jk}^H $, virtual transmit beamformers $ \tilde{v}_k = u_k$ and virtual receive beamformers $\tilde{u}_k = v_k$. The IA condition
%(\ref{diversity L IA}) is equivalent to the following IA condition for the dual interference channel:
%\begin{equation}\label{eqn: dual IA condition with U}
%\begin{split}
%& \tilde{u}_k^H \tilde{H}_{kj} \tilde{v}_j =0, \quad \forall\ k\neq j, 　　　　　　　\\
%& \tilde{u}_k^H \tilde{H}_{kk} \tilde{v}_k  \neq 0, \quad \forall\ k.  　\\
%\end{split}
%\end{equation}
%In the dual interference channel, the transmit dimension $N_r$ is smaller than the receive dimension $N_t$.

Due to the symmetry between the transmit and receive beamforming vectors in the IA condition \eqref{diversity L IA},
we can assume without loss of generality that $N_t \leq N_r$, which then implies
\begin{equation}\label{N=Nt assumption}
N = \min\{N_r, N_t\} = N_t.
\end{equation}

%The rest of this section is organized as follows.
%In Section \ref{subsec: thm 2 analysis}, we discuss the difficulties in proving Theorem \ref{thm: main result, bound on K} and how we resolve these difficulties.
%In Section \ref{subsec: prelim}, we overview some results in field theory \cite{Field}
%and prove a useful lemma that identifies the solvability of a special class of system of polynomial equations.
%We prove Theorem \ref{thm: main result, bound on K} for the case $L=2$ in Section \ref{sec: proof of L=2, K bound by (LN)square}.
% The proof of Theorem \ref{thm: main result, bound on K} for general $L$ is given in Appendix \ref{appen: proof of Thm 2 for general L}.
% % is similar to the proof for $L=2$, and
%% $dim( \text{span}\{ V_1, \dots, V_K \} ) =2 $

%\subsection{Preliminary Analysis}\label{subsec: thm 2 analysis}

{\black
The proof of Theorem \ref{thm 1: DoF bound for generic extensions} can not be directly applied to prove Theorem \ref{thm: main result, bound on K}.
There are two difficulties in applying the induction analysis and Lemma \ref{claim: generic system doesn't have a no-zero-component-solution}.
First, Lemma \ref{claim: generic system doesn't have a no-zero-component-solution}  only applies to a system with generic coefficients,
% is based on Bernstein's Theorem, which
whereas the coefficients of the system (\ref{diversity L IA, part a}) are not generic. To resolve this difficulty, we use Lemma \ref{lemma: rational count}
that determines the infeasibility of a system where each equation has the form that a generic parameter equals a rational function of variables.
}

%---------------------------------------------------------------------------------
% Therefore, we need other algebraic tools to determine the solvability of (\ref{diversity L IA, part a}).
%Reference \cite{MeisamIA} has analyzed the solvability of a special type of systems. Specifically, they show that if each equation takes the form that a generic parameter equals a polynomial function of the variables, then a necessary condition for the solvability is that the number of equations does not exceed the number of equations. We generalize this result to a system where each equation has the form that a generic parameter equals a rational function of variables. See Section \ref{subsec: prelim} for more details.
%---------------------------------------------------------------------------------
% \textbf{Difficulty 2}: As mentioned in Section (\ref{subsec: thm 1 analysis}), we can use induction analysis to implicitly utilize
%the system of inequalities (\ref{diversity L IA, part b}). The induction analysis applies to the case that a subset $S$ of $v_j$'s
%have common zero components, since these $v_j$'s satisfy an IA condition with lower ambient dimension.

The second difficulty has to do with the use of induction analysis. For the IA condition (\ref{diversity L IA}), we need to consider the case
that a subset $S$ of $v_j$'s lie in a $N'$-dim linear space, where $N' < N$.
To this end, we can express $v_j$ as $v_j= D \bar{v}_j, j\in S$,
where $D \in \mathbb{C}^{N \times N'}, \bar{v}_j \in \mathbb{C}^{N' \times 1} $.
Then $ u_k^H H_{kj} v_j = u_k^H H_{kj}D \bar{v}_j = u_k^H \bar{H}_{kj} \bar{v}_j, \forall k,j \in S$, where the new channel matrix $ \bar{H}_{kj} = H_{kj}D
= \sum_l \tau_{kj}^l A_l D \in \mathbb{C}^{N_r \times N'}$. At the first glance, it seems that $\{ u_j, \bar{v}_j\}_{j\in S}$ satisfies a new IA condition with
 different building blocks $A_l D, l=1,\dots, L$ and a lower ambient transmit dimension $N'$.
 However, the coefficients $\{\tau_{kj}^l\}$ can not be viewed as generic since they are not independent of $D$, the basis matrix of
 $\text{span}\{v_j, j\in S \}$ (recall that $ v_j$'s are design variables that depend on the channel coefficients $\{ \tau_{kj}^l\}$).  As a result, the induction hypothesis can not be applied to the lower dimensional IA problem.
 To resolve this difficulty, we use a ``lifting'' technique. Specifically, we show that the original IA condition (\ref{diversity L IA}) implies a
 ``lifted'' IA condition in a higher dimensional space.
 %-------------------------------------Deleted--------------------------
  % Thus, to upper bound the number of users $K$ such that (\ref{diversity L IA}) is feasible, we only need to bound $K$ such that the lifted IA condition is feasible.
  %---------------------------------------------------------------
  As will be seen, the induction analysis can be applied to the lifted IA condition.

%----------------------Deleted (Replaced)----------------------------------------------
% However, the new ``channel matrix'' $\bar{H}_{kj}$ depends on $D$, the basis matrix of $\text{span}\{v_j, j\in S \}$ which can be designed by us.
 % $\bar{H}_{kj}$'s cannot be viewed as the channel matrices for an interference channel.
 %% and $u_k^H \bar{H}_{kj} \bar{v}_j =0$ cannot be viewed as an IA condition.
 %-------------------------------------------------------------------------------------
% {\black
% Such a difficulty does not exist when proving Theorem \ref{thm 1: DoF bound for generic extensions}: the basis matrix $D$ there consists of $0$'s and $1$'s,
% thus does not contain any design variable. }

\subsection{ Proof of Theorem \ref{thm: main result, bound on K} for $L=2$}\label{sec: proof of L=2, K bound by (LN)square}

%%%The following proposition is a weaker version of Theorem \ref{thm: main result, bound on K}.
%%%\begin{proposition}\label{prop: weaker result, bound on K by (LN)square}
%%%Consider a $K$-user interference channel with the channel matrices $ H_{ij} = \tau_{ij}^1 A_1 + \dots + \tau_{ij}^L A_L  \in \mathbb{C}^{N_r \times N_t}$,
%%% where $(A_1,\dots, A_L) \in \Psi $ as defined in (\ref{full rank guarantee}). Define $N = \min \{N_t, N_r \}.$
%%% If for a positive measure of $(\tau_{ij}^l)$,
%%%the DoF tuple $(d_1,\dots, d_K) = (1,1,\dots, 1) $ is achievable via vector space IA, then
%%% \begin{equation}
%%% K \leq \frac{(L+1)}{2}N(N+1).
%%%\end{equation}
%%%\end{proposition}
%%% \emph{Proof of Proposition \ref{prop: weaker result, bound on K by (LN)square} when $L=2$ }:

The proof of Theorem \ref{thm: main result, bound on K} consists of two steps: first,
we ``lift'' the IA condition to a higher dimensional IA condition; second, using Lemma \ref{lemma: rational count},
we prove the bound for the ``lifted'' IA condition by induction on $N$.

In this subsection, we prove Theorem \ref{thm: main result, bound on K} for the case $L=2$. The case of general $L$ can be treated similarly and is given in Appendix \ref{appen: proof of Thm 2 for general L}.
As mentioned in Section \ref{subsec: thm 2 analysis}, without loss of generality we can assume \eqref{N=Nt assumption}.
We can also assume $\tau_{ij}^1 = 1, \forall\ i,j $ since multiplying $H_{ij}$ by a nonzero scalar $1/{\tau_{ij}^1}$
does not affect the IA condition (\ref{eqn: single beam IA condition}).
For simplicity, we further denote $\tau_{ij}^2$ as $\tau_{ij}$. Now the channel model becomes
 \begin{equation}\nonumber%\label{eqn: diversity 2 channel model, simplified}
 H_{ij} =  A_1 + \tau_{ij} A_2.
 \end{equation}
Denote $f(N)$ as the maximal $K$ such that IA is feasible; more specifically,
\begin{equation}
 \begin{split}
f(N) \triangleq \max \{K \;|\; & \text{for a positive measure of } (\tau_{ij}), \\
                       &\text{ the DoF tuple } (d_1,\dots, d_K) = (1,1,\dots, 1) \text{ is achievable via vector space IA} \} \\
     = \max \{K \;|\; & \text{for a positive measure of } (\tau_{ij}), \exists \;\{v_k\}_{k=1,\dots, K}, \; s.t. \text{ IA condition } (\ref{eqn: single beam IA condition}) \text{ holds } \}. \\
  \end{split}
 \end{equation}
 Note that $f(N)$ depends on the receive dimension $N_r$ and the building blocks $A_1, A_2$.
What we need to prove is: for any $N_r$ and $A_1,A_2 \in \Psi$ as defined in (\ref{full rank guarantee}), we have
\begin{equation}
f(N) \leq 2N + \frac{N^2}{4}.
\end{equation}

Bounding $f(N)$ directly by the IA condition is difficult. Instead, we consider a lifted IA condition:
\begin{equation}\label{eqn: L=2 lifted IA condition}
\begin{split}
\left( \begin{matrix} v_k \\ \tau_{kk}v_k \end{matrix} \right) \;\; &   \Perp \;\; \text{ span} \left\{  \left( \begin{matrix} v_j \\ \tau_{kj}v_j \end{matrix}  \right) \left. \;\right| \; j\in \{1,2,\dots, K \} \backslash \{ k\}   \right\}, \quad \forall\ 1 \leq k \leq K, \\
  v_k & \in \mathbb{C}^{N \times 1}\backslash \{ 0 \}, \quad \forall\ 1\leq k \leq K.\\
\end{split}
\end{equation}
% \text{ span}_{j\in \{1,2,\dots, K \} \backslash \{ k\}} \left\{  \left( \begin{matrix} v_j
Define $g(N)$ as the maximal $K$ for the lifted IA condition to hold; more specifically,
\begin{equation}
g(N)  = \max \{K  \;| \;  \text{for a positive measure of } (\tau_{ij}), \exists \;\{v_k\}_{k=1,\dots, K}, \; s.t. \text{ IA condition } (\ref{eqn: L=2 lifted IA condition}) \text{ holds} \}. \\
\end{equation}
Note that $g(N)$ does not depend on $N_r, A_1, A_2$.
The following lemma shows that $f(N)$ is upper bounded by $g(N)$.
\begin{lemma}\label{L=2, lemma: f(N)<=g(N)}
For any $N_r, N $ and any $N_r \times N$ building blocks $A_1, A_2 \in \Psi$ as defined in (\ref{full rank guarantee}), we have
\begin{equation}
f(N) \leq g(N).
\end{equation}
\end{lemma}
\emph{Proof of Lemma \ref{L=2, lemma: f(N)<=g(N)}:}
We only need to prove: for any given $(\tau_{ij})$, if $\{v_k\}_{k=1}^{K}$ satisfies the IA condition (\ref{eqn: single beam IA condition}),
 then $\{v_k\}_{k=1}^{K}$ satisfies the lifted IA condition (\ref{eqn: L=2 lifted IA condition}).
  We prove by contradiction. Assume $\{v_k\}_{k=1}^{K}$ does not satisfy (\ref{eqn: L=2 lifted IA condition}), then the independence condition does not hold for some $k$, i.e.
  \begin{equation} \label{dependence IA, L=2}
  \exists \lambda_1, \dots, \lambda_K, \; s.t. \quad   v_k            =  \sum_{j\neq k} \lambda_j v_j,                  \quad
                                                        \tau_{kk} v_k  =  \sum_{j\neq k} \lambda_j \tau_{kj}v_j.
\end{equation}
 Left multiplying the $l¨$th equation of (\ref{dependence IA, L=2}) by $A_l, l=1,2$ and summing up  yields
 $$
H_{kk}v_k = \sum_{j\neq k} \lambda_j H_{kj}v_j,
 $$
where we have used the equation $H_{kj} = A_1 + \tau_{kj}A_2$.
This contradicts the assumption that $\{v_k \}$ satisfies the IA condition (\ref{eqn: single beam IA condition}).    \QED

According to Lemma \ref{L=2, lemma: f(N)<=g(N)}, we only need to prove
\begin{equation}\label{ineqn: L=2, bound on g}
g(N) \leq 2N + \frac{N^2}{4}.
\end{equation}
We prove (\ref{ineqn: L=2, bound on g}) by induction on $N$.
For the basis of the induction, we consider the case $N = 1$. We need to prove $K \leq  2N+{ N^2 }/{4}  = 2.25$ or equivalently $K\le 2$.
Assume $K \geq  3$ and for a positive measure of $(\tau_{kj})$, the lifted IA condition (\ref{eqn: L=2 lifted IA condition}) has a solution $\{v_k\}_{k=1}^K$.
Since scaling each $v_j$ by a nonzero factor ${1}/{v_j}$ does not affect the linear independence condition, we have
 \begin{align}
 & \left( \begin{matrix} v_k \\ \tau_{kk}v_k \end{matrix} \right) \;\;    {\Huge \Perp} \;\; \text{ span} \left\{  \left( \begin{matrix} v_j \\ \tau_{kj}v_j \end{matrix}  \right) \left. \;\right| \; j\in \{1,2,\dots, K \} \backslash \{ k\}   \right\}, \quad \text{for }\; k=1,  \notag \\
\Longleftrightarrow \quad & \left( \begin{matrix} 1 \\ \tau_{11} \end{matrix} \right) \;\;   {\Huge \Perp} \;\; \text{ span} \left\{  \left( \begin{matrix} 1 \\ \tau_{12} \end{matrix} \right) , \left( \begin{matrix} 1 \\ \tau_{13} \end{matrix} \right)  \right\}.   \label{L=2, N=1 independence}
\end{align}
The linear independence relation (\ref{L=2, N=1 independence}) can not hold for a positive measure of $(\tau_{11},\tau_{12},\tau_{13})$,  a contradiction. Therefore, (\ref{ineqn: L=2, bound on g})
holds for $N=1$.

Suppose (\ref{ineqn: L=2, bound on g}) holds for $1,\dots, N-1$. We will prove (\ref{ineqn: L=2, bound on g}) holds for $N$.
To this end, suppose $K$ satisfies that for a positive measure of  $(\tau_{ij})$, $\exists \; v_k \in \mathbb{C}^{N\times 1}, k=1,\dots, K,$ such that the lifted IA condition (\ref{eqn: L=2 lifted IA condition}) holds.
Let us introduce the receive beamforming vector $u_k = \left( \begin{matrix} u_k^1 \\ u_k^2 \end{matrix} \right) \in \mathbb{C}^{2N\times 1} $, where $u_k^1, u_k^2 \in \mathbb{C}^{N \times 1}$, and transform the lifted IA condition (\ref{eqn: L=2 lifted IA condition}) to the following zero-forcing type IA condition:
\begin{subequations}  \label{eqn: L=2 lifted IA with U}
\begin{align}
u_k^H   \left( \begin{matrix} v_j \\ \tau_{kj}v_j \end{matrix} \right) &  =0, \quad\Longleftrightarrow \quad (u_k^1)^H v_j + \tau_{kj} (u_k^2)^H v_j = 0, \quad\forall\ k\neq j,
\label{L=2, lifted IA (a)} \\
 u_k^H \left( \begin{matrix} v_k \\ \tau_{kk}v_k \end{matrix} \right)  &  \neq 0,\quad \forall\ k.  \label{L=2, lifted IA (b)}
\end{align}
\end{subequations}
Define a set of transmitter-receiver pairs as
$$ \Omega = \left\{ (k,j) \;|\; 1\leq k\neq j \leq K, \; (u_k^1)^H v_j = (u_k^2)^H v_j = 0  \right\}.$$ The complement of $\Omega$ with respect to
$ \{(k,j)  \;|\; 1\leq k\neq j \leq K \}$ is given by
$$\Omega^c = \left\{ (k,j) \;|\; 1\leq k\neq j \leq K,  \; (u_k^2)^H v_j \neq 0  \right\} . $$
In addition, define
$$ \Omega_k = \left\{j \mid  (k,j) \in \Omega \right\} = \left\{ j \mid j\in \{1,\dots, K \}\backslash \{ k \}, \;  (u_k^1)^H v_j = (u_k^2)^H v_j = 0  \right\}.$$
Then $\Omega = \bigcup_{1\leq k \leq K} \Omega_k$.

Notice that for a positive measure of $(\tau_{ij})$, the lifted zero-forcing type IA condition (\ref{eqn: L=2 lifted IA with U}) has a solution $\{u_k, v_k\}_{k=1,\dots, K}$. In other words,
each $(\tau_{ij})$ corresponds to at least one solution $\{u_k, v_k\}_{k=1,\dots, K}$ which defines a collection of sets $\{ \Omega_k \}_{k=1,\dots, K} \subseteq \{1,\dots, K \}^K$. Since there are finitely many
subsets of $\{1,\dots, K \}^K$, we must have one collection of sets $\{ \Omega_k \}_{k=1,\dots, K}$ that corresponds to a positive measure of $( \tau_{ij} )$.
Denote this set of $\{ \tau_{ij} \}$ (which has a positive measure) as $\mathcal{H}_0$.
%% Each $(\tau_{ij})$ corresponds to at least one subset $\Omega \subseteq \{1,\dots, K \} \times \{1,\dots, K \}$. There are finitely many subsets of
%% $\{1,\dots, K \} \times \{1,\dots, K \}$, thus there exists at least one subset $\Omega_0$,
% such that a positive measure of $(\tau_{ij})$ correspond to $\Omega_0$. For simplicity, we still denote $\Omega_0$ as $\Omega$.

Consider the collection of sets $\{\Omega_k \}_{1\leq k \leq K}$ that corresponds to $( \tau_{ij} ) \in \mathcal{H}_0$.
We will bound $|\Omega|$ by induction   and bound $|\Omega^c |$ by Lemma \ref{lemma: rational count}. Specifically,
to derive an upper bound on $|\Omega|$, we use the definition of $\Omega_k$ to obtain
%% possible choices of $\{ \Omega_k \}_{k=1}^K$,
% For any $k\in \{ 1,\dots, K\}$,
% By the definition of $\Omega_k$, we have
%\begin{subequations}\label{ortho: U_k and V_j}
%\begin{align}
%    &
\[
\mathcal{U}_k \triangleq \text{span}\{ u_k^1, u_k^2 \} \perp \mathcal{V}_k \triangleq \text{span} \{ v_j \;|\; j\in \Omega_k \}, \quad \forall \; 1\leq k \leq K,
  \label{L=2, ortho: U_k and V_j in Omega_k}
\]
which implies
\begin{equation}
 \text{dim}( \mathcal{V}_k ) + \text{dim} ( \mathcal{U}_k )  \leq N, \quad \forall \; 1\leq k \leq K.  \label{ineq: L=2, sum dim of U,V bounded by N}
\end{equation}
Since $u_k = \left( \begin{matrix} u_k^1 \\ u_k^2  \end{matrix} \right) $ is a nonzero vector, $\text{dim} ( \mathcal{U}_k ) \geq 1 $.
Then $$ p_k \triangleq \text{dim}( \mathcal{V}_k ) \leq N - \text{dim} ( \mathcal{U}_k )  \leq N-1.  $$
Although all $(\tau_{ij})$ in $\mathcal{H}_0$ correspond to the same collection of sets $\{\Omega_k \}_{k=1,\dots, K}$, they may correspond to different
sets of dimensions $\{ p_k \}_{k=1,\dots, K}$.
 Using a similar argument as before, we can show that there exist a positive measure of $( \tau_{ij} )$ in $\mathcal{H}_0$  which corresponds to the same set of dimensions $\{ p_k \}_{k=1,\dots, K}$.
These $(\tau_{ij})$ form a subset of $\mathcal{H}_0$, denoted as $\mathcal{H}$.

Consider the collection of dimensions $\{ p_k \}_{k=1,\dots, K}$ that corresponds to $( \tau_{ij} ) \in \mathcal{H}$.  We prove that
\begin{equation}\label{Omega k bound}
|\Omega_k | \leq g(p_k) \leq 2 p_k + \frac{p_k^2}{4}, \quad \forall \; 1\leq k \leq K.
\end{equation}

Let $D \in \mathbb{C}^{N \times p_k}$ be the basis matrix of $ \mathcal{V}_k $, and suppose $v_j  = D \bar{v}_j, \forall\ j\in \Omega_k $,
 where $\bar{v}_j \in \mathbb{C}^{p_k}\backslash \{ 0\}$. From the lifted IA condition (\ref{eqn: L=2 lifted IA condition}), we have
\begin{subequations}
\begin{align}
\notag   & \left( \begin{matrix} v_i \\ \tau_{ii}v_i \end{matrix} \right) \;\;    \Perp \;\; \text{ span} \left\{  \left( \begin{matrix} v_j \\ \tau_{ij}v_j \end{matrix}  \right) \left. \;\right| \; j\in \Omega_k \backslash \{ i\}   \right\}, \quad \forall\ i\in \Omega_k, \\
 \Longleftrightarrow \quad & \left( \begin{matrix} D \bar{v}_i \\ \tau_{ii}D \bar{v}_i \end{matrix} \right) \;\;    \Perp \;\; \text{ span} \left\{  \left( \begin{matrix} D \bar{v}_j \\ \tau_{ij}D \bar{v}_j \end{matrix}  \right) \left. \;\right| \; j\in \Omega_k \backslash \{ i\}   \right\}, \quad \forall\ i\in \Omega_k,
 \label{L=2, part a of induc lifted IA} \\
\Longrightarrow \quad & \left( \begin{matrix} \bar{v}_i \\ \tau_{ii} \bar{v}_i \end{matrix} \right) \;\;    \Perp \;\; \text{ span} \left\{  \left( \begin{matrix} \bar{v}_j \\ \tau_{ij} \bar{v}_j \end{matrix}  \right) \left. \;\right| \; j\in \Omega_k \backslash \{ i\}   \right\}, \quad \forall\ i\in \Omega_k.
\label{L=2, induc lifted IA}
\end{align}
\end{subequations}
The last step can be proved by contradiction. If (\ref{L=2, induc lifted IA}) does not hold, then the LHS (left-hand side) of (\ref{L=2, induc lifted IA})
belongs to the space on the RHS (right-hand side) of (\ref{L=2, induc lifted IA}). Multiply both sides by
 $ \left( \begin{matrix} D  & 0 \\ 0 &  D \end{matrix} \right) $, we obtain that the LHS of (\ref{L=2, part a of induc lifted IA})
belongs to the space on the RHS of (\ref{L=2, part a of induc lifted IA}), which contradicts (\ref{L=2, part a of induc lifted IA}).

Now for a positive measure of $( \tau_{ij} )_{i,j \in \Omega_k }$, there exist $\bar{v}_i \in \mathbb{C}^{p_k}\backslash \{ 0\}, i\in \Omega_k $ that satisfy (\ref{L=2, induc lifted IA}).
By the definition of $g(\cdot)$, $|\Omega_k | \leq g(p_k) $. Using the induction hypothesis, $g(p_k) \leq 2 p_k + \frac{p_k^2}{4}$. Thus, (\ref{Omega k bound})
is proved.
Summing up (\ref{Omega k bound}) for $k=1,\dots, K$, we have
\begin{equation}\label{L=2, estimate Omega}
|\Omega | = \sum_{k=1}^K |\Omega_k | \leq  \sum_k \left( 2p_k + \frac{ p_k^2}{4} \right) .
\end{equation}

Next, we provide an upper bound on $|\Omega^c |$. Equation (\ref{L=2, lifted IA (a)}) implies that
\begin{equation}\label{eqn: L=2, tau expressed as rational}
\tau_{kj} = - \frac{ ( u_k^1 )^H v_j }{ (u_k^2)^H v_j } , \quad \forall\ (k,j)\in \Omega^c.
\end{equation}
The system of equations (\ref{eqn: L=2, tau expressed as rational}) in variables $u_i^1,u_i^2,v_i, i=1,\dots, K$ (parameterized by $(\tau_{kj})$) has $|\Omega^c|$ equations.

We compute the number of free variables in $\{u_k, v_k \}_{1\leq k \leq K}$.
Since scaling $v_j  $ does not affect the system of equations (\ref{eqn: L=2, tau expressed as rational}), we can scale each $v_j$ to make
one of its entries to be $1$. Therefore, the number of variables in $\{ v_i\}_{1\leq i \leq K}$ is $K(N-1)$.%
%\footnote{A more rigourous argument is as follows. Suppose we scale the $t_j$'th entry of $v_{j} $ to be one. Then each $(\tau_{kj})$ corresponds to a set of positions $\{ t_j\}$.
%Again, a positive measure of $(\tau_{kj})$ correspond to the same set of positions $\{t_j \}$; for these $(\tau_{kj}) $, (\ref{eqn: L=2, tau expressed as rational}) can be transformed to a new system with $K(N-1)$ variables in $\{ v_i\}_{1\leq i \leq K}$. A similar argument can be applied to counting
%variables in $\{u_k \}$ as well.}
To count the free variables in $u_k$, notice that the condition (\ref{L=2, ortho: U_k and V_j in Omega_k}) implies that
$(u_k^1)^Hv_j=0$ for all $j\in \Omega_k$. Since $\text{span}\{v_j:j\in\Omega_k \}$ has dimension $p_k$, it follows that
$p_k$ entries of $u_k^1$ can be written as linear functions of the remaining $N-p_k$ entries of $u_k^1$ (with coefficients being the rational functions of $\{v_j:j\in\Omega_k \}$). Similarly, $p_k$ entries of $u_k^2$ can be represented as linear functions of the remaining $N-p_k$ entries of $u_k^2$, with coefficients being some rational functions of $\{v_j:j\in\Omega_k \}$. Substituting these linear functions into the right hand sides of \eqref{eqn: L=2, tau expressed as rational} yields a new representation of each $\tau_{kj}$ as a rational function of the $2(N-p_k)$ free variables in $u^1_k$, $u^2_k$ as well as the $K(N-1)$ variables in $v_j$'s.
 Because of the homogeneity of these rational functions over the $2(N-p_k)$ free entries of $u_{k}$, we can further scale $u_{k} $
to make one of these entries to be $1$.
Thus, the number of free variables in $u_k^1, u_k^2$ is $2(N-p_k) - 1$.
In summary, the number of free variables in $\{u_k, v_k \}_{k=1}^K$ is
$$ K(N-1) + \sum_k \left( 2(N-p_k) - 1 \right) = K(3N-2) - 2 \sum_k p_k. $$

% The number of variables in $\{u_i^1,u_i^2,v_i\}_{1\leq i \leq K}$ is $K(2N+N) = 3KN$.
For a positive measure of $(\tau_{kj})_{(k,j) \in \Omega^c}$, the rational system (\ref{eqn: L=2, tau expressed as rational}) has a solution $\{u_k^1,u_k^2,v_k\}_{1\leq k \leq K}$
such that $ (u_k^2)^H v_j \neq 0 $.
It follows from Lemma \ref{lemma: rational count} that
the number of equations should not exceed the number of variables, i.e.
\begin{equation}\label{L=2, estimate Omega^c}
|\Omega^c| \leq  K(3N-2) - 2 \sum_k p_k.
\end{equation}

With the bounds on $|\Omega|$ and $|\Omega^c|$, we can now provide an bound on $K$.
Summing up (\ref{L=2, estimate Omega^c}) and (\ref{L=2, estimate Omega}) yields
\[
 K(K-1) = |\Omega| + |\Omega^c| \leq \sum_k \left( 2p_k + \frac{p_k^2}{4} \right)   + K(3N-2) - 2 \sum_k p_k
 \]
 implying
\begin{equation}
K(K-1) \leq K(3N-2) + \sum_k \frac{p_k^2}{4}
\quad\mbox{or equivalently}\quad
K \leq 3N-1 + \frac{1}{K}\sum_k \frac{p_k^2}{4}  \label{L=2, adding Omega and Omega^c}.
\end{equation}
% and use $0 \leq p_k \leq N-1 $, we have
%\leq K(3N-2) + K \frac{ (N-1)(N - 2)}{2} \\
%  \Longrightarrow \quad  & K \leq  3N -1 + \frac{ (N-1)(N - 2)}{2}  = \frac{N(N+3)}{2}. \label{induction step, K bound}
%Thus, $ g(N) \leq \frac{N(N+3)}{2}$, i.e. (\ref{ineqn: L=2, bound on g}) holds for $N$.
Thus, if $p_k \leq N-2,\ \forall\ k$, then (\ref{L=2, adding Omega and Omega^c}) leads to
\begin{equation}\label{eq:case1}
K \leq 3N -1 + \frac{1}{K}\sum_k \frac{p_k^2}{4} \leq 3N - 1 + \frac{(N-2)^2}{4} = 2N + \frac{N^2}{4}
\end{equation}
as desired.

It remains to consider the case $p_k=N-1$ for some $k$.
We need to prove the following claim.
\begin{claim}\label{L=2, claim of bounding p_k}
If there exists $k$ such that $p_k = N-1$, then
\begin{equation}\label{if p=N-1, bound K}
K \leq |\Omega_k| + 2.
\end{equation}
\end{claim}
\emph{Proof of Claim \ref{L=2, claim of bounding p_k}}:
 According to (\ref{ineq: L=2, sum dim of U,V bounded by N}), $1\leq dim(\mathcal{U}_k) \leq N - p_k = 1$, thus $dim(\mathcal{U}_k) = 1$.
Then $u_k^1 $ is parallel to $u_k^2$. Without loss of generality, we assume $ u_k^1 = \gamma u_k^2 $; then $\mathcal{U}_k = span\{ u_k^2\}$.

According to the IA condition (\ref{L=2, lifted IA (a)}), we have
$$ 0= (u_k^1)^H v_j + \tau_{kj} (u_k^2)^H v_j = (u_k^2)^H v_j( \gamma + \tau_{kj} ) , \quad \forall\ j \in \{1,\dots, K \}\backslash\{k\} . $$
Since $\gamma$ can be equal to at most one $-\tau_{kj}$, we have that $\gamma + \tau_{kj} \neq 0$ for at least $K-2$ $j$'s. Therefore,
$$ 0 = (u_k^2)^H v_j \Longleftrightarrow u_k^2 \perp v_j  $$ holds for at least $(K-2)$ $j$'s. Hence, $ \Omega_k = \{j \;|\; \mathcal{U}_k \perp v_j \}$ has
at least $(K-2)$ elements, which proves (\ref{if p=N-1, bound K}). \QED

To complete the induction step, suppose $p_k = N-1$. Using (\ref{if p=N-1, bound K}) and (\ref{Omega k bound}), we have
$$
K \leq |\Omega_k| + 2 \leq \frac{p_k^2}{4} + 2p_k +2 = \frac{(N-1)^2}{4} + 2(N-1) +2 < \frac{N^2}{4} + 2N
$$
as desired. Combining this with \eqref{eq:case1} yields $K \leq 2N + \frac{N^2}{4}$, which further implies $g(N) \leq 2N + \frac{N^2}{4}$ holds for $N$. This completes the induction step, so that (\ref{ineqn: L=2, bound on g}) holds for any $N$.
Finally, combining (\ref{ineqn: L=2, bound on g}) and Lemma \ref{L=2, lemma: f(N)<=g(N)}, we obtain $f(N)\leq g(N) \leq 2N + \frac{N^2}{4}$. \QED

\text{ }\newline
% \newpage
\appendix
\par\noindent
{\Large\bf Appendix}

%----------------------Move to Appendix -------------------------------------------------------
\section{ Bernstein's Theorem and Dimensionality Counting Argument }\label{appen: Bernstein's thm}

%In this appendix, we clarify the use of Bernstein's theorem in the IA context.
%for two reasons.
%First, a fact that may not be explicitly recognized is that Bernstein's theorem
%Second, it has several different versions in the algebraic geometry literature, and some of them may not be useful for the interference alignment problem.

%Consider a system of polynomial equations
%\begin{equation}\label{eqn: general poly system}
%\mathcal{P}: \quad f_i(x_1, \dots, x_n) = 0,\quad i=1,\dots, m,
%\end{equation}
%where $f_i, i=1,\dots, m$ are polynomials in variables $x_1, \dots, x_n$.
%A tempting conjecture is the following: % (used in \cite[Section IV.B]{SongIA}):
%\begin{equation}\label{misconcept 2}
%\mbox{A polynomial system } \mathcal{P} \mbox{ with generic coefficients is solvable only if }   m \leq n.
% \end{equation}
%% independently randomly generated from continuous distributions
%Here, we assume each polynomial in $\mathcal{P}$ has fixed monomial terms and the coefficients of these monimials are generic
%  (e.g. independently generated from continuous random distributions).
%The statement (\ref{misconcept 2}) is incorrect in general. One simple counterexample is the following:

%%------------------Move to Appendix ------------------------------------------------------------------------------------
In this appendix, we clarify the use of Bernstein's theorem in the IA context where it was first cited in \cite{YetisIA} as follows.
Suppose $f_i$ is a polynomial function with $m_i$ monomials, and $c_{ij}$ is the coefficient of the $j$'th monomial, $ j\in \{1,2,\dots, m_i \}$. The number of common solutions of polynomials $f_i,\ \forall\ i$ is simply the number of solutions of the system equations $f_i=0,\ \forall\ i$.
%The version of Bernstein's theorem stated in \cite{YetisIA} is the following:
%----------------Bernsein's thm in Yetis -----------------------
%\begin{claim}\label{claim in Yetis, wrong Bernstein}
\par\noindent
{\bf Bernstein's theorem \cite[Theorem 4]{YetisIA}}
\emph{Given $n$ polynomials $f_1,\dots, f_n \in \mathbb{C}[x_1,\dots, x_n]$ with common solutions in $(\mathbb{C}^*)^n$, let $P_i$ be the Newton polytope of $f_i$ in $\mathbb{R}^n$. For independent random coefficients $c_{ij}, \forall\ i\in \{1,\dots, n \}$ and $\forall\ j\in \{1,2,\dots, m_i \}$, the number of common solutions is exactly equal to the mixed volume
of Newton polytopes, $MV(P_1,\dots, P_n)$.
}
%\end{claim}

We will not define the Newton polytope $P_i$ and mixed volume $MV(P_1,\dots, P_n)$ here; interested readers can refer to \cite{YetisIA} or \cite{usingAG} for the definitions and more information.
The definition of the mixed volume may be useful for the achievability of DoF, though the difficulty seems to lie in how to determine whether the mixed volume is nonzero \cite{YetisIA} (see some related work in \cite{gonzalez2012feasibility,gonzalez2013finding}); nevertheless, these are beyond the scope of this paper.
For our purpose of deciding the infeasibility of IA systems, we only need to know that the mixed volume $MV(P_1,\dots, P_n)$ is a finite number for any polynomials $f_1,\dots, f_n$. In other words, we are interested in the following corollary of Bernstein's theorem.
\begin{coro}\label{coro: finiteness in C*n}
  A generic system of polynomial equations with $n$ equations and $n$ variables has a finite number of solutions in $(\Cs)^n$.
 \end{coro}

% and its extension such as the stable mixed volume).
% They may be useful for achievability, though the difficulty seems to lie in how to compute the mixed volume and its extension such as the stable mixed volume.
We emphasize that in the above Bernstein's theorem, the term ``common solutions'' should be interpreted as ``common solutions in $(\mathbb{C}^*)^n$'', that is, ``\emph{strictly nonzero} common solutions (i.e., with no zero entries)''. In fact, the number of ``common solutions in ${\C}^n$'' of a generic system can be infinite; see Example 4.4. % ; see \cite[Theorem 3.1]{Emiris1995},\cite[Thereom 3.2]{sturmfels}.
 %%%------------------Deleted; ------------------------------------------------------------------------------------
 Bernstein's theorem has been applied to IA problems in \cite{YetisIA, ShiIA,BengIA,SongIA} to derive performance bounds.
 For the original application in the constant MIMO IC \cite{YetisIA}, the distinction between solutions in $\mathbb{C}^n$ and $(\Cs)^n$ can be ignored due to the structureless property of the channel (see Section \ref{subsec: proper and feasible} for discussion).
 For other problems in \cite{ShiIA,BengIA,SongIA}, this distinction cannot be ignored and the results in these references
 require the artificial assumption that the IA solutions are strictly nonzero, though this assumption is not explicitly stated in these references.

 An important clarification regarding the use of Bernstein's theorem is: many versions of Bernstein's theorem in the algebraic geometry literature do not imply Corollary \ref{coro: finiteness in C*n}, thus do not imply the results in \cite{YetisIA, ShiIA,BengIA,SongIA} (even under the additional assumption
 of restricting to $(\Cs)^n$) and Corollary \ref{coro: finiteness in C*n}.
 % Lemma \ref{claim: generic system doesn't have a no-zero-component-solution}.
% Yet another confusion may arise from
% In popular version of Bernstein's theorem (see e.g. \cite[Theorem 1]{SturmCn},\cite[Theorem1.1]{BKKroot},\cite[Theorem 7.1.4]{dickenstein2005solving}), though % it does not explicitly make the finiteness assumption.
 In particular, \cite[Theorem 1]{SturmCn},\cite[Theorem1.1]{BKKroot} and \cite[Theorem 7.1.4]{dickenstein2005solving} state that the number of isolated zeros in $(\Cs)^n$ of any system $\mathcal{P}$ (possibly non-generic) is upper bounded by $MV(P_1,\dots, P_n)$, and the bound is exact for generic choices of the coefficients.
% %-----------------------------------------------------------------------
%  This result seems to imply that the number of solutions in $(\Cs)^n$ of a non-generic system is no more than that of a generic system, which is a finite number. Such an interpretation is incorrect since
   %-----------------------------------------------------------------------
  Note that the number of ``isolated'' solutions in $(\mathbb{C}^*)^n$ is not equal to the number of solutions in $(\mathbb{C}^*)^n$.
  %which means that Bernstein's theorem may also be applied to non-generic systems such as the constant MIMO IC in multi-beam case.
For example, the system $x_1 + x_2 = 1, (x_1 + x_2)^2 = 1$ has infinitely many strictly nonzero solutions $(x_1,x_2) = (t, -t), \forall t \neq 0$,
but none of these solutions is ``isolated''.
This example shows that a finite number of \emph{isolated} solutions in $(\mathbb{C}^*)^n$ does not imply
a finite number of solutions in $(\mathbb{C}^*)^n$.
 Therefore, the version of Bernstein's theorem involves ``isolated roots'' does not imply the desired result Corollary \ref{coro: finiteness in C*n}.
 % should be understood as ``the number of solutions in $(\mathbb{C}^*)^n$ when this number is finite''.
 % A confusion may arise from the version of
Another version of Bernstein's theorem in \cite[p. 346, (5.4)]{usingAG} starts with ``Given Laurent polynomials $f_1,\dots, f_n$ over $\C$ with finitely many common zeros in $(\mathbb{C}^*)^n$'', which seems to be making the assumption that the number of solutions in $(\mathbb{C}^*)^n$ is finite \footnote{ To be more precise, the statement in \cite[p. 346, (5.4)]{usingAG} can be understood in two different ways, where the second way is that the finiteness assumption
 is made only to the non-generic case. Though algebraic geometry experts may be able to tell which way is correct by resorting to the proof of Bernstein's theorem, a reader in IA field probably can not.}.
However, we are interested in the question \emph{whether the finiteness assumption holds},  % in order to determine the infeasibility of IA systems,
which is not answered by this version of Bernstein's theorem.

Because of these possible confusions, we recommend the following version of Bernstein's theorem.
\par\noindent %
{\bf Another version of Bernstein's theorem \cite[Theorem 3.1]{Emiris1995}}\\
\emph{Consider polynomials $f_1,\dots, f_n \in \mathbb{C}[x_1,\dots, x_n]$ with Newton polytopes $P_1,\dots, P_n$ in $\mathbb{R}^n$:}
\par\noindent
(a) \emph{For generic choices of the coefficients in $f_1,\dots, f_n$, the number of common solutions of $f_1=\dots = f_n = 0$ in $(\mathbb{C}^*)^n$ equals the mixed volume $MV(P_1,\dots, P_n)$.}
\par\noindent
(b) \emph{For a specific specialization of the coefficients, the number of common solutions in $(\mathbb{C}^*)^n$ is either infinity, or does not exceed $MV(P_1,\dots, P_n)$.
}

 % while for a non-generic system, the number of solutions in $(\mathbb{C}^*)^n$ is either infinity or upper bounded by the mixed volume.
% This assumption is necessary since for non-generic systems with an equal number of equations and variables, the number of solutions could be infinite.

% \subsection{ Exploring the Number of Solutions with Zero Components  }\label{subsec: explore C^n solution}
One may ask whether Bernstein's theorem can be extended to consider the number of solutions in $\mathbb{C}^n$ as opposed to $(\mathbb{C}^*)^n$.
This motivates Corollary \ref{a sufficient condition for the counting argument} which provides conditions on when a generic overdetermined system has no solution.
We briefly discuss the connection of Corollary \ref{a sufficient condition for the counting argument} and prior art.
The extension of Bernstein's theorem to $\mathbb{C}^n$ has been studied extensively; see \cite{BKKroot,SturmCn,Roj1996} and the references therein.
However, these results usually consider the number of isolated solutions and/or assume that the system has a finite number of solutions in $\mathbb{C}^n$.
% most effort has been been made on the root counting under the assumption that
For example, \cite[Theorem 2]{SturmCn} states that the number of \emph{isolated} zeros of a generic system equals the stable mixed volume \emph{provided that
the system has finitely many roots}.
For the question of whether the finiteness assumption holds for a generic square system (i.e. the same number of equations and variables), a simple sufficient condition is provided in \cite[Lemma 5]{SturmCn}, and a necessary and sufficient condition is provided in \cite[Lemma 3]{Roj1996}.
Nevertheless, these two results can not be directly applied to prove that a generic overdetermined system has no solution under the conditions of these results.
 %Corollary \ref{a sufficient condition for the counting argument}.
% in \cite[Lemma 3]{Roj1996}, both on when a ``square system'' (i.e. with $n$ variables and $n$ equations) has a finite number of solutions in $(\C)^n$.
% which is not useful to derive DoF upper bounds as we are interested in \emph{whether the finiteness assumption holds} (again, they may be useful
% for the achievability of DoF).
 % though the difficulty seems to lie in how to easily determine whether the mixed volume is nonzero. and its extension such as the stable mixed volume).
% when the finiteness assumption holds
% However, our goal is not to determine whether the system has a finite number of solutions, but to determine whether the system has no solution.
% As we mentioned in Section \ref{subsec: correct form of Bernstein thm},
%---------------------Delete ---------------------------------------------------------------------------------
%We are interested in exploring the conditions under which an overdetermined system has no solution in $\mathbb{C}^n$, since for such systems a simple dimensionality counting will provide a DoF upper bound.
%The genericity is not enough, since a generic overdetermined system may have infinitely many solutions in $\mathbb{C}^n$ (e.g. Example 4.2).
%------------------------------------------------------------------------------------------------------
% which provides a condition on when a generic overdetermined system has no solution.
 % to show
One may argue that if a subsystem has a finite number of solutions in $\mathbb{C}^n$, then these solutions can not satisfy the remaining
polynomial equations with generic coefficients, thus the system is not solvable.
A simple counterexample to this argument is that $(x_1, x_2,x_3) = (0,0, 1)$ satisfies a polynomial equation $ a x_1 + b x_2 x_3 + c x_1 x_3 = 0$ for generic $a,b,c$.
To make this argument work, we need to assume, again, that these finite number of solutions are in $(\mathbb{C}^*)^n$.
Therefore, Corollary \ref{a sufficient condition for the counting argument} can only be derived from Bernstein's theorem, not from the previous results \cite[Lemma 5]{SturmCn} and \cite[Lemma 3]{Roj1996}.

\section{Proof of Theorem \ref{thm 1: DoF bound for generic extensions}: General Case (MIMO IC)}\label{subsec: thm 1 MIMO proof}
{\black
In this subsection, we present the proof of Theorem \ref{thm 1: DoF bound for generic extensions}. As mentioned before, the proof for the MIMO IC case is not much harder than the SISO IC case, and the major difference is that we only consider the supports of $u_k, v_k$ in SISO IC, while in MIMO IC we consider both the supports and the ``block-supports'' of $u_k ,v_k$.
}

In an $M_t \times M_r$ MIMO IC with $T$ channel extensions, the channel matrix $$ H_{kj} = diag(H_{kj}^1,\dots, H_{kj}^T) $$ is a $M_r T \times M_t T$ block diagonal matrix, where each block $H_{kj}^l = \left( H_{kj}^l(p,q) \right)$ is a $M_r \times M_t$ matrix. % with generic entries.
The IA condition is given as follows:
\begin{subequations}\label{MIMO generic IA}
\begin{align}
& u_k^H H_{kj} v_j =0, \quad \forall\ 1\leq k\neq j \leq K, \label{MIMO generic IA, part a}\\
& u_k^H H_{kk} v_k  \neq 0, \quad \forall\ 1\leq k \leq K, \label{MIMO generic IA, part b}
\end{align}
\end{subequations}
where $v_j  \in \mathbb{C}^{ M_t T \times 1}, u_k \in \mathbb{C}^{ M_r T \times 1}$ are beamformers.

Fix $M_t, M_r$  and define  $f(T)$ as
\begin{equation}\label{MIMO, definition of f(T), max K}
%\begin{split}
f(T) \triangleq \max \left\{K \;|\; (\ref{MIMO generic IA}) \text{ is solvable for a positive measure of} \left( H_{kj}^l(p,q) \right)_{1\leq k,j \leq K, 1\leq l \leq T , 1 \leq p\leq M_r, 1\leq q \leq M_t} \right\}.   %\right\}.\right. \\  \left.
%\end{split}
\end{equation}

To prove part (a) of Theorem \ref{thm 1: DoF bound for generic extensions}, we only need to prove the following bound:
\begin{equation}\label{MIMO IC bound}
f(T) \leq (M_t + M_r)T -1, \quad \forall T \geq 1.
\end{equation}
We will do so by using an induction analysis.

For the basis of the induction ($T=1$), $f(1) \leq M_t + M_r -1$ holds according to Proposition \ref{prop: constant MIMO, single-beam}. Now suppose (\ref{MIMO IC bound}) holds for any positive integer that is smaller than $T$. We will prove (\ref{MIMO IC bound}) holds for $T$.
Suppose $K$ satisfies that (\ref{MIMO generic IA}) has a solution $\{ \tilde{u}_k, \tilde{v}_k \}_{k=1,\dots, K}$ for a positive measure of $\left( H_{kj}^l(p,q) \right)$.
Denote the support of a vector $x$ as $$ \mathrm{supp}(x) \triangleq \{j \mid x_j \neq 0 \}. $$
Each $\left( H_{kj}^l(p,q) \right)$ corresponds to (at least) one collection of supports $ \{ \mathrm{supp}(\tilde{u}_k), \text{supp}(\tilde{v}_k) \}_{k=1,\dots,K}$. Since there are finitely many possible choices for the collection of supports,
 it follows that there exist a positive measure of $\left( H_{kj}^l(p,q) \right)$ which corresponds to the same collection of supports $\{R_k, S_k \}_{k=1,\dots, K}$, where $R_k \subseteq \{1,\dots, M_r T \}, S_k \subseteq \{1,\dots, M_t T \}$.
%  $\{ R_k, S_k \}_{k=1,\dots, K}$, where $R_k, S_k \subseteq \{1,\dots, T \}$.
Denote this set of $\left( H_{kj}^l(p,q) \right)$ (which has a positive measure) as $ \mathcal{H} $ .
Denote $\{ \tilde{u}_k, \tilde{v}_k \}_{k=1,\dots, K}$ as the solution of (\ref{MIMO generic IA}) for some $\left( H_{kj}^l(p,q) \right) \in \mathcal{H} $.

We divide $\tilde{u}_k$ and $\tilde{v}_k$ into $T$ blocks: $\tilde{u}_k = \begin{bmatrix}
  \tilde{u}_{k}^1 \\ \vdots \\ \tilde{u}_{k}^T
\end{bmatrix} ,  v_k = \begin{bmatrix}
  \tilde{v}_{k}^1 \\ \vdots \\ \tilde{v}_{k}^T
\end{bmatrix} $, where each block $\tilde{u}_k^l = \begin{bmatrix}
  \tilde{u}_{k1}^l \\ \vdots \\ \tilde{u}_{k M_r}^l
\end{bmatrix} \in \mathbb{C}^{M_r \times 1} ,$ $v_k^l = \begin{bmatrix}
  \tilde{v}_{k1}^l \\ \vdots \\ \tilde{v}_{k M_t}^l
\end{bmatrix} \in \mathbb{C}^{M_t \times 1} $.
%We divide $u_k$ and $v_k$ into $T$ blocks: $u_k = \begin{bmatrix}
%  u_{k}^1 \\ \vdots \\ u_{k}^T
%\end{bmatrix} ,  v_k = \begin{bmatrix}
%  v_{k}^1 \\ \vdots \\ v_{k}^T
%\end{bmatrix} $, where each block $u_k^l = \begin{bmatrix}
%  v_{k1}^l \\ \vdots \\ v_{k M_r}^l
%\end{bmatrix} \in \mathbb{C}^{M_r \times 1} ,$ $v_k^l = \begin{bmatrix}
%  v_{k1}^l \\ \vdots \\ v_{k M_t}^l
%\end{bmatrix} \in \mathbb{C}^{M_t \times 1} $.
For a vector with $T$ blocks $x = \begin{bmatrix}
  x^1 \\ \vdots \\ x^T
\end{bmatrix} $, define the block-support of $x$ as
$$ \text{B-supp}(x) = \{ t \;|\;  x^t  \neq 0 \}. $$
Obviously, the support of $x$ determines its block-support (the reverse is not true). Since each $\left( H_{kj}^l(p,q) \right) \in \mathcal{H}$ corresponds to the same collection of supports $\{ R_k,S_k \} = \{ \text{supp}(\tilde{u}_k) , \text{supp}(\tilde{v}_k)  \}$, $\left( H_{kj}^l(p,q) \right) \in \mathcal{H}$ also corresponds to the same collection of block-supports % where $\{ \tilde{u}_k ,\tilde{v}_k\}$ is the corresponding solution,
$\{ \text{B-supp}(\tilde{u}_k) , \text{B-supp}(\tilde{v}_k) \} \triangleq  \{ R_k^{\mathrm{b}},S_k^{\mathrm{b}} \} $.
Define $\Omega$ as % a set of transmitter-receiver pairs as
\begin{align}
\notag \Omega \triangleq & \left\{ (k,j) \;|\; 1\leq k\neq j \leq K, \; \text{B-supp}(\tilde{u}_k) \cap \text{B-supp}(\tilde{v}_j) = \emptyset  \right\}
            \\  = & \left\{ (k,j) \;|\; 1\leq k\neq j \leq K, \; R_k^{\mathrm{b}} \cap S_j^{\mathrm{b}} = \emptyset  \right\}.
\end{align}
 The complement of $\Omega$ in
$ \{(k,j)  \;|\; 1\leq k\neq j \leq K \}$ is
$$\Omega^c = \left\{ (k,j) \;|\; 1\leq k\neq j \leq K,  \; R_k^{\mathrm{b}} \cap S_j^{\mathrm{b}} \neq \emptyset  \right\} . $$
Furthermore, we denote $a_k, \ b_k$ as the number of nonzero blocks of $\tilde{u}_k, \tilde{v}_k$ respectively, i.e.
\begin{equation}\label{ak, bk def}
a_k \triangleq |\text{B-supp}(\tilde{u}_k)| = |R_k^{\mathrm{b}}|, \;\; b_k \triangleq |\text{B-supp}(\tilde{v}_k)| = |S_k^{\mathrm{b}}|, \quad \forall k.
\end{equation}
Since $\tilde{u}_k\neq 0, \tilde{v}_k\neq 0, \forall k$, it follows that $a_k\ge1$ and $b_k\ge1$.

We will bound $|\Omega^c |$ by Lemma \ref{claim: generic system doesn't have a no-zero-component-solution} and bound $|\Omega|$ by the induction hypothesis.
We first provide an upper bound on $|\Omega^c |$.
Since scaling does not affect the solutions of (\ref{MIMO generic IA}), we can scale each $\tilde{u}_k,\tilde{v}_k$ to make one entry of them to be one. For simplicity, we still denote the scaled version of $\{ \tilde{u}_k, \tilde{v}_k \}_{k=1,\dots, K}$ as $\{ \tilde{u}_k, \tilde{v}_k \}_{k=1,\dots, K}$.
After scaling, (\ref{MIMO generic IA, part a}) becomes a new system of polynomial equations with $K(K-1)$ equations and $ K( M_t T +   M_r T -2 )$ variables, denoted as $\mathcal{P}$. Then the system $\mathcal{P}$ has
a solution $\{ \tilde{u}_k, \tilde{v}_k \}_{k=1,\dots, K}$.

The beamforming vectors $ \tilde{u}_k, \tilde{v}_k, k=1,\dots, K$ can be concatenated to form a vector in $\mathbb{C}^{ K( M_t T +   M_r T -2 ) }$
(after discarding the $2K$ one's that are generated by scaling).
Let $J$ be the support of this concatenated vector. Then $J$ is a subset of $\{ 1,\dots, K( M_t T +   M_r T -2 )\}$ determined by the supports
of $\tilde{u}_k, \tilde{v}_k, k=1,\dots, K$ (excluding the positions of the $2K$ one's).
% Since $\tilde{u}_k$ has $a_k$ nonzero blocks,
% the size of its support is no more than $a_k M_t$ (equals $a_kM_t$ in SISO case); thus it contributes at most $ (a_k M_r - 1)$ positions to $J$. Similarly, each $\tilde{v}_k$ contributes at most $(b_k M_t -1)$ positions to $J$.
% Note that the supports of $u_k,v_k$ are different from the block-supports of $u_k, v_k$.
% determined by the block-supports of .
Therefore, the size of $J$ is upper bounded as (in SISO case, the inequality becomes equality)
\begin{equation}\label{MIMO, count J}
\begin{split}
 |J| & = \sum_k \left( |\mathrm{supp}(\tilde{u}_k)| + |\mathrm{supp}(\tilde{v}_k)| \right) - 2K  \\
     & \leq \sum_k \left(|\text{B-supp}(\tilde{u}_k)| M_r + |\text{B-supp}(\tilde{v}_k)| M_t \right) -2K \\
     & = \sum_k (a_k M_r + b_k M_t) -2K,  % \sum_k (a_k M_r -1) + \sum_k (b_k M_t -1) =
     \end{split}
\end{equation}
where the last equality follows from \eqref{ak, bk def}.

Consider the system of equations $\mathcal{P}_J$, which is obtained by restricting $\mathcal{P}$ to the subset of variables $J$.
We claim that the number of nonzero polynomial equations in $\mathcal{P}_J$ is exactly $|\Omega^c|$, i.e.
\begin{equation}\label{MIMO, count PJ}
|\mathcal{P}_J| = |\Omega^c|.
\end{equation}

For any $k\neq j$, $u_k^H H_{kj} v_j = \sum_l ( u_{k}^l)^H H_{kj}^{l} v_{j}^l = \sum_{l,p,q} (u_{kp}^l)^* H_{kl}^{l}(p,q) v_{jq}^l  $. Then we have
\begin{align}
\notag       &    u_k^H H_{kj} v_j=0  \text{ does not become a zero equation in } \mathcal{P}_{J}  \\
\notag  \Longleftrightarrow \quad      &   \exists \; l,p,q,\; \st \; (\tilde{u}_{kp}^l)^* H_{kl}^{l}(p,q) \tilde{v}_{jq}^l \neq 0    \text{ (since the support of } \{\tilde{u}_i, \tilde{v}_i\} \text{ is } J ) \\
\notag   \Longleftrightarrow \quad      &   \exists \; l,p,q, \;\st \; \tilde{u}_{kp}^l \neq 0, \tilde{v}_{jq}^l \neq 0  \\
\notag  \Longleftrightarrow  \quad & \exists  \; l \in \text{B-supp}(\tilde{u}_k) \cap \text{B-supp}(\tilde{v}_j)   \\
\notag  \Longleftrightarrow  \quad & \text{B-supp}(\tilde{u}_k) \cap \text{B-supp}(\tilde{v}_j) \neq \emptyset  \\
\notag  \Longleftrightarrow  \quad & (k,j) \in \Omega^c.
\end{align}
Therefore, $|\Omega^c|$ equals the number of nonzero equations in $\mathcal{P}_J$, which proves (\ref{MIMO, count PJ}).

 Since $\mathcal{P}_J$ has a solution in $(\mathbb{C}^*)^{|J|}$ for a positive measure of $\left( H_{kj}^l(p,q) \right) \in \mathcal{H} $, by Lemma \ref{claim: generic system doesn't have a no-zero-component-solution}, we have  % using a similar argument as the proof of (\ref{PJ bounded by J})
\begin{equation}\label{MIMO, PJ bounded by J}
 |\mathcal{P}_J|  \leq |J|.
\end{equation}
Plugging (\ref{MIMO, count J}) and (\ref{MIMO, count PJ}) into (\ref{MIMO, PJ bounded by J}), we obtain an upper bound on $|\Omega^c|$:
\begin{equation}\label{bound on Omega^c, MIMO}
|\Omega^c| \leq \sum_k (a_k M_r + b_k M_t) -2K.
\end{equation}

Next, we provide an upper bound on $|\Omega|$. Define
$$ \Omega_k = \left\{j \;| \; (k,j) \in \Omega \right\} =
\left\{ j \; | j\in \{1,\dots, K \}\backslash \{ k \}, \;  \text{B-supp}(\tilde{u}_k) \cap \text{B-supp}(\tilde{v}_j) = \emptyset   \right\}.$$
Then $ \Omega =  \bigcup_{k=1,\dots, K} \{ (k,j) \mid j \in \Omega_k \}  $ and $  |\Omega| =  \sum_{k=1}^K |\Omega_k|$.  % \bigcup_{k=1,\dots, K} \{ \Omega_k \} $.
We claim that
\begin{equation}\label{MIMO, bound Omega_k}
|\Omega_k | \leq f( T - a_k), \quad k=1,\dots, K.
\end{equation}
Without loss of generality, assume $\text{B-supp}(\tilde{u}_k) = \{1,2,\dots, a_k \}$.
Since $ \text{B-supp}(\tilde{v}_j)\cap \text{B-supp}(\tilde{u}_k) = \emptyset, \forall j\in \Omega_k$, we have
that the first $a_k$ blocks of $\tilde{v}_j$ are zero, i.e.
\begin{equation}\label{MIMO, v outside support of u equals zero}
\tilde{v}_{j}^1 =\dots =\tilde{v}_{j}^{a_k}=0, \quad \forall j\in \Omega_k.
 \end{equation}
  Consider a restriction of the beamformers and channel matrices to the lower dimensional space for $(T-a_k)$ channel uses. Specifically, define
\begin{align}
\notag   x_j =  \begin{bmatrix}
  \tilde{u}_j^{a_k+1} \\ \vdots \\ \tilde{u}_{j}^T
\end{bmatrix}, \;\;  y_j = \begin{bmatrix}
  \tilde{v}_{j}^{a_k+1} \\ \vdots \\ \tilde{v}_j^T
\end{bmatrix}, \quad \forall\ j\in \Omega_k , \\
\notag   \hat{H}_{ij} = \text{diag}\left( H_{ij}^{a_k + 1}, \dots, H_{ij}^T  \right)  , \quad \forall\ i,j \in \Omega_k .
\end{align}
 It follows from (\ref{MIMO, v outside support of u equals zero}) that $ \tilde{u}_i^H H_{ij} \tilde{v}_j = x_i^H \hat{H}_{ij} y_j , \forall\ i,j \in \Omega_k$. Since $\{ \tilde{u}_k, \tilde{v}_k \}_{k=1,\dots, K}$ satisfies the IA condition (\ref{MIMO generic IA}), we have that $\{ x_k,y_k \}_{k=1,\dots, K}$ satisfies
\begin{equation}\label{MIMO, reduced dimension IA condition}
\begin{split}
x_i^H \hat{H}_{ij} y_j = 0,    & \quad \forall\ i\neq j \in \Omega_k, \\
x_j^H \hat{H}_{jj} y_j \neq 0, & \quad \forall\ j \in \Omega_k.
\end{split}
\end{equation}
Note that (\ref{MIMO, reduced dimension IA condition}) is the IA condition for a $M_t\times M_r$ MIMO IC with $(T-a_k)$ channel extensions and $|\Omega_k|$ users.
For a positive measure of $\left( H_{kj}^l(p,q) \right)$ (in $\mathcal{H}  $), (\ref{MIMO, reduced dimension IA condition}) has a solution $\{ x_k,y_k \}_{k=1,\dots, K}$.
By the definition of $f(\cdot)$ in (\ref{MIMO, definition of f(T), max K}), we have $|\Omega_k| \leq f(T-a_k)$, which proves (\ref{MIMO, bound Omega_k}).

%Similar to (\ref{bound of f(T-a_k)}), using induction hypothesis and $f(0)=0$, we have
Since $a_k\ge1$, it follows from the induction hypothesis that $f(T-a_k) \leq (M_r+M_t)(T-a_k)-1$ when $a_k < T$. Since $f(0)=0$, we have $f(T-a_k) = 0 =(M_r+M_t)(T-a_k)$ when $a_k = T$. In summary, we have
\begin{equation}\label{MIMO, bound of f(T-a_k)}
f(T-a_k) \leq (T-a_k)(M_r + M_t), \quad k=1,\dots, K.
\end{equation}
Combining (\ref{MIMO, bound Omega_k}) and (\ref{MIMO, bound of f(T-a_k)}), we obtain
\begin{equation}\label{MIMO, Omega bound with a_k}
|\Omega_k | \leq (T-a_k)(M_r + M_t), \quad k=1,\dots, K.
\end{equation}
Similarly, we have
\begin{equation}\label{MIMO, Omega bound with b_k}
|\Omega_k | \leq (T-b_k)(M_r + M_t), \quad k=1,\dots, K.
\end{equation}
Multiplying (\ref{MIMO, Omega bound with a_k}) by $\frac{M_r}{M_t + M_r}$ and (\ref{MIMO, Omega bound with b_k}) by $\frac{M_t}{M_t + M_r}$,
and summing up them for $k=1,\dots , K$ yields
\begin{equation}\label{MIMO, final bound of Omega_k}
|\Omega_k| \leq T(M_r + M_t) - (a_k M_r + b_k M_t), \quad k=1,\dots, K.
\end{equation}
Summing up (\ref{MIMO, final bound of Omega_k}) for $k=1,\dots, K$ and applying the relation $  |\Omega| =  \sum_{k=1}^K |\Omega_k|$, we obtain
\begin{equation}\label{bound on Omega, MIMO}
| \Omega | \leq  KT(M_r + M_t) - \sum_k (a_k M_r + b_k M_t).
\end{equation}

Finally, combining the bounds of $|\Omega|$ and $|\Omega^c|$ (c.f. (\ref{bound on Omega^c, MIMO}) and (\ref{bound on Omega, MIMO})), we obtain the following bound
\begin{equation}\nonumber
              K(K-1) = |\Omega^c| + |\Omega| \leq  \sum_k(a_k M_r + b_k M_t) -2K +  KT(M_r + M_t) - \sum_k (a_k M_r + b_k M_t) ,
\end{equation}
which simplifies to $K-1 \leq -2 + T(M_r + M_t)$, or equivalently $K  \leq T(M_r + M_t) -1.$
 Thus, $f(T) \leq  T(M_r + M_t) -1 $ holds for T. This completes the induction step, so that (\ref{MIMO IC bound}) holds for any $T$, as desired.

%%%%%%%%%%%%%%%%%%%%%%%%%%%%%%%%%%%%%%%%%%%%%%%%%%%%%%%%%%%%%%%%%%%%%%%%%%%%%%%%%%%%%%%%%%%%%%%%%%%%%%%%%%%%%%%%%%%%%%%%%%%%%%%%%5
%%%%%%%%%%%%%%%%%%%%%%%%%%%%%%%%%%%%%%%%%%%%%%%%%%%%%%%%%%%%%%%%%%%%%%%%%%%%%%%%%%%%%%%%%%%%%%%%%%%%%%%%%%%%%%%%%%%%%%%%%%%
%\subsection{ Proof of Theorem \ref{thm: main result, bound on K} for $L=2$}\label{sec: proof of L=2, K bound by (LN)square}
\section{Proof of Theroem \ref{thm: main result, bound on K} for General $L$ }\label{appen: proof of Thm 2 for general L}
Similar to the proof of Theorem \ref{thm: main result, bound on K}  for $L=2$ in Section \ref{sec: proof of L=2, K bound by (LN)square}, the proof for general $L$ also consists of two steps: first,
we ``lift'' the IA condition to a higher dimensional IA condition; second, using Lemma \ref{lemma: rational count},
we prove the bound for the ``lifted'' IA condition by induction on $N$.

As mentioned in Section \ref{subsec: thm 2 analysis}, without loss of generality we can assume \eqref{N=Nt assumption}.
We first eliminate the receive beamforming vectors from the IA condition \eqref{diversity L IA} to obtain
%For the single-beam case, the IA condition (\ref{eqn: IA condition with U}) becomes
\begin{equation}\label{eqn: single beam IA condition, original}
\begin{split}
H_{kk} v_k \;\; &   \Perp \;\; \text{ span}_{j\in \{1,2,\dots, K \} \backslash \{ k\}} \left\{  H_{kj}v_j   \right\}, \quad \forall\ 1\leq k \leq K, \\
 & H_{kk} v_k \in \mathbb{C}^{N_r \times 1}\backslash \{ 0 \}, \quad \forall\ 1\leq k \leq K, \\
\end{split}
\end{equation}
where the notation $  \Perp $ signifies linear independence.
The IA condition (\ref{eqn: single beam IA condition, original}) is solvable iff the IA condition (\ref{diversity L IA}) is solvable.
Furthermore,
%We claim that
%\begin{equation}\label{change Hv to v in IA}
%v_k \in \mathbb{C}^{N \times 1}\backslash \{ 0 \}  \Longleftrightarrow H_{kk} v_k \in \mathbb{C}^{N_r \times 1}\backslash \{ 0 \}.
%\end{equation}
%If $v_k = 0$, then $ H_{kk} v_k = 0$. Consider the case $v_k \in \mathbb{C}^{N \times 1}\backslash \{ 0 \}$.
since the basis matrices $(A_1,\dots, A_L) \in \Psi$ satisfy the condition (\ref{full rank guarantee}) and $N_r \leq N$, it follows that $H_{kk}$ has full column rank. Thus the condition $H_{kk} v_k\neq 0$ can be equivalently stated as $v_k\neq 0$. %Thus (\ref{change Hv to v in IA}) is proved.
Consequently, the IA condition (\ref{eqn: single beam IA condition, original}) can be further simplified as
\begin{equation}\label{eqn: single beam IA condition}
\begin{split}
H_{kk} v_k \;\; &   \Perp \;\; \text{ span}_{j\in \{1,2,\dots, K \} \backslash \{ k\}} \left\{  H_{kj}v_j   \right\}, \quad \forall\ 1\leq k \leq K, \\
 &  v_k \in \mathbb{C}^{N \times 1}\backslash \{ 0 \}, \quad \forall\ 1\leq k \leq K. \\
\end{split}
\end{equation}

%----------------------Deleted -----------------------------------------------------
%We prove Theroem \ref{thm: main result, bound on K} for general channel diversity order $L$.
%The channel matrix
%$H_{ij} = \tau_{ij}^{1} A_1 + \dots + \tau_{ij}^L A_L$, where $A_1,A_2,\dots, A_L \in \Psi$
%are fixed $N_r \times N$ matrices and $N \leq N_r$.
%---------------------------------------------------------------------------
%the coefficients $\tau_{i,j}^l$ are generic, and

%The IA condition is the following:
%\begin{equation}\label{Appen eqn: single beam IA condition}
%\begin{split}
%H_{kk} v_k \;\; &   \Perp \;\; \text{ span}_{j\in \{1,2,\dots, K \} \backslash \{ k\}} \left\{  H_{kj}v_j   \right\}, \;\; \forall k, \\
% & v_k \in \mathbb{C}^{N \times 1}\backslash \{ 0 \}, \;\; \forall k. \\
%\end{split}
%\end{equation}
% _{i,j=1,\dots, K; l=1,\dots, L}
%

Denote $f(N)$ as the maximal $K$ such that IA is feasible; more specifically,
\begin{equation}\nonumber
 \begin{split}
f(N)      = \max \{ \;K \;|\; & \text{for a positive measure of } \left(\tau_{ij}^l\right)_{1\leq i,j \leq K, 1\leq l \leq L}, \exists \; \{v_k\}_{1\leq k \leq K}, \; \st \text{ IA condition } (\ref{eqn: single beam IA condition}) \text{ holds} \; \}.
%% = \max \{K \;|\; & \text{for a positive measure of }  \left(\tau_{ij}^l\right), \\
%%                       &\text{ the DoF tuple} (d_1,\dots, d_K) = (1,1,\dots, 1) \text{ is achievable via vector space IA } \} \\
  \end{split}
 \end{equation}
  Note that $f(N)$ depends on the receive dimension $N_r$ and the building blocks $A_1, \dots, A_L$.
Then what we need to prove is: for any $N_r$ and $A_1,\dots, A_L \in \Psi$ as defined in (\ref{full rank guarantee}), we have
\begin{equation}
f(N) \leq LN + \frac{N^2}{4}.
\end{equation}

Define a lifted IA condition
\begin{equation}\label{eqn: general L lifted IA condition}\nonumber
\begin{split}
\left( \begin{matrix} \tau_{kk}^1 v_k \\ \vdots \\ \tau_{kk}^L v_k \end{matrix} \right) \;\; &   \Perp \;\; \text{ span} \left\{  \left( \begin{matrix}
\tau_{kj}^1 v_j \\ \vdots \\  \tau_{kj}^L v_j \end{matrix}  \right) \left. \;\right| \; j\in \{1,2,\dots, K \} \backslash \{ k\}   \right\}, \forall k, \\
  v_k &  \in \mathbb{C}^{N \times 1}\backslash \{ 0 \}, \forall k .\\
\end{split}
\end{equation}
% \text{ span}_{j\in \{1,2,\dots, K \} \backslash \{ k\}} \left\{  \left( \begin{matrix} v_j
Then we define $g(N)$ as the maximal $K$ for the lifted IA condition (\ref{eqn: general L lifted IA condition}) to hold; more specifically,
\begin{equation}
g(N)  = \max \{ \;K  \;| \;  \text{for a positive measure of } \left(\tau_{ij}^l\right), \exists \;\{v_k\}_{1\leq k \leq K}, \; s.t. \text{ IA condition } (\ref{eqn: general L lifted IA condition}) \text{ holds} \; \}. \\
\end{equation}
Note that $g(N)$ does not depend on $N_r, A_1,\dots, A_L$.
%% The following lemma shows that $f(N)$ is upper bounded by $g(N)$.
\begin{lemma}\label{general L, lemma: f(N)<=g(N)}
For any $N_r, N $ and any $N_r \times N$ building blocks $A_1, \dots, A_L \in \Psi$ as defined in (\ref{full rank guarantee}), we have
\begin{equation}
f(N) \leq g(N).
\end{equation}
\end{lemma}

\emph{Proof of Lemma \ref{general L, lemma: f(N)<=g(N)}:}

We only need to prove: for any given $\left(\tau_{ij}^l \right)$, if $\{v_k\}_{k=1}^{K}$ satisfies the IA condition (\ref{eqn: single beam IA condition}),
 then $\{v_k\}_{k=1}^{K}$ satisfies the lifted IA condition (\ref{eqn: general L lifted IA condition}).
  We prove by contradiction. Assume a set of nonzero vectors $\{v_k\}_{k=1}^{K}$ satisfies the IA condition (\ref{eqn: single beam IA condition}), but does not satisfy (\ref{eqn: general L lifted IA condition}), then for some $k$ the linear independence condition does not hold, i.e.
  \begin{equation}\label{L equations of dependence IA}
 \begin{split}
  \exists \lambda_1, \dots, \lambda_K, \; s.t. \quad & \tau_{kk}^1 v_k          =   \sum_{j\neq k} \lambda_j \tau_{kj}^1  v_j,                 \\
                                                     &       \quad    \quad    \quad        \vdots                                                  \\
                                                     &  \tau_{kk}^L v_k      =       \sum_{j\neq k} \lambda_j \tau_{kj}^L v_j.
\end{split}
\end{equation}
 Left multiplying the $l$th equation of (\ref{L equations of dependence IA}) by $A_l$ and summing up these $L$ equations,
 we obtain
 $$
H_{kk}v_k = \sum_{j\neq k} \lambda_j H_{kj}v_j,
 $$
where we use the equation $H_{ij} = \tau_{ij}^{1} A_1 + \dots + \tau_{ij}^L A_L$.
This contradicts the assumption that $\{v_k \}$ satisfies the IA condition (\ref{eqn: single beam IA condition}).    Q.E.D.

Remark: It can be easily verified that $f(N) = g(N)$ for the $1\times L$ SIMO IC with $N$ constant extensions.

According to Lemma \ref{general L, lemma: f(N)<=g(N)}, we only need to prove
\begin{equation}\label{ineqn: general L, bound on g}
g(N) \leq LN + \frac{N^2}{4}.
\end{equation}
We prove (\ref{ineqn: general L, bound on g}) by induction on $N$.
For the basis of the induction, we consider the case $N = 1$. We need to prove $K \leq  LN+\frac{ N^2 }{4}  = L + \frac{1}{4}$ or equivalently $K\le L$.
Assume $K \geq  L+1$ and for a positive measure of $\left(\tau_{ij}^l\right)$, the lifted IA condition (\ref{eqn: general L lifted IA condition}) has a solution $\{v_k\}_{k=1}^K$.
Since scaling each $v_j$ by a nonzero factor $1/v_j$ does not affect the linear independence condition, we have
 \begin{align}
 & \left( \begin{matrix} \tau_{kk}^1 v_k \\ \vdots \\ \tau_{kk}^L v_k \end{matrix} \right) \;\;    {\Huge \Perp} \;\; \text{ span} \left\{  \left( \begin{matrix} \tau_{kj}^1 v_j \\  \vdots \\ \tau_{kj}^L v_j \end{matrix}  \right) \left. \;\right| \; j\in \{1,2,\dots, K \} \backslash \{ k\}   \right\}, \quad \text{for }\; k=1, \dots, K.  \notag \\
\Longrightarrow \quad & \left( \begin{matrix} \tau_{11}^1 \\ \vdots \\ \tau_{11}^L \end{matrix} \right) \;\;   {\Huge \Perp} \;\; \text{ span} \left\{  \left( \begin{matrix} \tau_{12}^1 \\ \vdots \\ \tau_{12}^L \end{matrix} \right) , \dots, \left( \begin{matrix}  \tau_{1,L+1}^1 \\ \vdots \\ \tau_{1,L+1}^L \end{matrix} \right)  \right\}.   \label{general L, N=1 independence}
\end{align}
Since there are $L$ vectors in the righthand side of the above relation and the ambient dimension is $L$, this linearly independence relation (\ref{general L, N=1 independence}) can not hold for a positive measure of $\left(\tau_{ij}^l \right)$, a contradiction. Therefore, (\ref{ineqn: general L, bound on g})
holds for $N=1$.

Suppose (\ref{ineqn: general L, bound on g}) holds for $1,\dots, N-1$. We will prove that (\ref{ineqn: general L, bound on g}) holds for $N$.
To this end, suppose $K$ satisfies that for a positive measure of  $\left(\tau_{ij}^l \right)$, $\exists \; v_k \in \mathbb{C}^{N\times 1}, k=1,\dots, K,$ such that the lifted IA condition (\ref{eqn: general L lifted IA condition}) holds.
Let us introduce the receive beamformer $u_k = \left( \begin{matrix} u_k^1 \\ \vdots \\ u_k^L \end{matrix} \right) \in \mathbb{C}^{LN\times 1} $,
 where $u_k^1, \dots, u_k^L \in \mathbb{C}^{N \times 1}$, and transform the lifted IA condition (\ref{eqn: general L lifted IA condition}) to the following zero-forcing type IA condition:
\begin{subequations}  \label{eqn: general L lifted IA with U}
\begin{align}
u_k^H   \left( \begin{matrix} \tau_{kj}^1 v_j \\  \vdots \\ \tau_{kj}^L v_j \end{matrix}  \right) &  =0
\quad\Longleftrightarrow \quad \tau_{kj}^1 (u_k^1)^H v_j + \dots + \tau_{kj}^L (u_k^L)^H v_j = 0, \quad\forall\ k\neq j,
\label{general L, lifted IA (a)} \\
 u_k^H  \left( \begin{matrix} \tau_{kk}^1 v_k \\  \vdots \\ \tau_{kk}^L v_k \end{matrix}  \right) &  \neq 0,\quad \forall\ k.  \label{general L, lifted IA (b)}
\end{align}
\end{subequations}

Define a set of transmitter-receiver pairs as
$$ \Omega = \left\{ (k,j) \;|\; 1\leq k\neq j \leq K, \; (u_k^1)^H v_j = \dots = (u_k^L)^H v_j = 0  \right\}.$$
The complement of $\Omega$ with respect to
$ \{(k,j): 1\leq k\neq j \leq K \}$ is given by
$$\Omega^c = \left\{ (k,j) \;|\; 1\leq k\neq j \leq K,  \; \exists  \; l,  \;s.t.  \; (u_{k}^l)^H v_j \neq 0  \right\} .  $$
In addition, define
$$ \Omega_k = \left\{j \;| \; (k,j) \in \Omega \right\} = \left\{ j \; | j\in \{1,\dots, K \}\backslash \{ k \}, \;  (u_k^1)^H v_j = \dots = (u_k^L)^H v_j = 0  \right\}.$$
Then $ \Omega =  \bigcup_{k=1,\dots, K} \{ (k,j) \mid j \in \Omega_k \}  $ and $  |\Omega| =  \sum_{k=1}^K |\Omega_k|$.  % \bigcup_{k=1,\dots, K} \{ \Omega_k \} $. $\Omega = \bigcup_{1\leq k \leq K} \Omega_k$.

Notice that for a positive measure of $\left(\tau_{ij}^l \right)$, there exists $\{u_k, v_k\}_{k=1,\dots, K}$ that satisfies the lifted zero-forcing type IA condition (\ref{eqn: general L lifted IA with U}).
In other words, each $\left(\tau_{ij}^l \right)$ corresponds to at least one solution $\{u_k, v_k\}_{k=1,\dots, K}$ which defines a collection of sets $\{ \Omega_k \}_{k=1,\dots, K} \subseteq \{1,\dots, K \}^K$. There are finitely many
subsets of $\{1,\dots, K \}^K$, thus there exists one collection of sets $\{ \Omega_k \}_{k=1,\dots, K}$, such that a positive measure of $\left(\tau_{ij}^l \right)$ correspond to $\{ \Omega_k \}_{k=1}^K$.
Denote this set of $\left(\tau_{ij}^l \right)$(which has a positive measure) as $\mathcal{H}_0$.
%% Each $(\tau_{ij})$ corresponds to at least one subset $\Omega \subseteq \{1,\dots, K \} \times \{1,\dots, K \}$. There are finitely many subsets of
%% $\{1,\dots, K \} \times \{1,\dots, K \}$, thus there exists at least one subset $\Omega_0$,
% such that a positive measure of $(\tau_{ij})$ correspond to $\Omega_0$. For simplicity, we still denote $\Omega_0$ as $\Omega$.

Consider the collection of sets $\{\Omega_k \}_{1\leq k \leq K}$ that corresponds to $\left(\tau_{ij}^l \right) \in \mathcal{H}_0$.
We will bound $|\Omega|$ by the induction hypothesis and bound $|\Omega^c |$ by Lemma \ref{lemma: rational count}.
Specifically, to derive an upper bound on $|\Omega|$, we use the definition of $\Omega_k$ to obtain
\begin{equation}\label{general L, ortho: U_k and V_j in Omega_k}
      \mathcal{U}_k \triangleq \text{span}\{ u_k^1, \dots, u_k^L \} \perp \mathcal{V}_k \triangleq \text{span} \{ v_j \;|\; j\in \Omega_k \}, \quad \forall \; 1\leq k \leq K,
 \end{equation}
which implies
  \begin{equation}\label{ineq: general L, sum dim of U,V bounded by N}
  \text{dim}( \mathcal{V}_k ) + \text{dim} ( \mathcal{U}_k )  \leq N, \quad \forall \; 1\leq k \leq K.
\end{equation}
Since $u_k = \left( \begin{matrix} u_k^1 \\ \vdots \\ u_k^L  \end{matrix} \right) $ is a nonzero vector, $\text{dim} ( \mathcal{U}_k ) \geq 1 $.
Then $$ p_k \triangleq \text{dim}( \mathcal{V}_k ) \leq N - \text{dim} ( \mathcal{U}_k )  \leq N-1, \quad \forall \; 1\leq k \leq K.  $$

Although all $\left(\tau_{ij}^l \right)$ in $\mathcal{H}_0$ correspond to the same collection of sets $\{\Omega_k \}_{k=1,\dots, K}$, they may correspond to different
sets of dimensions $\{ p_k \}_{k=1,\dots, K}$.
 Using a similar argument as before, we can show that there exist a positive measure of $\left(\tau_{ij}^l \right)$ in $\mathcal{H}_0$ which corresponds to the same set of dimensions $\{ p_k \}_{k=1,\dots, K}$.
These $\left(\tau_{ij}^l \right)$ form a subset of $\mathcal{H}_0$, denoted as $\mathcal{H}$.

Consider the collection of dimensions $\{ p_k \}_{k=1,\dots, K}$ that corresponds to $( \tau_{ij} ) \in \mathcal{H}$. We prove that
\begin{equation}\label{Omega k bound, general L}
|\Omega_k | \leq g(p_k) \leq L p_k + \frac{p_k^2}{4}, \quad \forall \; 1\leq k \leq K.
\end{equation}

Let $D \in \mathbb{C}^{N \times p_k}$ be the basis matrix of $ \mathcal{V}_k $, and suppose $v_j  = D \bar{v}_j, \forall\ j\in \Omega_k $,
 where $\bar{v}_j \in \mathbb{C}^{p_k}\backslash \{ 0\}$. From the lifted IA condition (\ref{eqn: general L lifted IA condition}), we have
\begin{subequations}
\begin{align}
\notag   & \left( \begin{matrix} \tau_{ii}^1 v_i \\  \vdots \\ \tau_{ii}^L v_i \end{matrix}  \right) \;\;    \Perp \;\; \text{ span}
\left\{  \left( \begin{matrix} \tau_{ij}^1 v_j \\  \vdots \\ \tau_{ij}^L v_j \end{matrix}  \right) \left. \;\right| \; j\in \Omega_k \backslash \{ i\}   \right\}, \forall\ i\in \Omega_k, \\
 \Longleftrightarrow \quad & \left( \begin{matrix} \tau_{ii}^1 D \bar{v}_i \\  \vdots \\ \tau_{ii}^L  D \bar{v}_i \end{matrix}  \right) \;\;    \Perp \;\; \text{ span}
\left\{  \left( \begin{matrix} \tau_{ij}^1  D \bar{v}_j \\  \vdots \\ \tau_{ij}^L  D \bar{v}_j \end{matrix}  \right) \left. \;\right| \; j\in \Omega_k \backslash \{ i\}   \right\}, \forall\ i\in \Omega_k,
 \label{general L, part a of induc lifted IA} \\
\Longrightarrow \quad &  \left( \begin{matrix} \tau_{ii}^1 \bar{v}_i \\  \vdots \\ \tau_{ii}^L  \bar{v}_i \end{matrix}  \right) \;\;    \Perp \;\; \text{ span}
\left\{  \left( \begin{matrix} \tau_{ij}^1   \bar{v}_j \\  \vdots \\ \tau_{ij}^L   \bar{v}_j \end{matrix}  \right) \left. \;\right| \; j\in \Omega_k \backslash \{ i\}   \right\}, \forall\ i\in \Omega_k.
\label{general L, induc lifted IA}
\end{align}
\end{subequations}

The last step can be proved by contradiction. If (\ref{general L, induc lifted IA}) does not hold, then the LHS (left-hand side) of (\ref{general L, induc lifted IA})
belongs to the space on the RHS (right-hand side) of (\ref{general L, induc lifted IA}). Multiply both sides by
 $ \left( \begin{matrix} D  & 0 & \cdots & 0 \\ 0  & D & \cdots & 0 \\ \vdots & \vdots & \ddots & \vdots \\ 0 & 0 & \cdots &  D \end{matrix} \right) \in
  \mathbb{C}^{LN \times Lp_k}$, we obtain that the LHS of (\ref{general L, part a of induc lifted IA})
belongs to the space on the RHS of (\ref{general L, part a of induc lifted IA}), which contradicts (\ref{general L, part a of induc lifted IA}).

Now for a positive measure of $( \tau_{ij}^l )_{i,j \in \Omega_k, 1\leq l \leq L }$, there exist $\bar{v}_i \in \mathbb{C}^{p_k}\backslash \{ 0\}, i\in \Omega_k $ that satisfy (\ref{general L, induc lifted IA}).
By the definition of $g(\cdot)$, $|\Omega_k | \leq g(p_k) $.
Using the induction hypothesis, we have $g(p_k) \leq Lp_k + \frac{p_k^2}{4}$.
Thus, (\ref{Omega k bound, general L}) is proved.

Summing up (\ref{Omega k bound, general L}) for $k=1,\dots, K$, we have
\begin{equation}\label{general L, estimate Omega}
|\Omega | = \sum_{k=1}^K |\Omega_k | \leq  \sum_k \left( L p_k + \frac{ p_k^2}{4} \right) .
\end{equation}

Next, we provide an upper bound on $|\Omega^c |$.
For each pair $(k,j) \in \Omega^c $, there exists $l_{kj} \in \{1,\dots, L \} $ such that $u_{k l_{kj}}^H v_j \neq 0$.
Then equation (\ref{general L, lifted IA (a)}) implies that
\begin{equation}\label{eqn: general L, tau expressed as rational}
\tau_{kj}^{l_{kj}} = - \frac{ \sum_{l \neq l_{kj} } \tau_{kj}^l u_{kl}^H v_j }{ u_{k l_{kj}}^H v_j } , \quad \forall\ (k,j)\in \Omega^c.
\end{equation}
The system of rational equations (\ref{eqn: general L, tau expressed as rational}) in variables $u_i^1,\dots, u_{i}^L,v_i, i=1,\dots, K$ (parameterized by
$\left( \tau_{kj}^l \right)$) has $|\Omega^c|$ equations.

We compute the number of free variables in $\{u_k, v_k \}_{1\leq k \leq K}$.
Since scaling $v_j  $ does not affect the system of equations (\ref{eqn: general L, tau expressed as rational}), we can scale each $v_j$ to make
one entry of it to be $1$. Therefore, the number of variables in $\{ v_i\}_{1\leq i \leq K}$ is $K(N-1)$.
%\footnote{A more rigourous argument is as follows. Suppose we scale the $t_j$'th entry of $v_{j} $ to be one. Then each $(\tau_{kj}^l)$ corresponds to a set of positions $\{ t_j\}$.
%Again, a positive measure of $(\tau_{kj}^l)$ corresponds to the same set of positions $\{t_j \}$; for these $(\tau_{kj}^l) $, (\ref{eqn: general L, tau expressed as rational}) can be transformed to a new system with $K(N-1)$ variables in $\{ v_i\}_{1\leq i \leq K}$. A similar argument can be applied to counting
%variables in $\{u_k \}$ as well.}
To count the number of free variables in $\{ u_k\}$, notice that the condition (\ref{general L, ortho: U_k and V_j in Omega_k}) implies that $u_k^1, \dots, u_k^L$ lie in $\mathcal{V}_k^{\perp}$ (the orthogonal complement of a fixed $p_k$-dim space $\mathcal{V}_k$).
Since $\text{span}\{v_j:j\in\Omega_k \}$ has dimension $p_k$, it follows that
% $p_k$ entries of $u_k^1$ can be written as linear functions of the remaining $N-p_k$ entries of $u_k^1$ (with coefficients being the rational functions of $\{v_j:j\in\Omega_k \}$). Similarly
for each $l \in \{1,\dots, L\}$, $p_k$ entries of $u_k^l$ can be represented as linear functions of the remaining $N-p_k$ entries of $u_k^l$, with coefficients being some rational functions of $\{v_j:j\in\Omega_k \}$. Substituting these linear functions into the right hand sides of \eqref{eqn: general L, tau expressed as rational} yields a new representation of each $\tau_{kj}^{l_{kj}}$ as a rational function of the $L(N-p_k)$ free variables in $u_k^l, l=1,\dots, L$ as well as the $K(N-1)$ variables in $v_j$'s.
Because of the homogeneity of these rational functions over the $L(N-p_k)$ free entries of $u_{k}$, we can further scale $u_{k} $
to make one of these entries to be $1$.
Thus, the number of free variables in $u_k$ is $L(N-p_k) - 1$.
In summary, the number of free variables in $\{u_k, v_k \}_{k=1}^K$ is
  \begin{equation}\label{var of u,v count}
  K(N-1) + \sum_k \left( L(N-p_k) - 1 \right) = K((L+1)N-2) - L \sum_k p_k.
  \end{equation}

% The number of variables in $\{u_i^1,u_i^2,v_i\}_{1\leq i \leq K}$ is $K(2N+N) = 3KN$.
For a positive measure of $\left(\tau_{kj}^l\right)_{(k,j) \in \Omega^c, 1 \leq l \leq L }$, the rational system (\ref{eqn: general L, tau expressed as rational}) has a solution $\{u_k, v_k\}_{1\leq k \leq K}$
such that $ u_{k l_{kj}}^H v_j \neq 0 $.
By Lemma \ref{lemma: rational count},
the number of equations should not exceed the number of variables (given in \eqref{var of u,v count}), i.e.
\begin{equation}\label{general L, estimate Omega^c}
|\Omega^c| \leq  K( (L+1)N-2) - L \sum_k p_k.
\end{equation}

With the bounds on $|\Omega|$ and $|\Omega^c|$, we can now provide an bound on $K$.
Summing up (\ref{general L, estimate Omega^c}) and (\ref{general L, estimate Omega}) yields
\[
 K(K-1) = |\Omega| + |\Omega^c| \leq \sum_k \left( L p_k + \frac{p_k^2}{4} \right)   + K((L+1)N-2) - L \sum_k p_k,
\]
implying
\begin{equation}
 K(K-1) \leq K( (L+1) N-2) + \sum_k \frac{p_k^2}{4} \text{  or equivalently  } K \leq (L+1)N-1 + \frac{1}{K}\sum_k \frac{p_k^2}{4}.  \label{general L, adding Omega and Omega^c}
\end{equation}
Therefore, if $p_k \leq N-2, \forall k$, then (\ref{general L, adding Omega and Omega^c}) leads to
\begin{equation}\label{eq:case1, general L}
K \leq (L+1)N -1 + \frac{1}{K}\sum_k \frac{p_k^2}{4} \leq (L+1)N - 1 + \frac{(N-2)^2}{4} = LN + \frac{N^2}{4},
\end{equation}
as desired.

It remains to consider the case $p_k=N-1$ for some $k$.
In this case we have the following claim.
\begin{claim}\label{general L, claim of bounding p_k}
If there exists $k$ such that $p_k = N-1$, then
\begin{equation}\label{if p=N-1, bound K, general L}
K \leq |\Omega_k| + L.
\end{equation}
\end{claim}
\emph{Proof of Claim \ref{general L, claim of bounding p_k}}:
 According to (\ref{ineq: general L, sum dim of U,V bounded by N}), $1\leq dim(\mathcal{U}_k) \leq N - p_k = 1$, thus $dim(\mathcal{U}_k) = 1$.
Then $u_k^1, \dots, u_k^L$ are parallel. Without loss of generality, we assume $u_k^L\neq 0$; then $\mathcal{U}_k = span\{ u_k^L\}$.
Suppose $ u_{k }^t = \gamma_t u_k^L, t=1,\dots, L-1 $.

According to the IA condition (\ref{general L, lifted IA (a)}), we have
\begin{equation}\label{expand zero forcing equation}
0= \sum_l u_{kl}^H v_j = (u_k^L)^H v_j \left( \gamma_1 \tau_{kj}^1 + \dots + \gamma_{L-1}\tau_{kj}^{L-1} + \tau_{kj}^L \right) , \; \forall\ j \in \{1,\dots, K \}\backslash\{k\} .
\end{equation}

Let $\Omega_k^c$ be the complement of $\Omega_k$ with respect to $\{1,\dots, K \}\backslash\{k\}$.
By the definition of $\Omega_k^c$, $j \in \Omega_k^c \Rightarrow (u_k^l)^H v_j \neq 0 $ for some $l$, which implies $(u_k^L)^H v_j \neq 0$.
Combing with \eqref{expand zero forcing equation}, we obtain
$$
0= \gamma_1 \tau_{kj}^1 + \dots + \gamma_{L-1}\tau_{kj}^{L-1} + \tau_{kj}^L, \; \forall j \in  \Omega_k^c.
$$
Since $(\gamma_1, \dots, \gamma_{L-1}, 1)$ can be orthogonal to at most $L-1$ (generic) vectors $\left(\tau_{kj}^1, \dots, \tau_{kj}^L \right)$,
we obtain that $|\Omega_k^c| \leq L-1$, which implies $K \leq |\Omega_k| + L.$ \QED
% By further restricting to a subset with positive measure if necessary, we can assume that $\left(\tau_{kj}^l \right) \in \mathcal{H}$
% implies any $L$ vectors $\left(\tau_{k j}^1, \dots, \tau_{k j}^L \right)$
%---------------------------Deleted ---------------------------------------------- ----------------
%we have that $\gamma_1 \tau_{kj}^1 + \dots + \gamma_{L-1}\tau_{kj}^{L-1} + \tau_{kj}^L \neq 0$ for at least $(K-1)-(L-1) = K-L$ $j$'s. Therefore, for at least $(K-L)$ $j$'s we have $ 0 = (u_k^L)^H v_j$, or equivalently, $u_k^L \perp v_j  $. Hence, $ \Omega_k = \{j \;|\; \mathcal{U}_k \perp v_j \}$ has
%at least $(K-L)$ elements, which proves (\ref{if p=N-1, bound K}).
%-------------------------------------------------------------------------------------------

To complete the induction step, suppose $p_k = N-1$ for some $k$. Using (\ref{if p=N-1, bound K, general L}) and (\ref{Omega k bound, general L}), we have
$$
K \leq |\Omega_k| + L \leq Lp_k +  \frac{p_k^2}{4} + L = L(N-1) + \frac{(N-1)^2}{4} + L < LN + \frac{N^2}{4},
$$
as desired.
Combining this with \eqref{eq:case1, general L} yields $K \leq LN + \frac{N^2}{4}$, which further implies $g(N) \leq LN + \frac{N^2}{4}$ holds for $N$.
This completes the induction step, so that (\ref{ineqn: general L, bound on g}) holds for any $N$.
Finally, combining (\ref{ineqn: general L, bound on g}) and Lemma \ref{general L, lemma: f(N)<=g(N)}, we obtain $f(N)\leq g(N) \leq LN + \frac{N^2}{4}$.

\section*{Acknowledgement}
We would like to thank Guy Bresler and David Tse for many fruitful discussions related to the subject of this paper. We are also grateful to Syed Jafar for the valuable discussions regarding the use of Bernstein's theorem.
% Prof. Gennady Lyubeznik for valuable discussions on algebraic geometry tools.

\bibliographystyle {IEEEbib}%  {IEEEtran} %
{\footnotesize
\bibliography{ref}

\begin{thebibliography}{10}

\bibitem{MadIA}
M.A. Maddah-Ali, A.S. Motahari, and A.K. Khandani,
\newblock ``Communication over {MIMO X} channels: Interference alignment,
  decomposition, and performance analysis,''
\newblock {\em IEEE Transactions on Information Theory}, vol. 54, no. 8, pp.
  3457--3470, Aug. 2008.

\bibitem{JafarIA}
V.R. Cadambe and S.A. Jafar,
\newblock ``Interference alignment and degrees of freedom of the {K}-user
  interference channel,''
\newblock {\em IEEE Transactions on Information Theory}, vol. 54, no. 8, pp.
  3425--3441, 2008.

\bibitem{Host_outerbound}
A.~Host-Madsen and A.~Nosratinia,
\newblock ``The multiplexing gain of wireless networks,''
\newblock in {\em International Symposium on Information Theory (ISIT)}, 2005,
  pp. 2065--2069.

\bibitem{RealIA}
A.~S. Motahari, S.~O. Gharan, M.-A. Mohammad-Ali, and A.~K. Khandani,
\newblock ``Real interference alignment: Exploiting the potential of single
  antenna systems,''
\newblock {\em arXiv preprint arXiv:0908.2282}, 2009.

\bibitem{GMK_MIMO}
A.~Ghasemi, A.S. Motahari, and A.K. Khandani,
\newblock ``Interference alignment for the {K} user {MIMO} interference
  channel,''
\newblock in {\em IEEE International Symposium on Information Theory
  Proceedings (ISIT)}, Jun. 2010, pp. 360--364.

\bibitem{YetisIA}
C.M. Yetis, T~Gou, S.A. Jafar, and A.H. Kayran,
\newblock ``On feasibility of interference alignment in {MIMO} interference
  networks,''
\newblock {\em IEEE Transactions on Signal Processing}, vol. 58, no. 9, pp.
  4771--4782, Sep. 2010.

\bibitem{MeisamIA}
M.~Razaviyayn, G.~Lyubeznik, and Z.-Q. Luo,
\newblock ``On the degrees of freedom achievable through interference alignment
  in a {MIMO} interference channel,''
\newblock {\em IEEE Transactions on Signal Processing}, vol. 60, no. 2, pp.
  812--821, Feb. 2012.

\bibitem{TseFeasibility}
G.~Bresler, D.~Cartwright, and D.~Tse,
\newblock ``Settling the feasibility of interference alignment for the {MIMO}
  interference channel: the symmetric square case,''
\newblock {\em arXiv preprint arXiv:1104.0888}, 2011.

\bibitem{Raz2012-IAcomplexity}
M.~Razaviyayn, M.~Sanjabi, and Z.-Q. Luo,
\newblock ``Linear transceiver design for interference alignment: Complexity
  and computation,''
\newblock {\em IEEE Transactions on Information Theory}, vol. 58, no. 5, pp.
  2896--2910, 2012.

\bibitem{TseDiversity}
G.~Bresler and D.~Tse,
\newblock ``3 user interference channel: Degrees of freedom as a function of
  channel diversity,''
\newblock in {\em Annual Allerton Conference on Communication, Control, and
  Computing (Allerton)}, Oct. 2009, pp. 265--271.

\bibitem{JafarACS}
V.~R. Cadambe, S.~A. Jafar, and C.~Wang,
\newblock ``Interference alignment with asymmetric complex signaling --
  {S}ettling the {H}{\o}st-{M}adsen--{N}osratinia conjecture,''
\newblock {\em IEEE Transactions on Information Theory}, vol. 56, no. 9, pp.
  4552--4565, 2010.

\bibitem{ShiIA}
C.~Shi, R.A. Berry, and M.L. Honig,
\newblock ``Interference alignment in multi-carrier interference networks,''
\newblock in {\em IEEE International Symposium on Information Theory
  Proceedings (ISIT)}. IEEE, 2011, pp. 26--30.

\bibitem{BengIA}
R.~Brandt, P.~Zetterberg, and M.~Bengtsson,
\newblock ``Interference alignment over a combination of space and frequency,''
\newblock in {\em IEEE International Conference on Communications Workshops
  (ICC)}, 2013, pp. 149--153.

\bibitem{BlindIA1}
S.A. Jafar,
\newblock ``Exploiting channel correlations - simple interference alignment
  schemes with no {CSIT},''
\newblock in {\em IEEE Global Telecommunications Conference (GLOBECOM)}, Dec.
  2010, pp. 1--5.

\bibitem{BlindIA2}
T.~Gou, C.~Wang, and S.A. Jafar,
\newblock ``Aiming perfectly in the dark-blind interference alignment through
  staggered antenna switching,''
\newblock {\em IEEE Transactions on Signal Processing}, vol. 59, no. 6, pp.
  2734--2744, Jun. 2011.

\bibitem{GouMIMO}
T.~Gou and S.A. Jafar,
\newblock ``Degrees of freedom of the {K} user {MIMO} interference channel,''
\newblock in {\em Asilomar Conference on Signals, Systems and Computers}, Oct.
  2008, pp. 126--130.

\bibitem{JafarAlgo}
K.~Gomadam, V.R. Cadambe, and S.A. Jafar,
\newblock ``Approaching the capacity of wireless networks through distributed
  interference alignment,''
\newblock in {\em IEEE Global Telecommunications Conference (GLOBECOM)}, Dec.
  2008, pp. 1--6.

\bibitem{Field}
P.~Morandi,
\newblock {\em Field and Galois theory},
\newblock Springer, 1996.

\bibitem{sturmfels}
B.~Sturmfels,
\newblock {\em Solving systems of polynomial equations}, vol.~97,
\newblock American Mathematical Society, 2002.

\bibitem{usingAG}
D.~A. Cox, J.~B. Little, and D.~O'Shea,
\newblock {\em Using algebraic geometry},
\newblock Springer New York, 1998.

\bibitem{gonzalez2012feasibility}
{\'O}scar Gonz{\'a}lez, Carlos Beltr{\'a}n, and Ignacio Santamar{\'\i}a,
\newblock ``On the feasibility of interference alignment for the k-user mimo
  channel with constant coefficients,''
\newblock {\em arXiv preprint arXiv:1202.0186}, 2012.

\bibitem{gonzalez2013finding}
{\'O}~Gonz{\'a}lez, I~Santamaria, and C~Beltr{\'a}n,
\newblock ``Finding the number of feasible solutions for linear interference
  alignment problems,''
\newblock in {\em Information Theory Proceedings (ISIT), 2013 IEEE
  International Symposium on}. IEEE, 2013, pp. 384--388.

\bibitem{SongIA}
S.~H. Song, X.~Chen, and K.~B. Letaief,
\newblock ``Achievable diversity gain of {K}-user interference channel,''
\newblock in {\em IEEE International Conference on Communications (ICC)}, 2012,
  pp. 4197--4201.

\bibitem{SturmCn}
B.~Huber and B.~Sturmfels,
\newblock ``Bernstein{'}s theorem in affine space,''
\newblock {\em Discrete $\&$ Computational Geometry}, vol. 17, no. 2, pp.
  137--141, 1997.

\bibitem{BKKroot}
T.~Li and X.~Wang,
\newblock ``The {B}{K}{K} root count in $\mathbb{C}^n$,''
\newblock {\em Mathematics of Computation of the American Mathematical
  Society}, vol. 65, no. 216, pp. 1477--1484, 1996.

\bibitem{dickenstein2005solving}
A.~Dickenstein and I.Z. Emiris,
\newblock {\em Solving polynomial equations},
\newblock Springer, 2005.

\bibitem{Emiris1995}
I.Z. Emiris and J.F. Canny,
\newblock ``Efficient incremental algorithms for the sparse resultant and the
  mixed volume,''
\newblock {\em Journal of Symbolic Computation}, vol. 20, no. 2, pp. 117--149,
  1995.

\bibitem{Roj1996}
J.~M. Rojas and X.~Wang,
\newblock ``Counting affine roots of polynomial systems via pointed newton
  polytopes,''
\newblock {\em Journal of Complexity}, vol. 12, no. 2, pp. 116--133, 1996.

\end{thebibliography}
}

%%--------------Bibliography------------------------------------------------------------------
%\begin{thebibliography} {Tn99}
%\normalsize
%\bibitem{JafarIA} Jafar IA.
%
%
%%\bibitem{} J. Wright, A. Ganesh, S. Rao, and Y. Ma. Robust principal component analysis: Exact recovery
%% of corrupted low-rank matrices. arXiv:0905.0233, 2009.
%
%\end{thebibliography}

\end{document}